\documentclass[useAMS,usenatbib]{mn2e}
\usepackage{epsfig}
\usepackage{graphicx}
\usepackage{amsmath}

\title[The confluence of the outer 
accretion disc with the inner edge of the dusty torus.]{The broad emission-line
  region: the confluence of the outer 
accretion disc with the inner edge of the dusty torus.}
 \author[Goad, Korista and Ruff]{M. R. Goad$^{1}$\thanks{E-mail: mrg@star.le.ac.uk} and
 K. T. Korista$^{2}$, A. J. Ruff$^{3}$.  \footnotemark[1]\\
$^{1}$Department of Physics and Astronomy, 
College of Science and Engineering, University of Leicester,  University Road, Leicester, LE1 7RH\\
$^{2}$Department of Physics, Western Michigan University,  Kalamazoo, Michigan 49008-5252, USA\\
$^{3}$ School of Physics, University of Melbourne, Parkville, VIC 3010, Australia}
\begin{document}
\date{Accepted 2012 July 26; Received 2012 July 26; in original form May 2012}
\pagerange{\pageref{firstpage}--\pageref{lastpage}} \pubyear{2012}
\maketitle
\label{firstpage}
\begin{abstract}
We have investigated the observational characteristics of a class of
broad emission line region (BLR) geometries that connect the outer
accretion disc with the inner edge of the dusty toroidal obscuring
region (TOR). We suggest that the BLR consists of photoionised gas of
densities which allow for efficient cooling by UV/optical emission
lines and of incident continuum fluxes which discourage the formation
of grains, and that such gas occupies the range of distance and scale
height between the continuum-emitting accretion disc and the dusty
TOR. As a first approximation, we assume a population of clouds
illuminated by ionising photons from the central source, with the
scale height of the illuminated clouds growing with increasing radial
distance, forming an effective surface of a "bowl".  Observer lines of
sight which peer into the bowl lead to a Type 1 Active Galactic Nuclei
(AGN) spectrum. We assume the gas dynamics are dominated by gravity,
and we include in this model the effects of transverse Doppler shift,
gravitational redshift and scale-height dependent macro-turbulence.

Our simple model reproduces many of the commonly observed phenomena associated
with the central regions of AGN, including : (i) the shorter than expected
continuum--dust delays (geometry), (ii) the absence of response in the core of
the optical recombination lines on short timescales
(geometry/photoionisation), (iii) an enhanced red-wing response on short
timescales (GR and TDS), (iv) the observed differences between the delays for
high- and low-ionisation lines (photoionisation), (v) identifying one of the
possible primary contributors to the observed line widths for near face-on
systems even for purely transverse motion (GR and TDS), (vi) a mechanism
responsible for producing Lorentzian profiles (especially in the Balmer and
Mg~{\sc ii} emission-lines) in low inclination systems (turbulence), (vii) the
absence of significant continuum--emission-line delays between the line wings
and line core (turbulence; such time-delays are weak for virialised motion,
and turbulence serves to reduce any differences which may be present), (viii)
associating the boundary between population A and population B sources
(Sulentic et~al. 2000) as the cross-over between inclination dependent
(population A) and inclination independent (population B) line profiles
(GR+TDS), (ix) provides a partial explanation of the differences between the
emission-line profiles, here explained in terms of their line formation radius
(photoionisation and/or turbulence), and (x) the unexpectedly high (but
necessary) covering fractions (geometry).

A key motivation of this work was to reveal the physical underpinnings of the
reported measurements of supermassive black hole (SMBH) masses and their
uncertainties. We have driven our model with simulated continuum light-curves
in order to determine the virial scale factor $f$, from measurements of the
simulated continuum--emission-line delay, and the width ($fwhm$, $\sigma_{l}$)
and shape ($fwhm/\sigma_{l}$) of the {\em rms\/} and {\em mean\/} line
profiles for the energetically more important broad UV and optical
recombination lines used in SMBH mass determinations. We thus attempt to
illuminate the physical dependencies of the empirically determined value of
$f$. We find that SMBH masses derived from measurements of the $fwhm$ of the
mean and rms profiles show the closest correspondence between the emission
lines in a single object, even though the emission line $fwhm$ is a more
biased mass indicator with respect to inclination. The predicted large
discrepancies in the SMBH mass estimates between emission lines at low
inclination, as derived using $\sigma_{l}$, we suggest may be used as a means
of identifying near face-on systems. Our general results do not depend on
specific choices in the simplifying assumptions, but are in fact generic
properties of BLR geometries with axial symmetry that span a substantial range
in radially-increasing scale height supported by turbulence, which then merge
into the inner dusty TOR.

\end{abstract}

\begin{keywords}
methods : numerical -- line : profiles -- galaxies : active -- quasars : emission lines
\end{keywords}

\section{Introduction}

% Large scale spectroscopic surveys and 

Intensive multi-wavelength monitoring campaigns of a handful of individual AGN
have radically altered our view of the broad emission-line region (hereafter,
BLR).  Correlated multi-wavelength continuum and emission-line variations
(reverberation mapping, hereafter RM) reveal that the BLR is small in size
($R_{\rm BLR} \propto L_{uv}^{0.5}$), and shows strong gradients in density
and/or ionisation parameter (e.g., Clavel et~al.\ 1991; Peterson et~al.\ 1991;
Krolik et~al.\ 1991). The gas dynamics are largely dominated by the central
super-massive black hole, a realisation which when coupled with estimates of
the BLR size and gas velocity dispersion, have enabled the determination of
virial black hole masses in $\approx40$~ nearby AGN (Peterson 2010).  The
derived (geometry dependent) virial scale factors appear approximately
constant amongst the emission lines in individual sources, and over many
seasons for which the mean source luminosity can differ considerably (see
Krolik et~al.\ 1991; Peterson \& Wandel 1999, 2000; Peterson et~al.\ 2004;
Kollatschny 2003; Bentz et~al.\ 2007; Denney et~al.\ 2010; Peterson 2010, and
references therein), consistent with a virialised velocity field within an
ionisation stratified BLR. Further, the mass estimates compare favourably (to
within a factor of a few) with those determined from stellar velocity
dispersion estimates in nearby AGN (e.g. Ferrarese et~al. 2001), and
importantly allow scaling relations (e.g., BLR size vs.\ luminosity) to be
derived enabling characterisation of black hole masses in AGN with only single
epoch spectra (e.g., see Park et~al.\ 2012).

A key result of previous RM campaigns is that the recovered response functions
for the optical recombination lines {\em do not reach a maximum at zero
delay\/} as might be expected for a spherically symmetric distribution of line
emitting gas (see e.g. Horne, Welsh \& Peterson 1991; Ferland et~al.\ 1992;
Horne, Korista \& Goad 2003; Bentz et~al.\ 2010), which when taken at
face-value suggests that {\em the BLR gas lies well away from the line of
sight to the observer\/}, prompting a number of geometry dependent
interpretations (for example, see Manucci, Salvati and Stanga 1992; Chiang and
Murray 1996;
Murray  and Chiang 1997; Bottorff et~al.\ 1997). Alternative explanations
%, as in, for example the ``nest'' model of Manucci, Salvati and
%Stanga 1992; the (thick) disc wind model of Murray and Chiang (1996), or the
%hydromagnetically-driven out-flow (e.g. Emmering, Blandford, \& Shlosman 1992)
%of Bottorff et~al.\ (1997) which who adopted a bowl-like geometry to explain
%the reverberation of the C~{\sc iv} broad emission line profile in NGC~5548
%(Korista et~al.\ 1995)\footnote{The response of C~{\sc iv} to the continuum
%variations was only partially temporally resolved during the 1993 {\em HST}
%campaign.}. 
%Plausible alternatives 
include optically-thick line emission (see e.g. Ferland, Shields, \& Netzer
1979; Ferland et~al.\ 1992; Goad 1995, Ph.D. thesis; O'Brien, Goad \&
Gondhalekar 1994) or the presence of low-responsivity gas in the inner BLR
(e.g. Sparke 1993; Goad, O'Brien \& Gondhalekar 1993). To break the model
degeneracy requires determination of $\Psi(v,\tau)$ (Welsh \& Horne 1991; Goad
et~al.\ 1995, Ph.D.\ thesis; Horne et~al.\ 2004).  Initial attempts were
limited to determining the ``time-averaged response'' as a function of several
bins in line of sight velocity within the Balmer, He~{\sc i} $\lambda$5876 and
He~{\sc ii} $\lambda$4686 emission line profiles (see e.g., Ulrich \& Horne
1996; Kollatschny \& Bischoff 2002; Kollatschny 2003; Kollatschny \& Zetzl
2010; Denney et~al.\ 2009, 2010; Bentz et~al.\ 2010a) with somewhat mixed
results.  However, improvements in data quality and the continued refinement
of inversion techniques have now begun to reveal structure in the 2-d response
functions of the optical recombination lines in Arp~151 (Bentz et~al.\ 2010b)
and NGC~4051 (Denney et~al.\ 2011). For Arp~151, $\Psi(v,\tau)$ shows that the
red-wing responds first and, significantly, {\em a deficit in response in the
core of the lines at short time-delays\/}.  These data have been used to
exclude models with radial outflow, as well as flattened discs and thick
spherical shell geometries, and demonstrate consistency with a warped disc
geometry.  On the other hand, Denney et~al. (2011) suggest that in NGC~4051,
$\Psi(v,\tau)$ for H$\beta$ is consistent with a disc-like BLR geometry with
the response weighted toward larger BLR radii, and possibly a scale-height
dependent turbulent component (\S\ref{turb_sec}).

%with claims for both radial (inflows and outflows) and virialised gas motion,
%and in some cases evidence for both types of motion at different times within
%a single source.

\subsection{The BLR outer radius}

%Recently, most effort has focused on tying down the  BLR outer radius.
%In the context of the unified model of AGN (Antonucci 1993; Urry \&
%Padovani 1995), the broad emission line region lies between the inner
%accretion disc responsible for the optical-UV-X-ray continuum (R
%$\sim$ 100~$R_{g}$), where $R_{g} \equiv GM_{BH}/c^{2}$, and the dusty
%obscuring torus (R $>$ 20,000~$R_{g}$). 
At high incident continuum photon fluxes, dust grains cannot survive and
atomic emission lines dominate the cooling of the gas.  Conversely, at low
incident fluxes heavy elements condense into grains which absorb much of the
incident continuum flux and become an important cooling agent of the gas,
leading to a diminution of the line emission.  Thus, as was first suggested by
Netzer \& Laor (1993), the outer edge of the BLR is likely largely determined
by the distance at which grains can survive (see also Nenkova et~al.\ 2008;
Mor \& Trakhtenbrot 2011; Mor \& Netzer 2012), marking the inner boundary of
the ``toroidal obscuring region'' (hereafter, TOR). Several studies have
concentrated on determining the distance to the hot dust, from measurements of
the delay between the UV/optical continuum and near-IR ($\sim$2~$\mu$m)
thermal emission (e.g., Minezaki et~al.\ 2004; Suganuma et~al.\ 2004, 2006;
Yoshii et~al.\ 2004; Kishimoto et~al.\ 2007; Koshida et~al.\ 2009). These
reveal that $\tau_{\rm dust}\propto L^{0.5}$, and is uncorrelated with black
hole mass.  However, surprisingly $\tau_{\rm dust}$ is a factor of $\sim$2--3
smaller than expected from the grain sublimation radius (Barvainis 1987;
Nenkova 2008; Suganuma et al. 2006; Kishimoto et~al.\ 2007) based on the
central source luminosity and the assumption that $\rm{R_{sub} \approx
c\tau_{dust}}$. For example, in NGC~5548 $c\tau_{\rm dust}\approx
50$~light-days (Suganuma et~al.\ 2004; 2006), to be compared with an expected
sublimation radius of $\sim$150 light days (Nenkova et~al.\ 2008; Eq.~1).
When compared with the lag measurements for the broad C~{\sc iii}] and Mg~{\sc
ii} emission-lines in this source 28 and 34--72 days (with significant
uncertainties), respectively (Clavel et~al.\ 1991), places the outer regions
of the BLR in the vicinity of the inner edge of the dusty torus.

Hu et~al.\ (2008), Zhu et~al.\ (2009), Shields et~al.\ (2010), and Mor \&
Netzer (2012) suggest that the outer BLR and inner torus overlap in an
``intermediate'' line region (ILR).
%Hu et~al.\ (2008a,b) and Ferland et~al.\ (2009) suggest that
%much of the optical Fe~{\sc ii} emission and part of the Balmer lines are
%emitted in an ILR containing clouds of large column density undergoing infall.
In addition, Landt et~al.\ (2011;conf.proc.), adopting an isotropic,
virialised velocity field for the BLR and black holes masses inferred from
reverberation campaigns, used their sample of type~1 AGN with IR spectra
(Landt et~al.\ 2008) to show that the outer edge of the BLR scales as
$\rm{L^{0.5}}$, and corresponds to incident continuum photon fluxes that are
consistent with that expected at or just interior to the hot dust radius.

Recent observational and theoretical studies of the TOR (e.g., H\"{o}nig
et~al.\ 2006; Tristam et~al.\ 2007; Nenkova et~al.\ 2008; Ramos Almeida
et~al.\ 2009) find that it is most likely clumpy, consisting of dusty
clouds. Given that there must exist a reservoir of gas feeding the central
accretion disc, it is natural to assume that the BLR consists of gas with a
range of density and incident continuum flux which maximises atomic line
emission over atomic continuum emission (the inner accretion disc) and grain
emission (TOR). The BLR might then be expected to span between the inner
accretion disc and the TOR both in radius and in scale height.

\subsection{Geometry-dependent time-delays}

If interpreted in terms of a characteristic ``size'' for the dusty torus, then
such small delays pose severe problems both for grain survival, requiring both
larger and more robust grains (e.g., graphite; Mor \& Trakhtenbrot
2011)\footnote{Even graphite grains with sublimation temperatures of
$\approx$1700K with typical sizes of 0.1 microns would have a sublimation
distance of $\approx 110 (L_{bol}/10^{44.3})^{0.5}$ light-days (Nenkova
et~al.\ 2008).} or significant shielding.  Furthermore, the reduction in
radius that this implies for the outer BLR poses a severe problem for (the
already large) BLR gas covering fractions (see e.g. Kaspi \& Netzer 1999 and
Korista \& Goad 2000) required by photoionisation models of the broad
emission-lines in this source, the outer radius of which was chosen to be
roughly coincident with the distance at which hot dust can survive.
% ($\sim 100 (L_{ion}/10^{44} erg~s^{-1})^{0.5}$ light-days).

The simplest way of reconciling the dust-delays with the expected dust
formation radius is to invoke a geometry for the dust which departs from
spherical symmetry as was recently proposed by Kawaguchi \& Mori (2010, 2011;
see also Liu \& Zhang 2011).  In their model anisotropic continuum radiation
(see e.g. Netzer 1987; but see Nemmen \& Brotherton 2010) acts through grain
sublimation to impose a strong polar angle dependence on the distance at which
grains can form (smaller at large polar angles), carving the TOR into a
bowl-shaped geometry of significant scale-height $\rm{H/R_x} \approx 1$, where
$\rm{R_x}$ is measured along the roughly coincident mid-planes of the
accretion disc and the dusty TOR. This places the dusty gas nearer to the line
of sight for observers with typical viewing angles of 10--40 degrees with
respect to the polar axis (for observers of type~1 AGN), and more distant from
the central continuum source (R), than a comparable ring of dust of radius
$\rm{R_x}$ lying within the mid-plane (for the same time-delay)\footnote{Such
a ring would have an observed delay of $\rm{R_x/c}$ for an observer with a
line of sight near the polar axis.}, producing dust-delays that are
substantially smaller than $\rm{R_{sub}}$ as computed from the observed
continuum.  Curiously, missing from their model is the BLR, which in the
context of the unified model of AGN (Antonucci 1993; Urry \& Padovani 1995),
lies somewhere between the inner accretion disc responsible for the
optical-UV-X-ray continuum (R $\la$ 100~$R_{g}$), where $R_{g} \equiv
GM_{BH}/c^{2}$, and the dusty obscuring torus (R $>$ 20,000~$R_{g}$). A BLR
with a significant scale height would cut off much of the interior of their
model dusty torus from view of the incident continuum, obviating the need to
minimise contributions to the hot dust response from very short time delays.
It would also relegate most of the hot dust emission to dusty gas having
greater scale heights than and lying at distances somewhat further from the
central light source than the outer BLR, with a covering fraction smaller than
one might infer from the overall structure of the TOR.  Landt et~al.\ (2011b)
and Mor \& Trakhtenbrot (2011) deduced covering fractions (typically
$\la$~10\% and $\approx 13$\%, respectively) for the gas emitting the thermal
emission from the hot grains, smaller than expected from type~1/type~2
statistics (Schmitt et~al.\ 2001; Hao et~al.\ 2005) and simple geometrical
considerations. The hot grains might lie along the top rim of a bowl-like
geometry of clouds constituting the BLR and the dusty TOR.

Spatially-resolved near-IR imaging have largely confirmed the above basic 
picture of the TOR (Nenkova et~al.\ 2008) -- clumpy with significant scale 
height (Krolik \& Begelman 1988) and interferometric ring radii that are 
typically a factor of $\sim$2 or so larger than $\rm{c\tau_{dust}}$ 
(H\"{o}nig et~al.\ 2010; Kishimoto et~al.\ 2007,2009a,b,2011; 
Raban et~al.\ 2009; Tristram et~al.\ 2007, 2009).  

%(ie. an anisotropic continuum source), the actual form of which is highly 
% uncertain\footnote{Compare the
%dimming function adopted by Kawaguchi \& Mori (2010,2011; Netzer 1987) with
%the milder forms presented in Nemmen \& Brotherton (2010) who used non-LTE
%accretion disc spectral models, which included detailed radiative transfer and
%relativistic effects (Hubeny et~al.\ 2000).}. 

\subsection{Profile shape as a discriminator of BLR geometry}

For nearby AGN a comparison between their RM black hole mass estimates and
those derived from their stellar velocity dispersion, indicate significant
differences between the virial scale factors of the two broad emission lines
often used, being smaller for H$\beta$ than for C~{\sc iv} (Decarli
et~al. 2008). Decarli et~al.\ (2008) argue that the smaller virial scale
factor for C{\sc iv} indicates a more flattened distribution for the gas
producing this line, while H$\beta$ originates in a more spherical
distribution, or equivalently a region with larger scale height (e.g., a thick
disc).  Sulentic et~al. (2000) suggested that the absence of a correlation
between the line widths of C~{\sc iv} and H$\beta$ in individual sources,
together with observed differences in their line profile shapes,
$fwhm/\sigma$, indicate that C~{\sc iv} is produced in a more flattened
configuration, while H$\beta$ originates from a region with significant scale
height. The absence of a correlation in their respective line widths, is then
explained in terms of an additional scale height-dependent turbulent velocity
component for H$\beta$ (see also Collin et~al. 2006). Significantly, recent
dynamical modelling of the nearby Sy 1 Mrk~50 by the Lick AGN Monitoring
Project (LAMP), is suggestive of a thick disc geometry for the H$\beta$ line
emitting region (Pancoast et al. 2012).

Fine et~al.\ (2008, 2010, 2011) provide further evidence against a
pure-planar/disc-like geometry for the line emitting gas. By combining line
velocity dispersion data from four large quasar spectroscopic surveys they
found that the dispersion in the distribution of broad-line widths is smaller
for C~{\sc iv} than Mg~{\sc ii}, and unlike Mg~{\sc ii} is only weakly
correlated with source luminosity.  While differences in the line widths of
these two lines are consistent with ionisation stratification within a
virialised gas, the small dispersion in the distribution of the line widths of
both lines argues against a pure-planar/disc-like geometry for the line
emitting gas.

%Line profile shape has also be used to discriminate between BLR geometries.
%Sulentic et~al. (2000) in the largest spectroscopic study of its kind,
%categorised AGN according to their measured line widths. Narrow (fwhm $<$ 4000
%km/s) Population A sources tend to display Lorentzian profiles with extended
%line wings and narrow cores, while the broader (fwhm$>$ 4000~km/s) population 
%B sources have profiles which are more Gaussian in appearance (see also
%Marziani et~al. 2010). Sulentic et~al. 2000 and Collin et~al. 2006 suggest 
%that differences between the 2 populations arise due to differences in 
%orientation and or Eddington rate, with Pop A sources comprising low mass
%high accretion rate sources or sources at relatively low line of sight 
%inclination.

In a separate study, Kollatschny and Zetzl (2011) modelled the broad H$\beta$
and C~{\sc iv} emission-line profiles of AGN with RM data, by convolving a
rotational velocity component representing the gravitational potential of the
central super-massive black hole, with a Lorentzian component, which they
associated with turbulence. While for a given line, the degree of turbulence
remains similar from one object to the next, they found that their model fits
require a larger turbulent velocity component for C~{\sc iv} than for
H$\beta$. Assuming that the magnitude of the turbulent component increases
with increasing scale-height, this argues for a larger scale height for C~{\sc
iv}, in stark contrast to the results above.

\subsection{Summary and a proposed geometrical model of the BLR}

There is substantive evidence that the BLR geometry is neither very flat nor
spherical, with predominantly virialised gas dynamics, and is largely absent
from regions lying near to our privileged line of sight down into the opening
of the TOR in the unified model of AGN. Nenkova et~al.\ (2008) suggest that
what divides the TOR from the BLR is the ability of grains to survive or not
within clouds moving through the high radiation energy density environments
found near the central engines of AGN. Following Nenkova et~al.\ (2008), we
propose a model in which the BLR bridges the gap between the inner
continuum-emitting accretion disc and the hot grain emitting inner TOR (Netzer
\& Laor 1993), both in spatial distance from the super-massive black hole and
also in characteristic scale height that grows with increasing distance.
Recent results from dynamical mass estimates for the mass of the central black
hole in the nearby Seyfert~1 galaxy Mrk~50 (Pancoast et~al. 2012), appear to
favour dynamical models in which the BLR has substantial scale-height (ie. a
thick disc geometry).  Together the three structures present an {\em
effective} surface of a bowl-like geometry -- roughly flat on the bottom and
scale height increasing with distance $R_x$.  The inner accretion disc and BLR
form the bottom and lower sides of the bowl, while the gas containing the hot
dust is then found on the upper reaches of the bowl.  A similar configuration
has also been suggested recently by Gaskell (2009).

In this work we explore the observational characteristics of bowl-shaped BLR
geometries comparing our detailed model calculations with observations taken
from the literature. The outline of this work is as follows. In section 2 we
present generic toy bowl-shaped BLR models, illustrating the dependence of the
steady-state 2-d and 1-d response functions and emission-line profiles on the
key parameters describing the model. In \S\ref{fiducial_model} we outline a
fiducial BLR geometry the parameterisation of which has been chosen to to be
broadly representative of the conditions within the well-studied Seyfert~1
galaxy NGC~5548.  Our model includes a treatment of the transverse Doppler
shift, gravitational redshift and scale-height dependent macroscopic
turbulence. In \S\ref{simulations} we present the results from photoionisation
model calculations of the radial surface emissivities for four broad emission
lines commonly used in measuring AGN black hole masses. We also describe the
model continuum light-curves, chosen to match the continuum variability
behaviour of NGC~5548, and used to drive our fiducial BLR geometry to produce
time-variable emission-line profiles and light-curves. The results from our
simulations are presented in \S\ref{results}. In \S\ref{discussion} we compare
the results of our simulations with observations presented in the literature.

\begin{figure}
%\resizebox{\hsize}{!}{\includegraphics[angle=0,width=8cm]{geom.eps}}
\resizebox{\hsize}{!}{\includegraphics[angle=0,width=8cm]{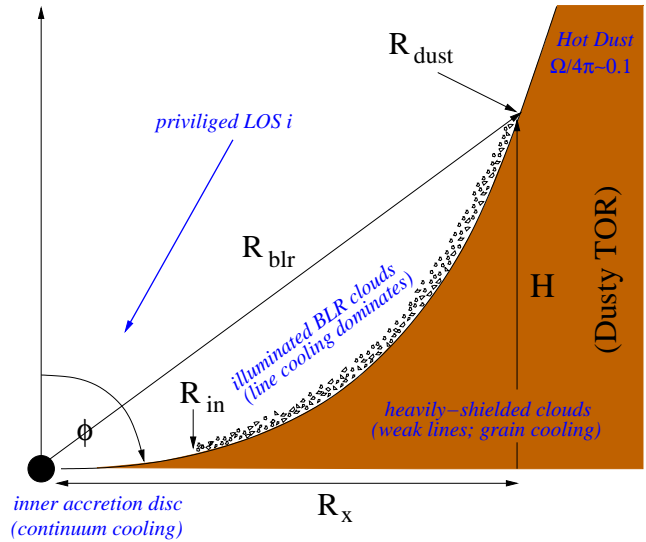}}
\caption{An illustration of the Bowl-shaped geometry. The BLR clouds occupy a
region where gas is strongly illuminated by the central continuum source and
line cooling dominates over either continuum cooling and grain cooling,
effectively bridging the region between the outer accretion disc and the hot
dust. The quoted hot dust covering fraction ($\Omega/4\pi\sim 0.1$) is taken
from Landt et~al. 2011b.}
\label{bowl}
\end{figure}

%%%%%%%%%%%%%%%%%%%%%%%%%%%%%%%%%%%%%%%%%%%%%%%%%%%%%%%%%%%%%%%%%%%%%%

\section{A Bowl-shaped BLR Geometry}

In order to investigate the properties (gas distribution, kinematics, 1-d and
2-d response, profile shape etc) of bowl-shaped geometries as a class, we
begin by first constructing a very simple toy model. For the purposes of
illustrating the concept, we initially set the inner radius at 1~lt-day and
outer radius at 10 lt-days and assume that the clouds lying near the surface
of the bowl radiate isotropically, and do not cover one another (ie., have a
direct view of the central continuum source). We parameterise the shape of the
bowl in terms of the BLR scale height $H$ (in units of lt-days), such that

\begin{equation}
H=\beta R_{\rm x}^{\alpha} \, ,
\end{equation}

\noindent

\noindent where $\beta$ and $\alpha$ are constants, $R_{\rm x}$ is the
projected radial distance (in lt-days) along the plane perpendicular to the
observers line of sight, ie. $R_{x} = R(\sin \phi)$, where $\phi$ is the polar
angle (see figure~\ref{bowl}).  In the limiting case of zero scale-height, the
normalisation constant $\beta$ is zero by definition, and the material lies
along the mid-plane of the disc.  For fixed (non-zero) $\beta$, the shape of
the bowl is determined by the power-law index $\alpha$. The bowl is convex for
$\alpha<1$, concave for $\alpha>1$, and cone-shaped for $\alpha=1$
(figure~2a). Note that $\phi$ is a function of radial distance $R$ unless
$\alpha=1.0$, for which $\phi={\rm constant}$. When viewed along the axis of
symmetry (down into the bowl), the spread in time-delays is minimised for
fixed $\beta$, when $\alpha=2$, i.e. parabolic. For $\alpha>2$, gas lying near
the surface of the bowl at larger radii will respond {\em before\/} gas at
small radii (figure~2b). Since the time-delay, $\tau$, at radial distance
$R$ is $\tau=R-H$, this condition will be met provided $\delta H > \delta R$
(or equivalently $d\tau/dR$ is negative). Note that in our model, we only
observe the inner surface of the bowl, the outer surface being largely
obscured from sight (since such sight lines would have to pass through the
dusty TOR, see e.g. figure~1).

\begin{figure}
%\resizebox{\hsize}{!}{\includegraphics[angle=0,width=8cm]{plot_bowl.ps}}
% plot_data.ps
\resizebox{\hsize}{!}{\includegraphics[angle=0,width=8cm]{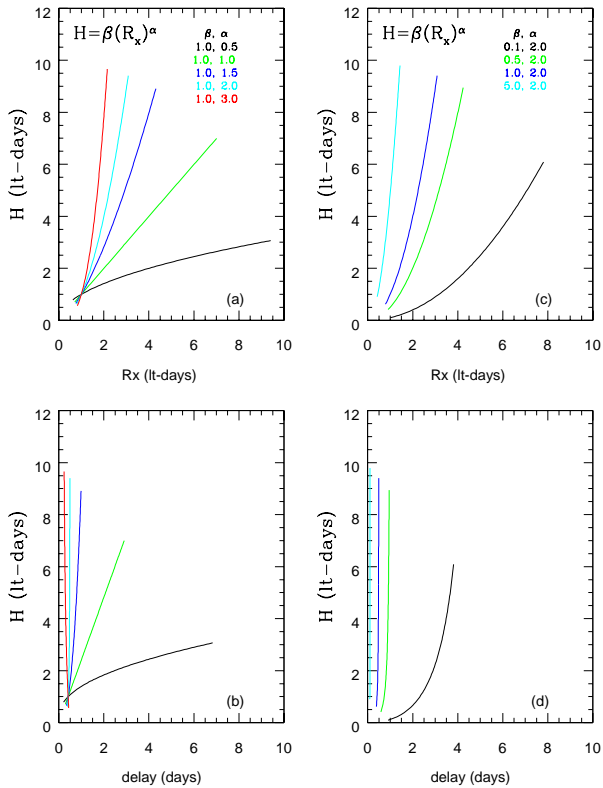}}
\caption{Bowl-shaped geometries with inner radius $R_{in} = 1$ lt-day, outer
radius $R_{out} = 10$ lt-days, inclination $i=0$, and scale height $H$
(lt-days) and indicating (i) the shape (upper panels), and (ii) the range in
expected delays (lower panels),
for fixed $\beta$ (figures~2a, 2b), and fixed $\alpha$ (figures 2c, 2d).}
\label{shape_delay}
\end{figure}

The toy models discussed above, and illustrated in figure~2 serve as a tool
for exploring how specific combinations of $\alpha$, $\beta$ determine the
bowl shape, and the expected spread in time-delay for a fixed $R_{out}$.
Since the formulism described leads to an infinite number of bowl shaped
geometries, we restrict the parameter space further, by fixing both the BLR
inner and outer radius ($R_{in}$, $R_{out}$), and the expected delay at the
outer radius, $\tau(R=R_{out})$ (for $i=0$).  Then, for a given choice of
$\alpha$, $\beta$ is adjusted to match the specified delay at the outer
boundary. In so doing, we can examine the properties of families of
bowl-shaped geometries, with fixed radial extent, and with shapes
differentiated by their chosen value of $\alpha$.

\subsection{The BLR velocity field}

\noindent For the velocity field, we assume that the gas motion is dominated
by the gravitational potential well in which it sits, and is largely circularised,
as might be expected in the presence of significant dissipative forces, such that:

\begin{equation}
v_{kep}^{2} = K \frac{ R_{\rm x}^{2} } { ( R_{\rm x}^{2}+\beta^{2} R_{\rm
    x}^{2\alpha} )^{3/2} } \, ,
\end{equation}

\noindent where $v_{kep}$ is the local Keplerian velocity, $K=GM_{\rm BH}$,
and $M_{\rm BH}$ is the mass of the central black hole. For the limiting case
of a geometrically thin disc (ie. $\beta=0$), $v_{kep}^{2} = K\frac{1}{R}$,
and the velocity field reduces to that expected for simple planar Keplerian
orbits. Note that in this formulism, there is no radial component to the
velocity field.  Significant radial motion, either in the form of turbulence,
or bulk radial motion may be introduced by the addition of an azimuthal
perturbation to the velocity field (see \S\ref{turb_sec}).  

While we do not exclude the possibility of infall, outflow (either centrifugal
or radiation pressure dominated), or line-driven wind contributions to the BLR
kinematics within individual AGN, we consider here the simplest possible and
likely underlying scenario, of a largely gravitationally-dominated
circularised flow.  In particular we do not here consider the effects of a
substantial wind contribution to the broad emission-line profiles of
especially the high ionisation resonance lines, such as C~{\sc iv} (see e.g.,
the disc-wind models of Chiang \& Murray 2006; Murray \& Chiang 2007), which
may be especially important in these lines in high Eddington rate AGN. We also
note that while outflows, as manifested in blue-shifted UV and X-ray
absorption lines (e.g., Elvis 2000) are common in AGN, their origin(s) is
presently uncertain, although the inner TOR remains a plausible candidate
(Pier \& Voit 1995; Mullaney et al. 2008). Indeed the velocities reported for
these outflows in Seyfert AGN (up to $\sim$1000 km~s$^{-1}$, e.g., Crenshaw et
al.  1999) roughly correspond with the expected gravitationally dominated
velocities at the purported distance of the TOR.  Moreover, in the sample of
AGN with reverberation mapping data (e.g., Kaspi et al. 2000; Richards et
al. 2011), there is no evidence for strong blue asymmetries in their C~{\sc
iv} emission-line profiles, normally taken to be a strong indicator for
outflows within the BLR.

%Additionally, we do not
%consider the effects of external line-broadening by fast moving electrons,
%since significant electron scattering would destroy the coherence between the
%continuum and emission-line variations.

Rather, we introduce macro-turbulent cloud motion (\S3.3) as a simple, and
often suggested, mechanism for providing a substantial radially-dependent
scale height for the BLR clouds, allowing the BLR gas to intercept a
significant fraction of the ionising continuum radiation necessary to produce
the observed line strengths, something which remains problematic for thin-disc
models of the BLR.  In justifying our assumption, we note that the TOR which
is the likely reservoir of gas that ultimately feeds the accretion disc,
itself has a very large scale-height, while dynamical models of the BLR based
on reverberation data appear to favour flared-disc geometries with substantial
opening angles above the disc mid-plane (e.g. Pancoast et~al. 2012, see \S3.3
for details).  Importantly in \S\ref{tds} we do investigate the effects of
both {\em transverse Doppler shift and gravitational redshift\/} on the
observed emission line profile and velocity delay map.  A particular choice of
$M_{BH}$, $R_{in}$, $R_{out}$, $\alpha$, and $\tau(R_{out})$ completely
describes the BLR geometry and velocity field.

We emphasise that we do not advocate a fixed inner and outer boundary for the
BLR nor do we mean to imply that the BLR forms the smooth inner surface of a
bowl-shaped geometry.  These are merely the simplest parameterisations of
something which is very likely far more complex.\footnote{The location of the
effective (line emissivity weighted) inner and outer BLR radii will vary in
response to the incident continuum flux, being generally forced to larger
radii in higher continuum flux states. The inner radius is set by some
combination of over-ionisation and emission line optical depth effects,
along with line-width visibility in the deep potential well, presuming the
availability of emitting gas. These adjustment time scales
should be fairly rapid compared to the central continuum variability time
scales. The outer radius is set by the effects of grain heating/cooling and
line destruction on emission line emissivity, and the grain vaporisation and
condensation time scales which are likely comparable to or longer than, typical
incident central continuum variability time scales.}  Rather we aim to explore
the observational consequences of assuming a bowl shaped BLR geometry in which
the gas dynamics are dominated by the central super-massive black hole, and
how these then impact on our interpretation of line profile shapes, correlated
continuum and emission-line variability, velocity resolved response functions,
black hole mass estimates and virial scale factors reported in the literature
for both individual sources and among the AGN population as a whole.

%In this spirit...
%Different AGN may differ in the details (e.g.,
%relative importance in radiation pressure, winds and such).

\section{A fiducial BLR geometry}\label{fiducial_model}

To illustrate the general properties of bowl-shaped BLR geometries we
construct a fiducial bowl-shaped geometry for the BLR for comparison with
observations. For expedience we choose a parameter set appropriate for the
Seyfert 1 galaxy NGC~5548. Our fiducial BLR geometry has a central black hole
mass $M_{BH}=1.0\times 10^{8}~M_{\odot}$. We set the inner radius at 200
gravitational radii ($\approx$1.14 light-days for our assumed black hole mass)
noting that the response time scales for He~{\sc ii} $\lambda$1640\AA\/ and
N~{\sc V} $\lambda$1240\AA\/ are $\sim$2~days in NGC~5548 (Korista et~al.\
1995).  We choose an outer BLR radius marking the upper rim of our bowl
geometry of 100 light-days, roughly speaking the graphite grain sublimation
radius (Nenkova et~al.\ 2008) for our chosen continuum normalisation (see
\S\ref{pi_model}), and a maximum time-delay at the outer radius (for $i=0$) of
$\tau=(R-H)/c=50$~days, similar to the dust reverberation time-delay measured
for this object.
% We adopt $\alpha=2$, and a maximum time-delay at the outer
% radius $\tau=(R-H)/c=50$~days ($\beta=1/150$) similar to the dust
% reverberation time-delay measured for this object. 
With these parameters, the source covering fraction as determined from the
polar angle to the bowl rim (60 degrees) is 50\% for our fiducial BLR geometry.

We populate the bowl surface with discrete line emitting entities (hereafter,
clouds) of fixed column density so that each cloud has an un-obscured view of
the continuum source and radiates energy in a manner approximating the phases
of the moon (see e.g., O'Brien et~al.\ 1995, Goad 1995, Ph.D.\ thesis); we
elaborate in \S 3.1. For simplicity, we ignore the effects of cloud--cloud
shadowing in assuming an effective bowl surface, but do account for
self-obscuration of the bowl by the outer rim (ie. lines of sight which pass
through the obscuring dusty torus) which for this geometry occurs at
inclinations $i > 45$ degrees for $\alpha \ga 2$ (the bowl geometry power law
index). We thus propose the bulk of the broad emission lines to form in gas
clouds lying along an effective surface which spans between the accretion disc
at small scale height and the TOR spanning a range in scale heights (see also
Czerny \& Hryniewicz 2011).  The dimly illuminated gas clouds beyond the broad
emission line region bowl surface are then likely dusty. The velocity field is
described by equation~4 (discussed further in \S 3.3), and importantly we
include the effects of Transverse Doppler Shift (hereafter TDS), Gravitational
Redshift (hereafter GR), and turbulence on the emergent line profile (see
\S\ref{tds} and \S\ref{turb_sec}).

Initially, we parameterise the radial surface line emissivity distribution
$F(r)$ as a simple power-law in radius ($F(r)\propto
r^{\gamma}$), with power-law index $\gamma=-1$ which is a fair approximation
to the expected radial emissivity distribution derived from photoionisation
calculations for several of the commonly observed UV and optical
emission-lines (e.g., figure~\ref{plot_flux_eta}).  We further assume that the
line-emitting gas responds linearly to variations in the incident ionising
photon flux $\Phi_{\rm H}$ (i.e., locally, the marginal response of the line
to continuum variations remains constant with time). The spatially dependent
line responsivity $\eta(r)$, is here defined as in Korista \& Goad (2004),
i.e.,

\begin{equation}
\eta(r)=\frac{\Delta \log F(r)_{line}}{\Delta \log \Phi_{\rm H}}
\end{equation}

\noindent where $F(r)_{line}$ is the radial surface line emissivity, and $\Phi_{\rm
H}$ is the incident hydrogen ionising continuum flux. For small amplitude 
continuum variations, this definition of responsivity converges to that given 
in Goad et~al.\ 1993. \footnote{Here we chose $\eta=1.0$ noting that for
$\eta(r,t)=constant$, the size of $\eta$ effects only the amplitude of the
line response. In reality, $\eta(r,t)=constant$ is a poor approximation for
most lines (see e.g., figure~\ref{plot_flux_eta} of this work, and Goad
et~al.\ 1993, and Korista \& Goad 2000, 2004).}  Since we observe a spread in
delays for lines of different species (see e.g., Clavel et~al.\ 1991), in any
given system, we focus here on bowl-shaped geometries parameterised by a
bowl-shape power-law index $\alpha$ in the range $1 \leq \alpha \leq 2.5$,
noting that in geometries with $\alpha<1$, gas at larger radii will not have a
direct view of the central continuum source, while gas at small radii on the
near side is obscured from sight even at relatively small viewing angles, a
consequence of the steepening of the bowl sides as $R$ decreases (increasing
$dH/dR_{x}$).

In figures~3--6, we show the spatial distribution of BLR gas (as viewed from
the azimuthal direction, R$_{z}$), the {\em emissivity-weighted\/} 2-d and 1-d
response functions and steady-state line-profiles for bowl-shape power-law
index $\alpha$ in the range $1.0 \leq \alpha \leq 2.5$ (with the bowl geometry
normalisation constant $\beta$ chosen to match the requirement that the
maximum time-delay at the outer radius when viewed face-on is 50 days), and
the fiducial geometry emission line profile as a function of the line of sight
observer inclination, $i=$5, 10, 30, and 45 degrees. At large line-of-sight
inclinations, the surface of the bowl becomes increasingly self-obscured since
such sight lines must first pass through the surrounding dusty torus
(the inclination at which self-obscuration first occurs decreases with
increasing $\alpha$).

%, compare figure~3, lower 4 panels).

As expected the form of the 2-d response functions are broadly similar to
those obtained for geometrically thin discs (the superposition of elliptical
response functions) with a few notable differences. Firstly, as $\alpha$
increases, the fraction of surface area that lies at larger radii increases,
so that the BLR response becomes increasingly weighted toward larger
time-delays and thus lower line of sight velocities (e.g., figure~4
left--right).  Furthermore, since rings at larger radii, are elevated with
respect to those closer in, there is a smaller offset in time-delay between
the centre of each ellipse than would be expected for a geometrically thin
disc.  This elevation of material out of the plane of the disc at larger
radii, results in bowl-shaped geometries displaying enhanced response at small
time-delays when compared to thin disc geometries with similar $R_{in}$,
$R_{out}$. We note that for fixed $R_{out}$, the surface area of the bowl
decreases with increasing scale height $H$, and increases with increasing
$\alpha$.

\begin{figure}
%\resizebox{\hsize}{!}{\includegraphics[angle=0,width=8cm]{geom_all_no_srgr_noturb.ps}}
\resizebox{\hsize}{!}{\includegraphics[angle=0,width=8cm]{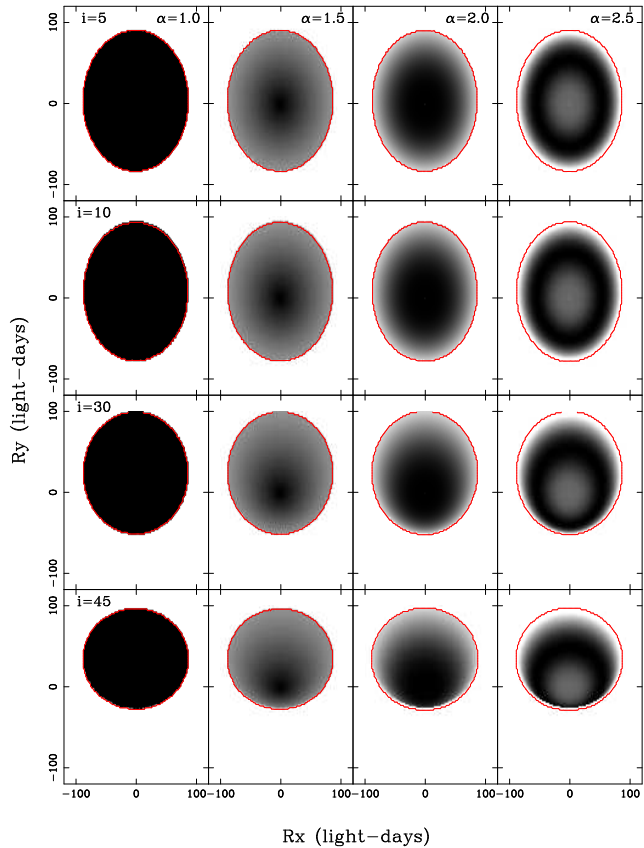}}
\caption{The spatial distribution of BLR clouds as viewed from the polar axis
($R_{z}$ direction). Individual panels show the expected cloud distribution
for bowl-shape power-law index $\alpha$ in the range $1.0 \leq \alpha \leq
2.5$, and inclination i=5,10,30 and 45 degrees. $\alpha=1$ represents a
special case wherein individual clouds populate the surface of a cone. For our
model, no shadowing is assumed to take place. The grey-scale represents the
observed intensity (here normalised to the peak), for our adopted power-law
emissivity distribution $F(r)\propto r^{\gamma}$, with power-law index $\gamma=-1$.}
\label{2d_proj}
\end{figure}

\begin{figure}
%\includegraphics[clip,width=140mm,angle=0]{plot_delay.ps}
%\resizebox{\hsize}{!}{\includegraphics[angle=0,width=8cm]{2d_resp_all_no_srgr_noturb.ps}}
%\resizebox{\hsize}{!}{\includegraphics[angle=0,width=8cm]{2d_resp_all_no_srgr_noturb_dark.ps}}
\resizebox{\hsize}{!}{\includegraphics[angle=0,width=8cm]{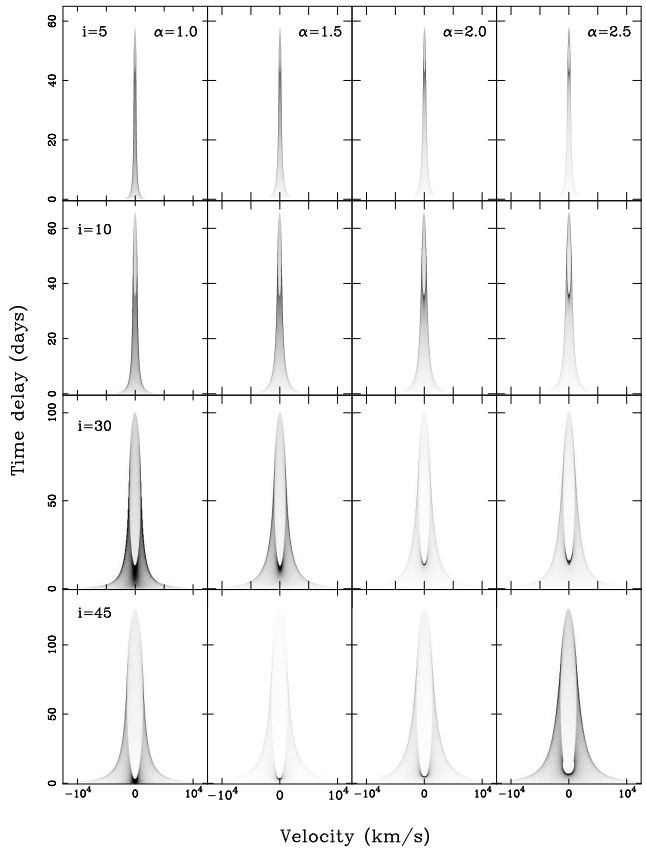}}
\caption{The corresponding 2-d response functions for figure~\ref{2d_proj}.}
\label{2d_resp}
\end{figure}

The increased weighting toward larger radii with increasing $\alpha$ is most
evident in the 1-d response functions. For $\alpha=1$, the 1-d response
function is essentially flat with increasing time-delay, falling after a
time-delay $R_{out}(1-\cos (\phi-i))$, corresponding to the light-crossing
time of that part of the BLR which lies closest to the line of sight. At large
$\alpha$, the increased weighting of the emissivity to larger radii (an area
effect) causes the 1-d response function to peak at $R_{out}(1-\cos(\phi-i))$,
moving towards smaller delays as $i$ increases. In the lower right panel of
figures 4 and 5 the occultation of gas close to the line of sight results in
the appearance of a secondary peak in the response function at small
time-delays.

In all cases, the emission-line profiles are double-peaked, and broadly
similar to those found for thin disc geometries. The locations of the peaks
are as for thin disc geometries determined by the velocity at the outer radius
(as given by equation~2), and the inclination $i$ (ie. $\pm v \sin i$).
Obscuration by the surrounding dusty torus at large line of sight inclinations
can be identified in the bottom right hand panels of figure 6, by the far
larger difference in height between the line peaks and line centre and the
U-shaped appearance of the line core, resulting from obscuration of
low-velocity gas at large BLR radii and lying close to the line of sight.

While the parameter space to be explored is clearly large, figures 4--6
indicate that the {\em emissivity-weighted\/} 2-d and 1-d response functions
and steady-state emission-line profiles are broadly similar for fixed observer
inclination. Thus for the remainder of this work, we explore the observational
characteristics of a single reference model, which we refer to as our fiducial
BLR geometry, parameterised by bowl-shape power-law index $\alpha=2$
(corresponding to a normalisation constant $\beta=1/150$, for R and H measured
in lt-days).

\subsection{The role of emission line anisotropy and bowl scale height}
For isotropically emitting gas orbiting in a geometrically thin disc, the
centroid of the 2-d response function and the emission-line profile 'shape' as
quantified by the ratio $fwhm/\sigma_{l}$ in the absence of transverse Doppler
shift (TDS) and gravitational redshift (GR) are independent of
inclination. Furthermore, for our adopted line radiation pattern for
individual clouds (which roughly speaking approximates a radiation pattern
similar to the phases of the moon, e.g. O'Brien et~al. 1994), increased
anisotropy increases the centroid of the 2-d response function as $i$
increases, but the profile shape remains unaltered (reduced emission on the
side nearest to the observer due to increased anisotropy in the line is
replaced by enhanced emission on the far-side at the same velocity -- front
back symmetry).

By contrast for bowl-shaped geometries, the centroid of the 1-d response
function increases with increasing observer line of sight inclination even for
fully isotropically-emitting clouds (since rings on the bowl are pivoted about
an axis running through the centre of the ring and the centre of the bowl). In
addition, increased emission line anisotropy (see also Ferland et~al.\ 1992)
reduces the centroid of the 1-d response function at small inclinations (by
reducing the contribution of gas at high elevations relative to that at low
elevations), while increasing the centroid at larger inclinations due to the
reduced contribution of gas closer to the observers line of sight.  However,
as for a thin disc geometry, front-back symmetry of our adopted line radiation
pattern, yields line profile shapes which are independent of inclination in
the absence of GR, TDS effects. Notably, for bowl-shaped BLR geometries, the
centroid of the response function is a stronger function of inclination when
compared with that for a standard thin disc geometry.

%The first of these illustrates the effect of transverse Doppler shift and
%gravitational redshift which result in significant velocity dispersion and
%strongly red asymmetric 2-d response functions and line profiles at low
%line-of-sight inclinations.

In practice we are mainly interested in geometries for which the line of sight
inclination $i$ allows an un-obscured view of the bowl surface ($0 \leq i <
\phi$), that is, over the edge of the TOR. The BLR will be obscured on the
near-side (closest to the line of sight) at larger inclinations either by
self-occultation by BLR gas lying at larger radial distances, or by the
surrounding TOR unless the line of sight to the central continuum source
covering fractions of these components are themselves low\footnote{The dusty
TOR might well extend to substantially greater scale heights in this geometry
than that demarcated by the location of the grain sublimation radius (e.g.,
see figure 1). In this case obscuration of the central regions can occur at
smaller observer inclination angles. It is also possible that such sight lines
are only partially obscured by dusty clouds, as in Nenkova et~al.\
2008.}. Indeed, low line of sight inclinations are suggested by a number of
studies. X-ray studies of Seyfert 1 galaxies suggest BLR inclinations of $i\la
30$ degrees (Nandra et~al.\ 1997, Nandra et~al.\ 1999, Tanaka et~al.\
1995). Disc model fits to the double-peaked emission-line objects favour
inclination angles of between 18--36 degrees (Eracleous et~al.\
1994). Furthermore, if the measured opening angles of ionisation cones
observed in type 2 AGN are representative of AGN in general, then type 1
objects must be viewed at angles of $i<35$--60~degrees (Wilson and Tsvetanov
1994).

%, a finding supported by CO observations of the molecular ring . 
% Storchi-Bergmann, Wilson & Baldwin 1992; Dopita et al. 1998).
% S.J. Curran - Differences in the dense gas between type 1 and type 2 Seyferts
% Astron. Astrophys. Suppl. Ser. Volume 144, Number 2, June I 2000 
% + torus aligned with disc to high degree though molecular ring may be offset
% in pa - not large in general (may obscure BLR in some cases).
% Sy 1s tend to be located in galaxies at low inclinations.
% have small inclinations

\begin{figure}
%\includegraphics[clip,width=140mm,angle=0]{plot_resp.ps}
%\resizebox{\hsize}{!}{\includegraphics[angle=0,width=8cm]{1d_resp_all_no_srgr_noturb.ps}}
\resizebox{\hsize}{!}{\includegraphics[angle=0,width=8cm]{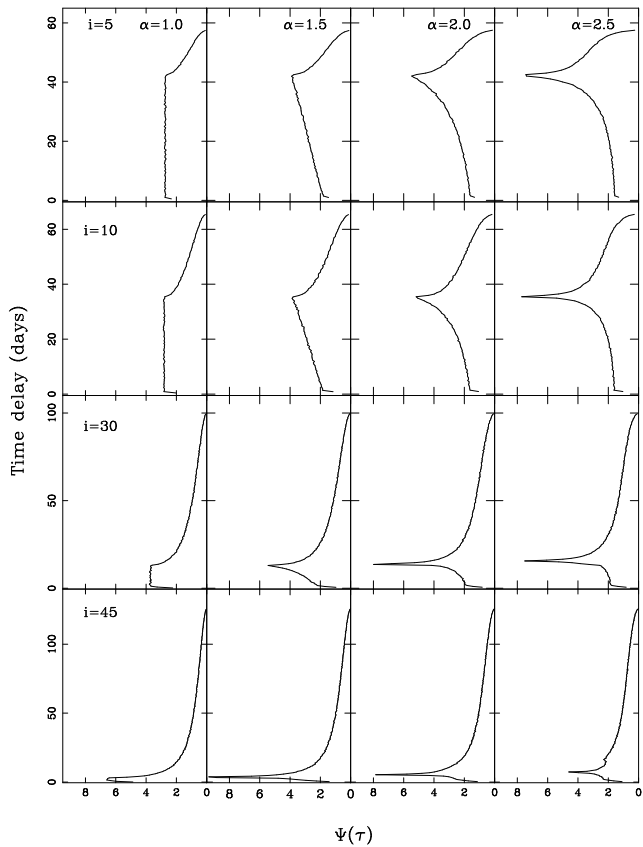}}
\caption{The corresponding 1d response functions for figure~\ref{2d_proj}.}
\label{1d_resp}
\end{figure}

\begin{figure}
%\includegraphics[clip,width=140mm,angle=0]{plot_prof.ps}
%\resizebox{\hsize}{!}{\includegraphics[angle=0,width=8cm]{prof_all_no_srgr_noturb.ps}}
\resizebox{\hsize}{!}{\includegraphics[angle=0,width=8cm]{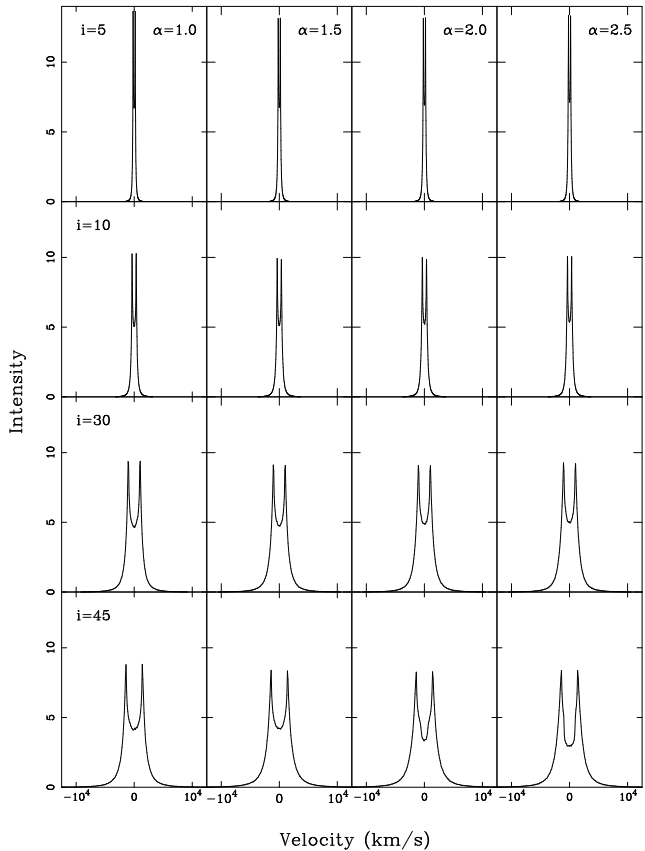}}
\caption{The corresponding line profiles for figure~\ref{2d_proj}.}
\label{1d_prof}
\end{figure}

%\begin{figure}
%\resizebox{\hsize}{!}{\includegraphics[angle=0,width=8cm]{plot_geom_1.ps}}
%\caption{The observed flux distribution for our fiducial toy model inclined at
%  an angle of 30 degrees wrt the observers line of sight and assuming a
%  power-law emissivity with slope -1, and isotropic emission.}
%\label{geom_1}
%\end{figure}

%\begin{figure}
%\begin{figure*}
%\resizebox{\hsize}{!}{\includegraphics[angle=0,width=8cm]{plot_geom_1_anis.ps}}
%\includegraphics[clip,width=140mm,angle=0]{plot_geom_1_anis.ps}
%\caption{As for fig\ref{geom_1n}, for the case of 100\% anisotropy. The same
%  intensity scale applies.}
%\label{2d_proj}
%\end{figure*}
%\end{figure}

%\begin{figure*}
%\includegraphics[clip,width=140mm,angle=0]{a1_i75.ps}
%\caption{An illustration of the effects of obscuration by the torus. Only
%those clouds indicated in yellow (ie viewed over the rim of the bowl), are observed}
%\label{geom_view}
%\end{figure*}

%\begin{figure}
%\resizebox{\hsize}{!}{\includegraphics[angle=0,width=8cm]{plot_dist.ps}}
%\caption{Distribution in delays.}
%\label{plot_dist}
%\end{figure}

\subsection{Transverse Doppler shift and Gravitational Redshift}\label{tds}

\begin{figure}
%\resizebox{\hsize}{!}{\includegraphics[angle=0,width=8cm]{sgr.ps}}
%\resizebox{\hsize}{!}{\includegraphics[angle=0,width=8cm]{sgr_dark.ps}}
\resizebox{\hsize}{!}{\includegraphics[angle=0,width=8cm]{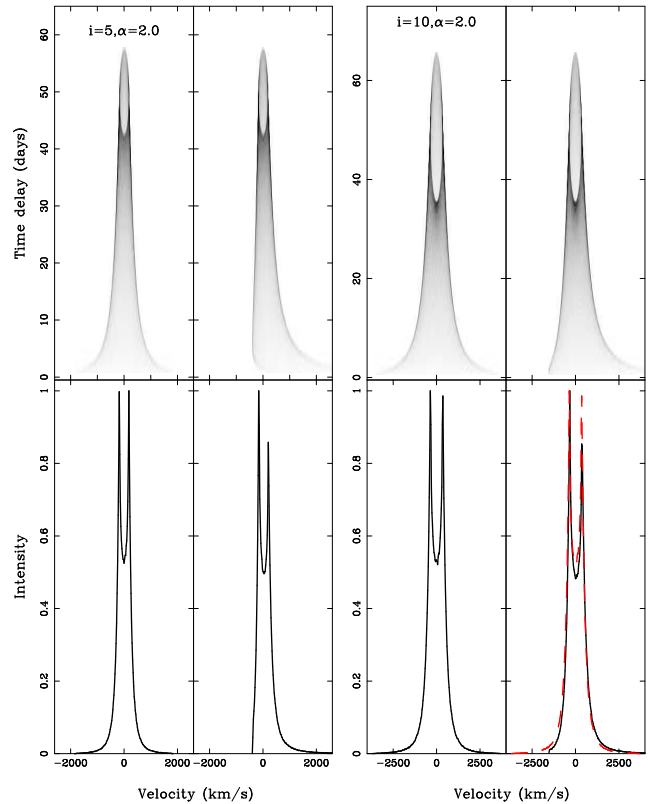}}
\caption{Upper panels -- 2-d response functions for our fiducial BLR geometry,
with (right-hand panels) and without (left-hand panels) the combined effects
of Transverse Doppler Shift and Gravitational Redshift.  Lower panels -- the
corresponding 1-d line profiles. The dashed line in the right-hand side of
panel 4 shows the same model without these effects (appearing in the left-hand
side of panel 4).}
\label{sgr}
\end{figure}

At the inner radius chosen for our fiducial BLR geometry (200 $R_{g}$), the
effects of transverse Doppler shift (hereafter, TDS) and gravitational
redshift (hereafter, GR) have a significant effect on the emergent line
profile, at low line of sight inclinations for the majority of lines, and in
particular for those lines which form preferentially at small BLR radii.
Using the standard formulism for each of these effects, we compare in
figure~\ref{sgr} the 2-d response functions and emission-line profiles for our
fiducial BLR geometry for inclinations i=5 and i=10 degrees, with (right
panels) and without (left panels) the combined effects of TDS and GR (the 1-d
response function is not shown here as it remains unaltered by the inclusion
of these effects).

%that is, the
%transverse Doppler shift is given by
%\begin{equation}
%\cos\theta_{obs} = \cos\theta_s -v/c 1 -v/c \cos\theta_s
%\end{equation}
%\noindent where $\theta_{s}$ is the angle between the velocity vector 
%of the source (ie emitting
%and the observers line of sight in the frame of the source, and $\theta_{obs}$
%is the corresponding angle in the observes frame of reference.

TDS and GR act to enhance the red wing response at small time delays (at the
expense of the blue-wing response), thereby creating a strong red-ward
asymmetry in the line profile. This effect is largest for small line of sight
inclinations, since for our model, in the absence of turbulence, the line of
sight velocity is zero at $i=0$, if the effects of TDS and GR are not
included.  GR effects will be most important for lines which form at small BLR
radii (deep within the gravitational potential). Similarly for our chosen
velocity field effects due to TDS will be stronger in lines which form at
small BLR radii (e.g. compare figure~\ref{turb} panels 5 and 6), where the
velocities are more extreme. As shown in figure~\ref{sgr} the strong red-blue
asymmetry introduced into the line profile is significant, as is the enhanced
red wing response and diminished blue-wing response in the 2-d response
function at small time-delays. At an inclination of 10 degrees, the importance
of these effects diminishes (figure~\ref{sgr} -- right panels, cf. dashed
line), and at 30 degrees are barely detectable ($fwhm/\sigma_{l} \approx
constant$ for $i>20$ degrees).
%One of the key aspects of GR and TDS is that together they set a minimum width
%to the emission line profile (set by the mass of the black hole and the radial
%line emissivity distribution) even for thin disc geometries observed face-on
%(ie. even for purely transverse motion)
%However, if these mildly relativistic effects are indeed important (largely
%determined by the mass of the black hole and the radial surface line
%emissivity distribution), then
A key result of this work, is that the effects of GR and TDS acting alone can
provide substantial width to the emission-line ($\sim$ several hundred
km~s$^{-1}$), even for flattened, thin- or thick disc geometries, with pure
planar Keplerian motion, viewed at low line-of-sight inclination.
\footnote {We do not consider here the effects of internal line broadening by
electron scattering within the BLR clouds. For the physical conditions extant
within the BLR gas, electron scattering optical depths of up to $\tau_{e}
\ge 0.1$ are expected (for cloud ionisation parameters $U>0.1$, and column
densities $N_{\rm H}> 10^{23}$~cm$^{-2}$; see Laor 2006). Electron scattering
optical depths of $\tau_{e} < 0.5$ can readily account for broadening of up to
a few hundred km~s$^{-1}$ in the {\em scattered fraction} of the line photons
emerging from such a cloud (perhaps contributing to reducing the number of
required clouds for smooth broad emission line profile wings; Arav et
al. 1997, 1998).  However, unless the electron scattering optical depths are
large over most of the BLR, its effect on the emission line profiles, even at
small inclinations, should be minor. }

In the absence of these effects line profile shapes as quantified by
$fwhm/\sigma_l$ are independent of inclination in thin disc
geometries. However, as we show in \S6.1, in the context of flattened BLR
geometries, when present GR and TDS introduce {\em a strong inclination
dependence to the line profile shape at low line of sight inclinations\/}
which matches both qualitatively and quantitatively the observed correlation
between $fwhm/\sigma_l$ and $\sigma_l$ among AGN reported in the literature
(e.g. figure~3 of Collin et~al. 2006).

While Mannucci, Salvati and Stanga (1992) considered relativistic effects when
modelling the 1-d response and emission-line profiles of disc- and nest-shaped
BLR geometries, the effects of TDS and GR on the form of the 2-d response
function have largely been overlooked\footnote{Eracleous \& Halpern (1994,
2003), considered relativistic effects when modelling the double-peaked broad
optical emission-line profiles of a sample of radio-loud AGN. Corbin (1997) 
considered these effects on the shapes of the line profiles, but not on the 
reverberation line response functions. Kollatschny (2003b) argued that 
the redshifted component of the rms profile in the Seyfert~1 galaxy Mkn~110 
was a result of gravitational redshift. From this he deduced
a black hole mass of $1.4\times 10^{8}~M_{\odot}$ and an inclination of i=19
degrees for this object.}. 
However, a prominent red-wing response is a key
observational feature of the recently recovered 2-d response function for the
optical emission-lines in the Seyfert~1 galaxy Arp~151 (Bentz et al. 2010b,
their figure 4), though the response on the shortest timescales remain
temporally unresolved. While these authors used the recovered response
function to rule out a number of simple models for the BLR in this object, the
effects of TDS and GR were not considered.  If instead the enhanced red-wing
response is a direct consequence of the combined effects of TDS and GR, then
potentially Arp~151 may be a system with a low line-of sight inclination, or a
system with significant emission at small BLR radii, for which the effects of
TDS and GR are both larger.

\subsection{Macro-turbulent cloud motion}\label{turb_sec}
% introduce turbulence

For material to accrete, significant angular momentum must be removed via
viscous dissipation outward through the disc. However, this process does not
require the disc to be geometrically thin. Indeed, the relative ionising
continuum and broad emission-lines fluxes require covering fractions for the
BLR gas of at least 10\%, and often substantially larger. This means that even
for thin disc BLR geometries, the disc must have significant scale height in
order to intercept sufficient ionising continuum necessary to explain the
observed line strengths.  Furthermore, moderately flared discs are predicted
by standard accretion disc models (Shakura \& Sunyaev 1973), and turbulence
remains the simplest mechanism for supplying the internal pressure necessary
to support the discs' vertical extent. While the predicted scale heights for
thin-disc geometries are rather modest ($H/R \sim 0.1$), in our model the BLR
is not physically attached to the disc, though indeed it may be the originator
of the BLR clouds.  Instead we model the BLR as an ensemble of line-emitting
clouds, and invoke macroscopic turbulent cloud motion as the mechanism which
provides the BLR geometry with the necessary scale-height to intercept a
substantial fraction of the ionising continuum. Recent dynamical models of the
continuum and broad emission-line variations in the Seyfert~1 galaxy Mrk~50
(Pancoast et~al. 2011), suggest flared discs with significant scale height
($\sim 0.5$ at the outer radius, for their assumed opening angle of $25\pm 10$
degrees) as a plausible geometry for this system.  We also note that the
surrounding TOR, which is the likely reservoir of material feeding the
black-hole, itself must have substantial scale-height in order to obscure the
BLR in type 2 objects (in the general orientation-dependent unified picture
for AGN).

While GR and TDS provide substantial line width in thin disc geometries even
for face-on objects, the line width is still largely determined by the black
hole mass and source inclination. Thus for NLSy1s, which are thought to be low
mass high accretion rate systems observed at low inclinations, a significant
turbulent component may still be required to ensure sufficient width in the
line.\footnote{We speculate that high Eddington rate sources may have TORs
with larger than typical covering fractions, necessitating smaller observed
viewing angles $i$ for un-obscured lines of sight.}

%Recently, Kollatschny \& Zetzl (2011) used a Keplerian velocity field
%convolved with a turbulent component in order to model the broad H$\beta$ and
%C~{\sc iv} emission-lines in a large sample of AGN, and found that though
%dependent on the line in question, the degree of turbulence was broadly
%similar from one object to the next. However, they also found that the broad
%C~{\sc iv} emission-line requires a larger scale height than broad H$\beta$.
%In stark contrast Decarli et al. demonstrated that based on line profile
%shapes alone, H$\beta$ arises in a more isotropic (ie. turbulent) velocity
%field than C~{\sc iv}.

In order to implement isotropic turbulence within the context of our model 
we add in quadrature a randomised velocity component $v_{\rm turb}$ which 
increases the local Keplerian velocity $v_{\rm kep}$ (equation~2) according to
\begin{equation}
v^{2} = v_{\rm kep}^2 \left[ 1+ \left ( \frac{ v_{\rm turb} }{ v_{\rm kep} }
  \right )^{2} \right ] \, ,
\end{equation}

\noindent and such that $|v_{\rm turb}|$ increases linearly with scale height
$H$, i.e., $|v_{\rm turb}|= b_{\rm turb}\left(\frac{H}{R_{x}}\right)v_{\rm kep}$, where
$b_{\rm turb}$ is a constant of order unity. This is
similar to the implementation of turbulence with BLR scale height suggested in
Collin et~al.\ (2006).  In practice, we draw at random a randomised (in
direction) velocity vector from a Gaussian distribution of the appropriate
width.  Using this formulism, the turbulence is zero both when $b_{\rm
turb}=0$ (turbulence switched off) and when $H=0$ (zero scale height). Thus in
the context of bowl-shaped BLR geometries, the contribution that turbulence
makes to the 2-d response function and line profile depends upon the
bowl-shape, and on the radial surface line emissivity distribution which together
determine the scale height at which a given line forms. Thus, in our model,
the effects of turbulence are minimised for lines formed near the base of the
bowl (small scale height).  For non-zero $H$ and sufficiently large $b_{\rm
turb}$, the randomised turbulent velocity component dominates over the planar
Keplerian motion.

\begin{figure}
%\resizebox{\hsize}{!}{\includegraphics[angle=0,width=8cm]{turb2.ps}}
%\resizebox{\hsize}{!}{\includegraphics[angle=0,width=8cm]{turb2_dark.ps}}
\resizebox{\hsize}{!}{\includegraphics[angle=0,width=8cm]{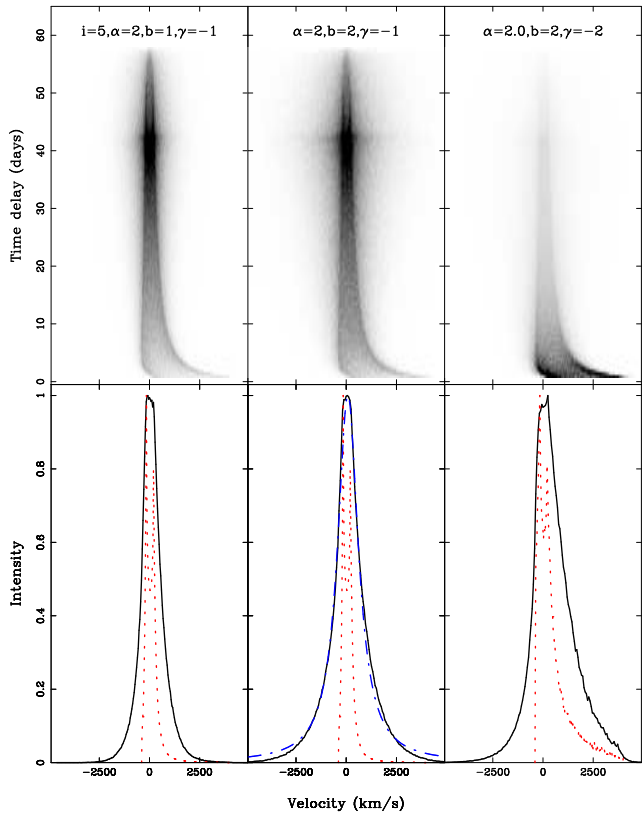}}
\caption{Upper panels - 2-d response functions for our fiducial BLR geometry
with a line-of-sight inclination 5 degrees and turbulence parameter $b_{\rm
turb}=1$, (left panel), $b_{\rm turb}=2$ (middle panel).  In panel 3 we show
$b_{\rm turb}=2$ for a power-law emissivity distribution with
$\gamma=-2$. Lines formed at smaller scale height (with steeper emissivity
distributions) are less affected by turbulence.  Also shown (lower panels) are
the steady-state line profiles (solid black lines), together with the expected
profile for no turbulence, $b_{\rm turb}=0$ (red dotted line). In the middle
lower panel we also show our best fit Lorentzian profile (blue dashed line)
with fit parameters, $I=1.025$ , $\Gamma=623.5$~km~s$^{-1}$, and $x_0 =
+82.2$~km~s$^{-1}$.}
\label{turb}
\end{figure}

In figure~\ref{turb} we illustrate the effects of turbulence on the 2-d
response function and emission-line profile for our fiducial BLR geometry
assuming a power-law emissivity distribution with power-law index $\gamma=-1$,
and including the effects of TDS and GR, for an inclination i=5 degrees and
turbulence parameter $b_{\rm turb}=1$ (left-hand panels), and $b_{\rm turb}=2$
(middle 2 panels).  Turbulence has a number of important
attributes. First, the lines are broader (as expected), and at low
inclinations the prominent shoulders/horns indicative of flattened BLR
geometries have disappeared even for moderate turbulence. Secondly, gas at
large radii now makes a significant contribution to the line wings, enhancing
the line wings relative to the line core. The effect of turbulence on line
shape depends upon the size of $b_{\rm turb}$ and on the line formation
radius. For example, as expressed in terms of the typically measured
parameters used to quantify line profile shapes, the ratio of the line
full-width at half maximum (hereafter $fwhm$) to the line dispersion
(hereafter $\sigma_{l}$, see e.g., Collin et al.\ 2006), then for $b_{\rm
turb}=0$, $\gamma=-1$ (dotted red line, lower left panel),
$fwhm/\sigma_{l}$=1.4, increasing to 1.5 for $b_{\rm turb}=1$ (solid black
line). For larger $b_{\rm turb}$, $fwhm/\sigma_{l}$=1.1 (lower middle panel,
solid black line). While in panel 3, where we have now increased the radial
power-law emissivity to $\gamma=-2$, $fwhm/\sigma_{l}$ increases from 1.0
without turbulence (dotted red line) to 1.4 with $b_{\rm turb}=2$ (solid black
line).

We note that the observed lack of response of the broad emission-line wings on
short time-scales in some objects has traditionally been explained by the
presence of a low responsivity component in the inner BLR (see e.g., Goad
et~al.\ 1993; Sparke 1993; Korista \& Goad 2004). Here, a similar effect
arises for a different reason, a significant fraction of the broad
emission-line wings, as observed at low inclinations, are now due to gas lying
at large BLR radii which responds on longer time scales. If turbulence
dominates the contribution to the emission-line response, then the core and
wings of the line will respond on similar timescales.

One potentially significant result of this work is that at low inclinations,
and in the presence of turbulence our model line profile displays a striking
resemblance to a Lorentzian function.  By way of illustration, we model the
line profile with a 3-parameter Lorentzian of the form
\begin{equation}
I(x) = I_{0}\left[ \frac{ \Gamma^{2} } { ( x-x_{0} )^{2} + \Gamma^{2} }
\right] \, ,
\end{equation}

\noindent where $I_{0}$ is the peak intensity, $x_{0}$ is the median velocity,
and $\Gamma$ represents the half-width at half maximum of the line. Our best
fit model shown in the lower middle panel of figure~\ref{turb} (dashed blue
line) with fit parameters, $I_{x=0}=1.025$, $\Gamma=623.5$~km~s$^{-1}$, and
$x_0 = +82.2$~km~s$^{-1}$ provides a remarkably good fit to the line profile,
failing only in the extreme line wings which would in any case be difficult to
measure in emission-line spectra due to their low contrast. The appearance of
narrow cores and extended broad wings as observed in some type 1 Seyferts (in
particular NLSy1s or ``Pop~A'' objects; see Collin et~al.\ 2006) may indicate
the presence of significant turbulence and low inclination ($< 20$ degrees) in
these objects. While many studies have modelled the Balmer and Mg{\sc ii}
emission line profiles with Lorentzians (e.g, Zamfir et~al.\ 2010, their
``Pop~A'' type~1 AGN), to our knowledge no explanation of why such profiles
should manifest themselves has ever been offered, heretofore.

For our model, turbulence is more significant for lines formed at larger BLR
radii (e.g., H$\beta$, Mg~{\sc ii}), and hence larger scale heights. This is
illustrated in figure~\ref{turb} -- right-hand panels, where we show the
effect of steepening the power-law emissivity distribution to a slope of
$-2$. The bulk of the line is now formed at small BLR radii, and thus small
scale heights where the effect of turbulence is low. Consequently, though the
turbulence parameter is large, a strong red-blue asymmetry, a result of TDS
and GR whose effects are larger at small BLR radii, can still be discerned in
the 2-d response function and emission-line profile. Thus at low line of sight
inclinations, we expect to observe strong differences between the 
broad emission-line profiles of low- and high-ionisation lines.

In summary, TDS, GR and turbulence play a significant role in determining the
shape of the 2-d response function and emission-line profile.  Lines formed
preferentially at small BLR radii will show an enhanced red-blue asymmetry in
their 2-d response function and line profile due to the combined effects of
TDS and GR (these effects will remain apparent provided the emission from the
turbulent component does not dominate the contribution to the line profile).
By contrast lines formed at large BLR radii will display more symmetric 2-d
response functions and line profiles (because the effects of GR and TDS are
diminished at larger BLR radii, and turbulence at larger scale height acts to
reduce the dependence of line profile shape on inclination by randomising the
velocity field).  If turbulence is significant these lines' emission-line
profiles will show enhanced wings relative to the line core. For low
inclination systems, turbulence produces Lorentzian profiles, especially in
lines forming at large radii (e.g., Balmer and Mg~{\sc ii}) and so substantial
scale height, similar to those seen in many NLSy1 (or ``Pop~A'') spectra. At
higher inclinations, the Keplerian velocity dominates over the turbulent
component, although the latter still affects especially the cores of the
emission-line profiles (see Figures~A1, A2).  In light of these findings, we
suggest that the previous association of an enhanced red-wing response with
in-falling gas or a Keplerian disc + hotspot (for example, as suggested for
the velocity field in the nearby NLSy1 Arp~151 (Bentz et al. 2010) and indeed
spherical or disc-wind models of the BLR (K\"onigl and Kartje 1994; Chiang and
Murray 1996; Murray and Chiang 1997), {\em may also in part be attributed to 
the presence of mildly relativistic gas in the inner BLR\/}, as we have
demonstrated here, as was previously suggested to account
for the enhanced red-wing response observed in broad H$\beta$ for the Sy1
galaxy Mrk~110 (Kollatschny 2003b).

We note here that the recovered 2-d response function and emission-line
profile for H$\beta$ in the NLSy1s NGC~4051 (Denney et al.\ 2011, their figure
4), Arp~151 (Bentz et al. 2010, their figures 2,3 and 4), and in the 2-d CCF
of Mkn~110 (Kollatschny 2003, figure~7), bares an intriguing resemblance to
that predicted for a bowl shaped geometry viewed close to face-on with a
significant scale height dependent turbulent component (e.g., compare their
figures with the left hand panel of Figures A1--A3 in Appendix A, for $i=10$
degrees). In Arp~151, an enhanced red-wing response is evident in all of the
Balmer lines.

{\em Significantly, GR, TDS and turbulence can on their own provide
substantial line-width even for purely transverse motion and may explain the
observed line widths in systems generally thought to be viewed close to
face-on as well as the break in the $fwhm/\sigma_l$ relation at small
$\sigma_l$ observed among the AGN population.}

%Finally, the absence of short timescale variability in the broad
%wings of some lines may have an alternative explanation than that commonly
%reported in the literature, low responsivity gas in the inner BLR (e.g. see
%Korista and Goad 2004), a consequence of the larger ionising photon flux at
%small radii.  A large line formation radius in the presence of a strong
%turbulent velocity field may produce a similar effect. Similarly, if
%turbulence dominates the velocity field, then in our model there will be
%little if any delay between variations in the line wings and line core.}

\section{Simulations}\label{simulations}

\subsection{Photoionisation calculations}\label{pi_model}

%\begin{figure}
%\resizebox{\hsize}{!}{\includegraphics[angle=0,width=8cm]{plot_fluxes_hden12.ps}}
%\caption{Radial emissivity distributions for selected low and high ionisation
%  lines. The red line indicates a power-law emissivity distribution with slope $-1$.}
%\label{emiss}
%\end{figure}

\begin{figure}
%\resizebox{\hsize}{!}{\includegraphics[angle=0,width=8cm]{plot_flux_eta.ps}}
\resizebox{\hsize}{!}{\includegraphics[angle=0,width=8cm]{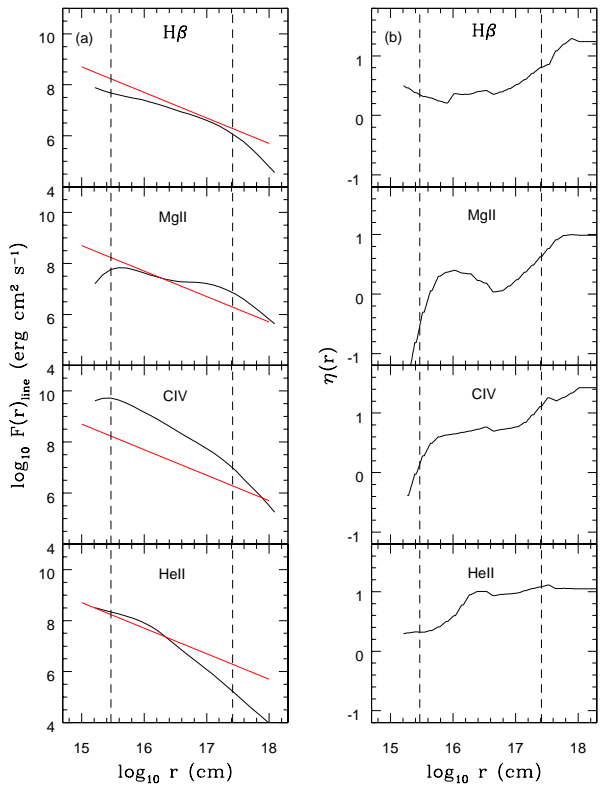}}
\caption{(a) Radial emissivity distributions for selected low and high
  ionisation lines. The red line indicates a power-law emissivity distribution
  with power-law index $\gamma=-1$. (b) Radial surface responsivity distributions
  for selected low and high ionisation lines. The dashed vertical lines
  indicate the BLR inner and outer radius for our fiducial BLR geometry.}
\label{plot_flux_eta}
\end{figure}

As in previous work, we adopt the most general model for the BLR gas, the
locally optimally emitting cloud model of Baldwin et al.\ 1995 (see also
Korista \& Goad 2000, 2001, 2004). In brief, this model assumes that there
exists a large population of BLR clouds with a broad range of gas density, gas
column density, and ionising source distance. The observed broad emission-line
spectrum arises by a process of natural selection, since for a given line only
clouds whose properties span a relatively narrow range of physical conditions
are efficient in re-processing the incident ionising continuum into particular
line emission (hence ``locally optimally emitting'', or LOC). This model not
only explains why the observed spectra of AGN are generally similar, but
significantly, requires minimal fine-tuning.

Using the photoionisation modelling code CLOUDY version c08.00 (Ferland
et~al. 1997, Ferland et~al.\ 1998), we generated a grid of photoionisation
models for simple slabs of gas (hereafter, clouds), each with a constant
density and an un-obscured view of the ionising continuum source. In this
simple approach contributions from the reprocessed continuum (the diffuse
component from the cloud) and the effects of cloud shadowing are ignored.
Moreover, we do not consider the complicating effects of micro-turbulence,
which would tend to broaden the intrinsic line profile far above the thermal
width and reduce the central line optical depth, thereby increasing the line
escape probability isotropically (in this case) and hence altering the
emergent spectrum (see e.g., Bottorff et~al. 2000). This would also act to
relax the requirement of large cloud numbers to explain the relative
smoothness of the observed broad emission-line profiles (Arav et~al. 1997;
1998). Of the four emission lines considered here (see
Figure~\ref{plot_flux_eta}), micro-turbulence would affect mainly H$\beta$ and
Mg~{\sc ii}, effectively moving their luminosity-weighted radii to smaller
values.

The full grid spans 7 decades in the gas hydrogen number density--hydrogen
ionising photon flux plane, $7 < \log_{10} n_{\rm H}$ (cm$^{-3}$)$ < 14$, and
$17 < \log_{10} \Phi_{\rm H}$ (cm$^{-2}$~s$^{-1}$)$ < 24$, stepped in 0.25
decade intervals in each dimension (see also Korista et~al.\ 1997). Since the
emitted spectrum is not all that sensitive to the cloud column density over
the range $22 < \log_{10} N_{\rm H}$ (cm$^{-2}$)$ < 24$, all clouds are
assumed to have a constant total hydrogen column density $\log_{10} N_{\rm
H}$(cm$^{-2}$)$ = 23$. The adopted elemental gas abundances are solar. The
adopted spectral energy distribution of the incident continuum is a modified
form of the Mathews \& Ferland (1987) generic AGN continuum.\footnote{The
effects of a polar-angle dependent ionising continuum shape on the detailed
emission line spectrum are explored in Appendix \ref{continuum_shape}.} For
the adopted mean ionising continuum source luminosity of NGC~5548,
$\log_{10}L_{ion}=44.14$~erg~s$^{-1}$, a hydrogen ionising photon flux
$\log_{10}\Phi=20.0$~photon/s, corresponds to a cloud--ionising source
distance of $R=15$~($H_{0}$/70~km/s)~light-days. 

Utilising the standard LOC gas density distribution (weighting)
function\footnote{This standard LOC distribution (weighting) function in
hydrogen gas density is similar to that found by Krause et al. (2012), for BLR
clouds confined by a magnetic field undergoing filamentary fragmentation.},
$g(n_{\rm H}) \propto n_{\rm H}^{-1}$, spanning the range in gas density $8 <
\log_{10} n_{\rm H}$ (cm$^{-3}$)$ < 12$, as described in Korista \& Goad
(2000; see also Baldwin et~al.\ 1995; Bottorff et~al. 2002), we sum the
emission along the density axis for each radius, producing a radial surface
line emissivity function (erg~cm$^{-2}$~s$^{-1}$) for each of the lines
considered.
%The inner radius we fix at 100~$R_{s}$ (where $R_{s}$, the Schwarzschild
%radius, is related to the black hole mass $M_{\rm BH}$ via $R_{s} = GM_{\rm
%BH}/c^{2}$, and $c$ is the speed of light).
% M_{BH}=6.5x10^7 Msun  (Rs=0.0074 lt-days)
%For NGC~5548 we adopt a value of $M_{\rm BH}=6.5\times10^{7}~M_{\odot}$ (Bentz
%et~al. 2007), for which 100~$R_{s}$ equates to an inner radius of $\approx
%0.74$~light-days. 
The radial surface line emissivity distributions are computed beyond the point
at which grains will form, i.e., beyond the point where the grain temperature
at the illuminated face of the cloud falls below the grain sublimation
temperature.  The radius at which this occurs ($\sim 100$~lt-days for our
adopted continuum and the hardiest of grains) marks the inner boundary of the
dusty TOR. For simplicity, we will assume the TOR to be opaque to UV/optical
photons, and thus the central regions are obscured along lines of sight that
pass through the TOR. The radial surface line emissivity distributions for
C~{\sc iv}~$\lambda$1549, Mg~{\sc ii}~$\lambda$2798, H$\beta$~$\lambda$4861
and He~{\sc ii}~$\lambda$4686 whose emission line profiles are commonly used
in measuring black hole masses, are shown in figure~\ref{plot_flux_eta}(a)
(solid black lines). For our fiducial model we assume a constant inward
fraction for each of the lines (approximately a luminosity weighted-average)
of 70\% for C~{\sc iv} and He~{\sc ii}, and 80\% for H$\beta$ and Mg~{\sc ii},
similar to the values determined for these lines over a broad range of
physical conditions (e.g. O'Brien et~al.\ 1994; Goad 1995, Ph.D. thesis;
Korista et~al. 1997a). For individual BLR clouds, we adopt a line radiation
pattern which approximates the phases of the moon (see e.g., O'Brien et al. 1994).

In figure~\ref{plot_flux_eta}(b) we show the corresponding radial surface
responsivities $\eta(r)$ (Eq. 3) for each line.  These are derived from the
radial surface line emissivities $F(r)$, by measuring locally the slope of
this distribution and dividing by $-2$. Since the radial surface line
emissivities are a function of the ionising photon flux and our choice of
weighting in the $\Phi_{\rm H}$, $n_{\rm H}$ plane, so too are the radial
surface line responsivities. However, some general trends do exist.  For
example, the responsivities of H$\beta$ and Mg~{\sc ii} are anti-correlated
with the incident ionising photon flux, being stronger at low incident fluxes
(in the outer BLR, where the optical depths are lower). This behaviour is also
expected for He~{\sc ii} within the inner BLR (see
figure~\ref{plot_flux_eta}). The C~{\sc iv} responsivity tends to align along
lines of constant ionisation parameter (anti-correlating with $U = \Phi_{\rm
H}/n_{\rm H}c$), but also gently declines for larger incident fluxes at
constant $U$ for gas densities $n_{\rm H} \geq 10^{10}$~cm$^{-3}$ (see also
Korista \& Goad 2004, their figure~4).

\subsection{The driving continuum light-curve}

A model that describes the BLR gas distribution, must be able to reproduce the
measured broad emission-line intensities and importantly their relative
response to variations in the ionising continuum. While a single spectrum
provides a ``snapshot'' of the physical conditions within the BLR, it is not a
``true'' representation of the current state of the line-emitting gas. This
arises because the BLR is physically extended and therefore a single spectrum
comprises contributions from gas in response to prior continuum states (due to
the finite light-travel time across the BLR) up to and including the continuum
state at the current epoch.  Consequently, in order to compare the general
characteristics of our fiducial bowl-shaped BLR model (see Table~1) with
observations, we must first drive it with a model continuum light-curve whose
variability power is representative of that determined for the source in
question, taking into account both sampling effects and the finite duration of
the light-curve (data sampling can be so sparse that the highest frequencies
present in the underlying PSD are only poorly sampled by the observations).
Our aim here is not to match the detailed response of individual lines, nor to
fit the emission-line spectrum at a single epoch (for the reasons outlined
above), rather here we aim to explore the gross properties of bowl-shaped
geometries by determining the time-averaged spectrum and the average
emission-line response timescale for the strongest UV and optical
recombination lines for comparison with observations.

Here we adopt the approach of Kelly et~al.\ (2009), and model the driving
continuum light-curve as a damped random walk (see Kelly et~al.\ 2009, and
Macleod et al.\ 2010, for details), which has been shown to be an extremely
good match to the variability of quasars (Kozlowski et al.\ 2010).  We set the
characteristic continuum variability timescale $\tau_{\rm char}$ to that
measured by Collier and Peterson\ 2001, $\tau_{\rm char} = 40$~days for
NGC~5548, with a total duration of 450 days and daily sampling.
%%If the observed variations are due to thermal
%%variations in the disc, then for the mass of the central black hole used here
%%$10^{8}M_{\odot}$, this equates to a disc viscosity $\alpha \approx 0.025$
%%{\bf [check this because we now use $10^{8}$]}.
%In order to simulate the light-curves, we first draw a random variable from a
%normal distribution with mean $\tau b$ and variance $\tau\sigma^{2}/2$, where
%$\tau$ is the characteristic timescale (relaxation time) and $b$...
The duration of our light curve is chosen to approximately match that of a
typical monitoring season noting that the first $\tau_{max}$ points must be
discarded since at prior times, the outer BLR has yet to respond to the
continuum variations.  

For each input continuum light-curve we compute the corresponding velocity
resolved emission-line light-curve $L(v,t)$, {\em assuming that locally, the
emission-line gas responds linearly to continuum variations\/}, i.e. a
locally-linear response approximation (see Goad et~al. 1993), so that while
$\eta(r)$ remains time-independent, we now take into account the radial
variation in line responsivity $\eta(r)$.  As shown in
Figure~\ref{plot_flux_eta}(b), and by Goad et~al. 1993, the emission-line line
responsivity shows a strong radial dependence. However, a locally linear
response has been shown to be a reasonable approximation {\em provided that
the amplitude of the continuum variations are small\/} (Goad, 1995, Ph.D.\
thesis, O'Brien et~al.\ 1995). For larger continuum variations the locally
linear response approximation is no longer valid and a full treatment of the
time-dependent non-linear effects is required.\footnote{We do not consider
changes in the emission-line response due to variations in ionising continuum
shape nor variations in the inner and outer BLR boundaries. In any case, it is
likely that local variations in $\eta(r)$ due to changes in $\Phi(\rm H)$ will
be the dominant non-linear effect for the majority of lines. A thorough
treatment of these non-linear effects in the context of bowl-shaped BLR
geometries will be explored elsewhere.}

For each combination of continuum--emission-line light-curve we compute the
peak and the centroid of their cross-correlation function (CCF, see
e.g., Gaskell \& Peterson 1987), for the latter adopting a threshold of 0.6
for the centroid calculation, as well as four different measures of the
velocity field, the $fwhm$ and dispersion of the mean and root-mean square
(hereafter rms) profiles, quantities that are commonly used by observers in
the estimation of black hole masses using the reverberation mapping technique
and/or scaling relations derived therein.  To ensure that the full range in
light-curve behaviour is sampled we repeat this process 1000 times and
thereafter compute probability distribution functions for each of the measured
quantities. In the following section we present the results of our simulations
and place them within the context of results reported in the literature.

\section{Results}\label{results}

Before exploring the effects of reverberation within the physically extended
BLR on measurements of the continuum--emission-line delay and emission-line
velocity dispersion, parameters typically used in virial mass estimates of the
central black hole, we start by investigating an idealised case in which the
predicted delay and velocity dispersion have been measured from the
instantaneous 1-d responsivity-weighted response function and variable line
profile. Such a situation would arise, for example, if the BLR were
illuminated by a delta-function pulse rise in the ionising continuum flux.  In
so doing, we aim to reveal the relationship between the physical properties of
our model (ie. our assumed geometry, emissivity, responsivity, velocity field,
turbulence, and inclination) and the measured quantities used in black hole
mass determinations.

The black hole mass estimates are here based on a virial product formed from
the mean square dispersion, $\sigma_{l}^{2}$, of the variable line profile and
the centroid, $\tau_{\rm cent}$, of the 1-d responsivity-weighted response
function, such that.
\begin{equation}
M_{BH} = \frac{R_{blr}\delta v^{2}}{G}  \, .
\label{vp}
\end{equation}

\noindent where $R_{blr} = c \tau_{\rm cent}$, and $\delta v^{2} =
\sigma_{l}^{2}$. 
Since the black hole mass is an input to our model, it is
straightforward to determine the virial scale factor $f$ required to force
agreement between the measured black hole mass and the input
value\footnote{For our fiducial BLR geometry, $\alpha=2$, $\beta=1/150$, the
denominator in equation~2 is 53\% larger ($(4/3)^{3/2}$) than for a standard
Keplerian velocity field at the outer radius. Hence in the absence of
turbulence, the velocities are lower than in the standard Keplerian model, and
thus the masses derived using equation~\ref{vp} are also lower, 
even before the effects of inclination are taken into account.}.

We have calculated virial masses, virial scale factors $f$, and line shapes
for two pairs of emission-lines, H$\beta$--Mg~{\sc ii}, and C~{\sc
iv}--He~{\sc ii}, chosen because they are representative of lines typically
measured in ground- and space-based AGN monitoring campaigns, and importantly
because they span a broad range in their characteristic line formation radii,
and thus will probe the largest range in variability behaviour for our
fiducial BLR geometry.  We use the calculated radial surface line emissivity
distributions of \S4.1 (see figure~9), and adopt a locally-linear response
approximation. We include the effects of anisotropic line emission, as
described near the end of \S4.1

\subsection{The importance of turbulence, line emissivity, and inclination 
on $M_{\rm BH}$ determinations}\label{tei}

\begin{figure}
%\resizebox{\hsize}{!}{\includegraphics[angle=0]{plot_f_new_jan2012.ps}}
%\resizebox{\hsize}{!}{\includegraphics[angle=0]{goad_fig14.ps}}
% now goad_fig10
\resizebox{\hsize}{!}{\includegraphics[angle=0]{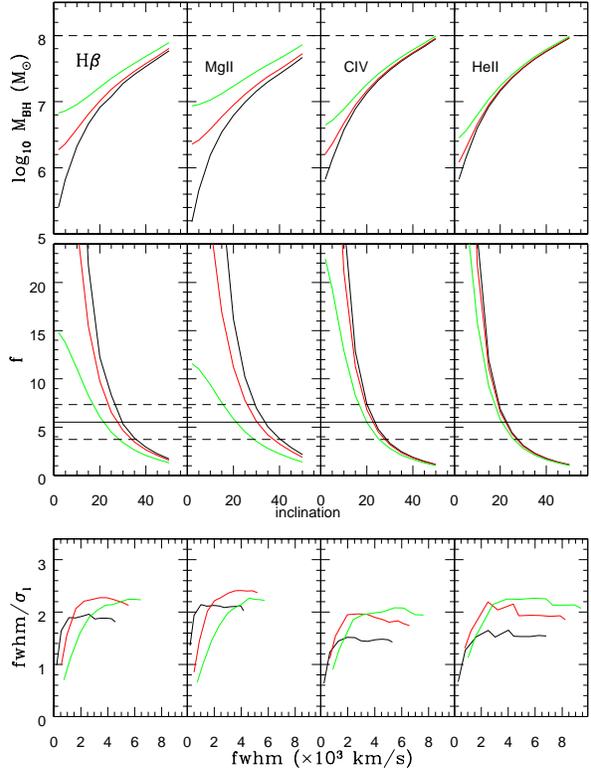}}
\caption{Panels 1--4 -- $M_{\rm BH}$ as a function of line of sight
inclination $i$ for each line for our fiducial BLR geometry and 3 values of
the turbulence parameter, $b_{\rm turb}=0$ (black line), $b_{\rm turb}=1$ (red
line) and $b_{\rm turb}=2$ (green line). Mass estimates are here based on
measurements of the mean-square dispersion of the steady-state (mean) line
profile and centroid of the 1-d responsivity-weighted response
function. Panels 5--8 -- the corresponding virial scale factors, $f$ also as a
function of inclination $i$. Panels 8--12 -- dependence of line shape
$fwhm/\sigma_{l}$ as a function of line $fwhm$ (for the same 3 values of
$b_{\rm turb}$).}
\label{plot_f_new}
\end{figure}

To assess the impact of turbulence on the derived black hole masses and virial
scale factor $f$, we have estimated $M_{\rm BH}$ for our fiducial BLR geometry
for each of the four lines, for line-of-sight inclinations spanning the range
2--50 degrees and using 3 values for the turbulence parameter, $b_{\rm
turb}=0$ (no turbulence), $b_{\rm turb}=1$ (moderate turbulence) and $b_{\rm
turb}=2$.  In figure~\ref{plot_f_new} we show the virial mass estimates using
equation~\ref{vp} for each of the four emission lines (upper 4 panels), as
well as the virial scale factor $f$ (middle 4 panels), required to reproduce
the input black hole mass, as a function of line-of-sight inclination
$i$. Individual colours indicate the degree of turbulence, $b_{\rm turb}=0$
(black line), $b_{\rm turb}=1$ (red line) and $b_{\rm turb}=2$ (green line).
We find that in all cases, the estimated black hole mass is lower than the
input mass, and is a strong function of inclination, being larger at larger
line-of-sight inclinations. The middle 4 panels of figure~\ref{plot_f_new}
indicate that for a given inclination, and in the absence of reverberation
effects, each line requires a different virial scale factor to recover the
input mass. The discrepancy between the emission-lines is largest for zero
turbulence and low line-of-sight inclinations.  Additionally, black hole mass
estimates are systematically lower and thus $f$-factors systematically higher
with turbulence switched off (solid black line). The discrepancy between the
virial mass estimate and the input mass can be larger than 2 dex for near
face-on systems with turbulence switched off.  For lines which form at large
BLR radii, corresponding to a larger scale height in our model, e.g. H$\beta$,
Mg~{\sc ii}, turbulence significantly enhances the black hole mass estimates
at low line of sight inclinations (solid green line), reducing the discrepancy
between the virial mass estimate and the input mass by more than 1 dex. This
arises because for our chosen velocity field, in the absence of turbulence,
the line width at small line-of-sight inclinations is a result of the combined
effects of TDS and GR only. Turbulence in our model acts to increase the local
velocity field at large scale heights thus broadening the emission-line and
thereby increasing the virial mass estimate and consequently lowering the
virial scale factor. Since in our model turbulence increases with increasing
scale height this effect is more pronounced in lines which form at large BLR
radii (e.g. H$\beta$, Mg~{\sc ii}). Furthermore, turbulence randomises the
direction of the velocity field and thereby acts to reduce the otherwise
strong dependence of the emission-line width on inclination, normally found
for planar Keplerian motion.

In the lower panel of figure~\ref{plot_f_new} we show the emission-line shape
($fwhm/\sigma_{l}$) as a function of line $fwhm$ (or equivalently
line-of-sight inclination $i$). This reveals a strong dependence of
emission-line shape on line $fwhm$, such that broader lines have more Gaussian
profiles, while narrower lines are more Lorentzian in shape.  In general
turbulence acts to soften the otherwise strong dependence of line shape on
inclination. However at low line-of-sight inclinations, turbulence dominates
over the planar Keplerian motion moving low velocity gas from the line core to
the line wings, resulting in profiles with narrower cores and extended line
wings ($fwhm/\sigma_{l} < 1$).  Once again, this effect is most pronounced in
lines which form at large BLR radii (e.g. H$\beta$, Mg~{\sc ii}).
Figure~\ref{plot_f_new} indicates that for lines formed at small scale heights
(C~{\sc iv} and He~{\sc ii}), the effect of turbulence on the derived $M_{\rm
BH}$ and $f$-factors is significantly smaller.

%\onecolumn

Onken et~al. (2004) showed that nearby AGN for which both reverberation
based and velocity dispersion based mass estimates are available, follow the
same $M_{BH}-\sigma$ relation as quiescent galaxies. From this they derived a
statistical estimate of the virial scale factor for all AGN with reverberation
mapping data, finding an average $f$-factor of $f=5.49\pm1.65$ for H$\beta$,
see also Table 2 of Collin et al. (2006), here indicated by the horizontal
solid and dashed lines (middle panel of figure~\ref{plot_f_new}).  Woo
et~al. (2010) provide an updated value for the virial scale factor $f$. Their
estimate $\log f = 0.72\pm0.1$ is based on matching reverberation masses of 24
AGN with those obtained from recent stellar velocity dispersion estimates of
the host galaxy, adopting the $M_{\rm BH}-\sigma$ relation of quiescent
galaxies from G\"ultelkin et al. (2009), and is consistent within the errors
with that found by Onken et al. (2004). We note that for our fiducial BLR
geometry, the inclusion of a scale height dependent turbulent component,
increases the range in inclination returning viral scale factors that are
consistent within the errors with the average $f$-factor for H$\beta$ reported
by Collin et~al (2006) and Woo et al. (2009).

\begin{figure}
%\resizebox{\hsize}{!}{\includegraphics[angle=0]{plot_disc_beta_feb2012.ps}}
%\resizebox{\hsize}{!}{\includegraphics[angle=0]{goad_fig15.ps}}
%now goad_fig11
\resizebox{\hsize}{!}{\includegraphics[angle=0]{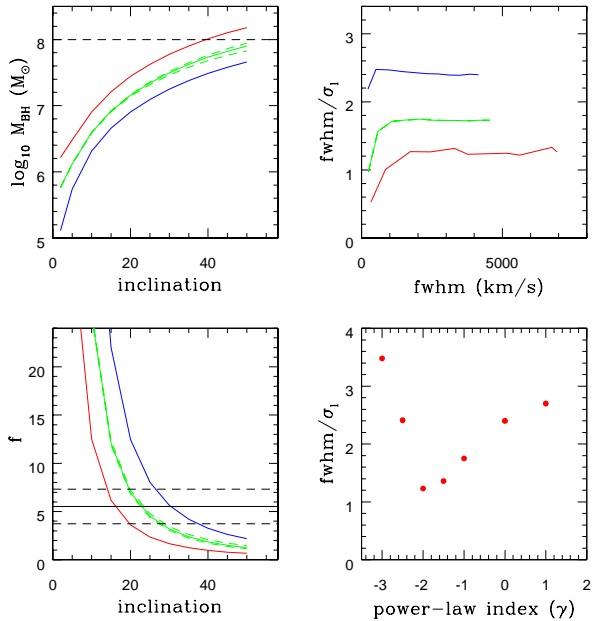}}
\vspace{-15mm}
\caption{Panel 1 (upper left) -- $M_{\rm BH}$ as function of inclination for a
geometrically thin disc and power-law emissivity distribution with power-law
index $\gamma=0$ (blue lines/symbols), $-1$ (green) and $-2$ (red). Also shown
(green dashed lines) is the effect of changing the line radiation pattern (for
$\gamma=-1$) from 50\% (isotropic)--100\% (anisotropic). Panel 2 (lower left)
- corresponding virial scale factors.  Panel 3 (upper right) - corresponding
line profile shapes as a function of $fwhm$. Panel 4 (lower right)
$fwhm/\sigma_{l}$ as a function of power-law emissivity index
$\gamma$. Differences in the ratio $fwhm/\sigma_{l}$ between lines may result
from differences in their radial surface emissivity distributions. N.B. a
geometrically thin disc
has a very small scale height by construction, and thus the turbulent velocity is
effectively zero even though we chose $b_{\rm turb}=2$.}
\label{plot_disc_beta}
\end{figure}

For completeness we show in the left hand panels of
figure~\ref{plot_disc_beta} the derived black hole mass and virial scale
factor, $f$, as a function of inclination for a geometrically thin disc with
the same inner and outer radius as our fiducial BLR geometry (1.14 and 100
light-days respectively), turbulence parameter $b_{\rm turb}=2$, and using
simplified power-law line emissivity distributions with power-law index
$\gamma=0$ (solid blue line), $-1$ (solid green line), and $-2$ (solid red
line). We also show (dashed green lines) the effect of changing the line
anisotropy from 50\% (ie. isotropic, $F_{inwd}=\frac{1}{2}F_{totl}$) to 100\%
(ie. anisotropic, $F_{inwd}=F_{totl}$). Line anisotropy increases the
emissivity-weighted radius by a factor $1+\frac{1}{2} \sin^{2} i$, (Goad 1995,
Ph.D.\ thesis, O'Brien et~al.\ 1994), and thus for the limited range of
inclinations studied here results in a marginal increase in the virial scale
factor for a geometrically thin disc.  As for our bowl-shaped geometry the
masses are systematically underestimated (except at large inclination and for
steep emissivity distributions -- red line, upper left-panel) and the
discrepancy between the calculated and input black hole mass decreases with
increasing inclination. The thin disc model also shows a significant change in
profile shape at small inclinations (figure~\ref{plot_disc_beta}, upper right
panel). Since a disc by construction has little scale height, the contribution
of the turbulent component is minimal (even though $b_{\rm turb}=2$). Thus for
a given radial surface line emissivity distribution, only GR and TDS can play
a significant role in modifying the line shape for thin-disc geometries at
small inclinations. We note that in the absence of these effects, the
emission-line shape is independent of inclination for our adopted
emission-line radiation pattern.  Of more significance is the strong
dependence of the mass estimates on the radial surface line emissivity
distribution, which indicates that {\em in the absence of turbulence, lines
which are preferentially formed at small BLR radii yield larger black hole
masses\/}, and thus smaller virial scale factors. A similar result was found
for our fiducial model (figure~\ref{plot_f_new} -- solid black lines,
cf. H$\beta$--Mg~{\sc ii} and C~{\sc iv}--He~{\sc ii}). We discuss this
further in the following section.

Finally, we note that unless turbulence dominates the velocity field, then for
a bowl-shaped BLR geometry, we expect to observe a strong $f$--$fwhm$
dependence in all lines regardless of where they form (due to the strong
dependence of $fwhm$ on inclination), and broadly similar to that found for
disc-like configurations (see Decarli et~al. 2008, their figure 6). Scatter in
the relation between different lines may point to differences in their radial
surface line emissivity distributions and/or the presence of a significant
turbulent velocity component.

\subsection{$fwhm/\sigma_{l}$ - the role of emissivity}\label{fwhm_sigma}

% steeper emissivity gives lower fwhm/\sigma

As noted by Collin et~al.\ (2006) the ratio of the zeroth and 2nd order
moments (mean and root mean square respectively) of the broad emission-line
profile can be used to place constraints upon the BLR geometry and kinematics.
Decarli et~al.\ (2008) found that in a sample of 36 AGN with roughly equal
numbers of radio-loud and radio-quiet objects, that the ratio
$fwhm/\sigma_{l}$ for H$\beta$ is closer to that derived for an isotropic BLR
geometry (wherein $fwhm/\sigma_{l}=2.35$), while the smaller values ($\approx
1$) measured for C~{\sc iv} are more suggestive of a flattened BLR
geometry. Combined with the strong correlation between their virial scale
factor $f_{d}$\footnote{Decarli et al.\ (2008) use a different definition of
the virial scale factor $f_{d}$, $v_{\rm blr}=f_{d} \times fwhm$, which using
our definition implies $f_{d} = \sqrt f$.}  and $fwhm$ found for both lines
(such a correlation is expected if line of sight velocity is a strong function
of inclination), and the absence of a correlation between the $fwhm$ of
H$\beta$ and C~{\sc iv}, Decarli et al.\ (2008) argue that their results are
consistent with C~{\sc iv} originating in a flattened BLR geometry, with
H$\beta$ originating in a geometrically thick disc with a significant
turbulent component. As in our model, the increased turbulence at large scale
height reduces the dependence of the H$\beta$ line width on
inclination and therefore could account for the reported absence of a
correlation between the H$\beta$ and C~{\sc iv} emission-line widths, the more
isotropic appearance of the H$\beta$ line profile, and if the turbulence is
large enough, the larger width of H$\beta$ relative to C~{\sc iv}\footnote{We
note that the number of objects for which {\em simultaneous\/} line width
comparisons between H$\beta$ and C~{\sc iv} have been made is small, and that
claims of a non-correlation between their respective line widths may be
premature.}.

However, we caution here that the ratio $fwhm/\sigma_{l}$ for fixed $R_{in}$,
$R_{out}$ has a strong dependence on the radial surface line emissivity
distribution. For example, if we approximate the radial surface line
emissivity distribution as a power-law in radius ($F(r)\propto r^{\gamma}$),
then for a geometrically thin disc, the ratio $fwhm/\sigma_{l}$ is a minimum
for $\gamma=-2$, and increases for both smaller and larger $\gamma$ (see
figure~\ref{plot_disc_beta}, lower right panel).  For $\gamma$ large and
positive, the emission is weighted toward the outer edge only, yielding
$fwhm/\sigma_{l}\approx 2.8$ as appropriate for a ring-like
distribution. Similarly, for negative $\gamma$, the inner radius dominates the
emission, while emission from rings at larger radii (and hence lower
velocity), tend to fill in the dip at line core (between the horns) producing
more rectangular-looking profiles (smaller $fwhm/\sigma_{l}$). In extreme
cases ($\gamma < -3$), this can lead to $fwhm/\sigma_{l} > 3$. For our
fiducial model, C~{\sc iv} has a steeper emissivity distribution than H$\beta$
and thus a smaller $fwhm/\sigma_{l}$ (cf. figure~\ref{plot_f_new} -- lower
panels 1 and 3, and figure~\ref{plot_disc_beta} -- upper right panel), which
may in part explain the differences found by Decarli et~al.(2008). A similarly
strong dependence of line shape on radial surface line emissivity distribution
can be found for other BLR geometries (see also Robinson 1995a,b)\footnote{In
support of this claim, we show in Figure~A2 the mean responsivity-weighted
(black solid line) and emissivity-weighted (red dotted line) emission-line
profiles, for both the low- and high-ionisation emission lines, for our
fiducial model with turbulence parameter $b_{\rm turb}=2$, and a range of line
of sight inclinations.}  Thus while differences in the ratio of
$fwhm/\sigma_{l}$ between H$\beta$--C~{\sc iv} may indicate differences in
their scale height, we suggest that differences in their radial surface line
emissivity distributions likely also play a significant role.  Evidence in
support of this claim comes from the large range (factor of a few) in
$fwhm/\sigma_l$ displayed by the H$\beta$ line in NGC~5548 during 13 years of
monitoring (Collin et~al. 2006). Since in an individual source neither the
black hole mass nor its inclination can change appreciably on such a short
timescale, this suggests that the observed variations in the Balmer line
profile shape for NGC~5548, over the 13 years of observation, are due to gross
changes in the radial surface line emissivity distribution in response to
large variations in the ionising photon flux within the physically extended
BLR, though we cannot rule out dynamical changes on longer baselines
($\sim$several years for NGC~5548).

\begin{figure}
%\resizebox{\hsize}{!}{\includegraphics[angle=0,width=8cm]{plot_tau_ll.ps}}
%\resizebox{\hsize}{!}{\includegraphics[angle=0,width=8cm]{goad_fig10.ps}}
%now fig 12
\resizebox{\hsize}{!}{\includegraphics[angle=0,width=8cm]{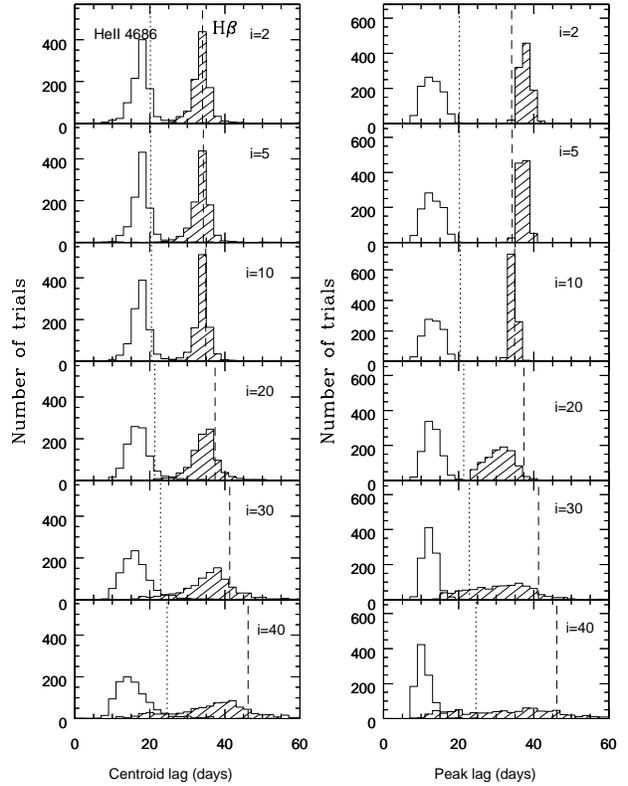}}
\caption{H$\beta$ (hashed lines) and He~{\sc ii} CCF centroid (left panels)
and peak (right panels) distribution functions (resulting from 1000 simulated
continuum--emission-line light-curves) for a range of BLR inclinations. Also
shown is the centroid of the 1-d response function for H$\beta$ (vertical
dashed line) and He~{\sc ii} (vertical dotted line) as determined from the
steady-state model.}
\label{plot_cent_hb}
\end{figure}

\begin{table}
% \centering
\caption{Our fiducial BLR model.}
  \begin{tabular}{@{}ccccccc@{}}
  \hline
$M_{BH}$ & $R_{min}$ & $R_{max}$ & $\tau(R_{max})$ & $\alpha$ &  $\beta$ &
  $b_{\rm turb}$ \\
($M_{\odot}$) & (lt-days) & (lt-days) & (lt-days) &  &   &   \\ \hline
$10^{8}$  & 1.14 & 100.0 & 50.0 & 2.0 & 1/150 & 2.0 \\
\hline
\end{tabular}
\end{table}

\subsection{The importance of Reverberation: Which measures of $R_{blr}-\delta v$ should be used?}

In order to assess the impact of continuum--emission-line reverberation on the
virial mass estimates and hence scale factors $f$, we have driven our fiducial
BLR geometry, with simulated continuum light-curves whose variability
characteristics have been designed to match those observed in the best studied
AGN, the Seyfert~1 galaxy NGC~5548. Our fiducial BLR model (see Table~1 for
details) utilises the computed radial surface line emissivity distributions for
each line (\S4.1), our prescription for emission-line anisotropy and line
radiation pattern (\S4.1), and a turbulence parameter
$b_{\rm turb}=2$.  All simulated emission-line light curves have been
calculated assuming a locally-linear response approximation.  For each of the
four lines considered (H$\beta$, Mg~{\sc ii}, C~{\sc iv} and He~{\sc ii}), we
compute their velocity resolved time-variable emission-line light-curve
$L(v,t)$, and their {\em mean\/} and {\em root mean square\/} (rms)
emission-line profiles.

% (due to differences in their characteristic line formation
%radii). We include the effects of anisotropic line emission assuming for each
%line a constant inward fraction (approximating a luminosity-weighted average)
%of 80\% for H$\beta$--Mg~{\sc ii}, and 70\% for C~{\sc iv}--He~{\sc ii}
%similar to the values determined for these lines over a broad range of
%physical conditions (e.g. O'Brien et~al.\ 1994; Goad 1995, Ph.D. thesis;
%Korista et~al. 1997a), and adopting a radiation pattern for the line emission
%which approximates the phases of the moon (e.g. O'Brien et~al. 1994).

\begin{figure}
%\resizebox{\hsize}{!}{\includegraphics[angle=0,width=8cm]{plot_tau2_ll.ps}}
%\resizebox{\hsize}{!}{\includegraphics[angle=0,width=8cm]{goad_fig11.ps}}
%now fig 13
\resizebox{\hsize}{!}{\includegraphics[angle=0,width=8cm]{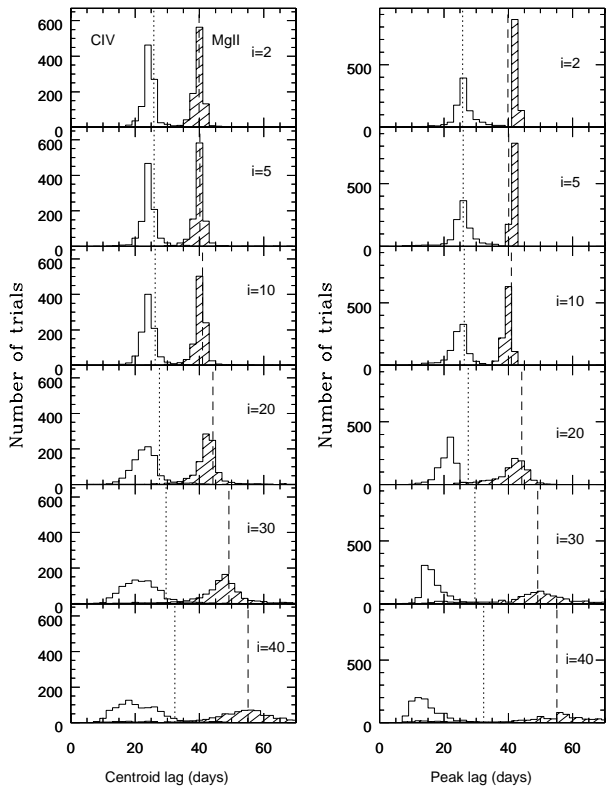}}
\caption{As for figure~\ref{plot_cent_hb} for Mg~{\sc ii} (hashed area) and C~{\sc iv}.}
\label{plot_cent_c4}
\end{figure}

Simulations were performed for a range of line of sight inclinations
i=2,5,10,20,30,40, and 50 degrees, and repeated 1000 times, in order to ensure
that the full range in continuum variability behaviour is sampled and to allow
the construction of probability distribution functions in each of the desired
quantities that we wish to measure. For each continuum -- emission-line
light-curve combination we have calculated the peak and centroid of the
cross-correlation function (figures~\ref{plot_cent_hb}--\ref{plot_cent_c4}).
Additionally, from the velocity resolved time-variable light-curve, we have
constructed mean and root-mean square profiles from which measurements of
their full width at half maximum ($fwhm$) and line dispersion ($\sigma_{l}$)
have been determined. Thus, from each simulation we have two estimates of the
continuum--emission-line delay (or characteristic size of the BLR), and four
estimates of the velocity dispersion (two from the mean profile and two from
the rms profile). From these we construct eight estimates of the BLR mass in
the standard fashion using equation~\ref{vp}.

%by a factor of $\approx 2.37$.}.

For a few individual objects, intensive ground and space-based spectroscopic
monitoring campaigns have been employed to determine black holes masses under
the assumption of virialised gas motion for broad emission-lines spanning a
broad range in ionisation state (e.g. Clavel et~al. 1991, Ulrich et~al. 1991,
Peterson et~al. 1991, Krolik et al. 1991). For NGC~5548 Peterson and Wandel
(1999) showed that while lines of differing ionisation state display
differences in both their line dispersion and response timescale they all
appear to follow (more or less) the same virial relation. Furthermore,
Peterson et al. 2004, and Collin et~al. 2006, using 13 years of monitoring
data of the optical emission-lines in NGC 5548 showed that differences in the
measured lags and line widths from one season to the next yielded virial
products which are effectively constant within the errors, demonstrating that
the reverberation masses while not necessarily accurate, are at least robust
(reproducible).  However, as noted by these authors, it is not immediately
obvious which of the measured quantities (CCF peak or centroid, $fwhm$,
$\sigma_{l}$ of the mean or rms profile) yield the most accurate black hole
masses, nor which of the emission-lines should be used in mass determinations.

% the lines are typically broader in mean spectra
Theoretical considerations suggest that measurements from rms profiles rather
than mean profiles should yield more accurate estimates of the virial product,
since they isolate the variable part of the emission-line from non-variable
components (Fromerth and Melia 2000).  However, because of the lower S/N of
rms profiles, a consequence of the small variability amplitude, measurements
from rms profiles tend to be more uncertain. Collin et al. (2006) argued that
provided the non-varying components can be isolated from the mean profile,
then the mean profile should yield estimates of the virial product of a
similar precision\footnote{Collin et~al. 2006, as with previous authors find
that mean profiles are typically 20\% broader than rms profiles, possibly due
to the presence of low responsivity gas in the inner BLR.}. Additionally,
Peterson et al. (2004) showed that the line dispersion $\sigma_{l}$ can be
measured with greater precision than the line $fwhm$ and provides a better
match to the virial relation, which suggests that estimates of the virial
product based on $\sigma_{l}$ should have smaller uncertainty.  Finally, the
centroid of the CCF, equivalent to the centroid of the 1-d response function
(or the emissivity-weighted radius for a linear response) is the generally
preferred quantity used for estimating the 'size' of the BLR since it is less
sensitive to the inner BLR radius than is the peak of the CCF (Peterson et
al. 1998), but see Welsh 1999, for an alternative viewpoint. Indeed, for
objects in which monitoring campaigns have been performed for multiple lines,
Peterson et~al. 2004 found that the tightest virial relation was found for
virial products determined from measurements of the centroid of the CCF
($\tau_{cent}$) and the line dispersion ($\sigma_{l}$) of the rms spectrum.

In summary, from an observational perspective virial products based on
measurement of the CCF centroid and the dispersion of the rms profile remain
the quantities of choice. However, it seems at least plausible that the choice
of these quantities may simply reflect limitations in the observational data,
be it signal--noise or sampling. Here we attempt to address the question as to
which of the measured quantities and for which lines yield the most accurate
black hole masses from a modelling perspective.

\subsubsection{CCF centroid or peak}

In terms of the measured delay, or characteristic size of the BLR, it has long
been known that the CCF peak (or lag) is a less reliable measure of the
characteristic BLR size than is the centroid, and tends to be biased towards
the inner BLR radius (see e.g. Edelson and Krolik 1988; P\'{e}rez, Robinson
and de la Fuente 1992; Melnikov and Shevchenko 2008) though see Welsh 1999 for
an alternative point of view. By contrast, the centroid of the CCF is directly
related to the centroid of the response function (the responsivity weighted
radius, Koratkar and Gaskell 1991) and thus should be a more accurate
representation of the 'size' of the region responding to continuum variations
(the correspondence is not exact even for the centroid and depends strongly on
the continuum variability behaviour exhibited during the observing campaign,
see e.g. P\'{e}rez, Robinson and de la Fuente 1992).

Figures~\ref{plot_cent_hb}--\ref{plot_cent_c4} show the distribution in the
measured delay in terms of both the centroid and peak delay (or lag) for
each line over a range of BLR inclinations. Also shown is the centroid of the
1-d responsivity-weighted response function (vertical dotted line for He~{\sc
ii} and C~{\sc iv}, vertical dashed line for H$\beta$ and Mg~{\sc ii}) for the
steady-state for the same inclination.  For low inclinations the CCF centroid
for all lines bar He~{\sc ii} tracks the centroid of the response function
reasonably well, while for larger inclinations ($i>10$ degrees), the measured
centroid displays larger scatter and is systematically smaller than the true
centroid in all lines. The discrepancy between the centroid of the 1-d
responsivity-weighted response function 
%(or equivalently the emissivity-weighted radius, assuming a linear response) 
and the centroid of the CCF is notably worse at large inclinations for He~{\sc
ii} and C~{\sc iv}, which in our model preferentially form at small BLR
radii. Furthermore while at low inclinations the peak of the CCF also tracks
the centroid of the response function reasonably well, it performs far worse
at larger inclinations. For He~{\sc ii} the peak delay is systematically
smaller than the centroid of the 1-d response function regardless of
inclination. Our adopted continuum variability timescale of 40
days is close to the centroid of the 1-d response function for H$\beta$ and
Mg~{\sc ii} respectively and is the most likely explanation for their improved
performance in tracking the continuum variations when compared to C~{\sc iv}
and He~{\sc ii}, particularly at large inclinations.

\begin{figure}
%\resizebox{\hsize}{!}{\includegraphics[angle=0,width=8cm]{plot_vir_new_hb+he2_ll.ps}}
% files are in plot_cent_hb etc plot_vir_new_hb
%\resizebox{\hsize}{!}{\includegraphics[angle=0,width=8cm]{goad_fig12.ps}}
%now fig 14
\resizebox{\hsize}{!}{\includegraphics[angle=0,width=8cm]{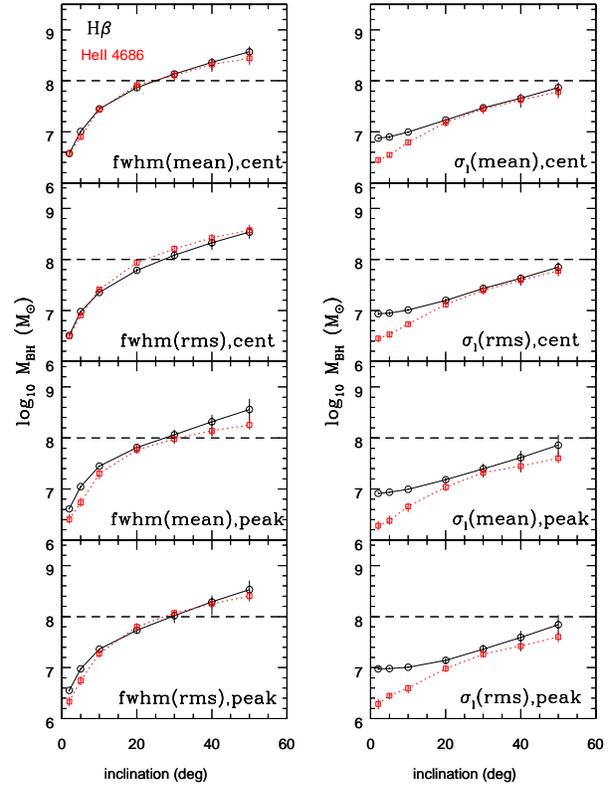}}
\caption{8 estimates of the black hole mass, $M_{BH}$ (median, and 1$\sigma$
confidence interval) based on the virial product $R_{blr}\delta v^{2}/G$ as a
function of inclination $i$ for our fiducial BLR model (see Table 1), and
based on 1000 simulated light-curves at each inclination. Our calculated
virial products (Eqtn. 6) employ two measures of $\delta v$, $fwhm$ and
$\sigma_{l}$, and two measures of $R_{blr}$ ($\equiv c\tau$), the CCF centroid
and CCF peak. Together these provide 4 estimates of $M_{BH}$ for both the {\em
mean\/} and {\em rms\/} profiles (giving 8 estimates in total). Individual
panels show results for H$\beta$ (open circles, solid black line), and He~{\sc
ii} $\lambda$4686 (open squares, red dashed line).}
\label{plot_vir_hb+he2}
\end{figure}

\begin{figure}
%\resizebox{\hsize}{!}{\includegraphics[angle=0,width=8cm]{plot_vir_new_mg2+c4_ll.ps}}
%\resizebox{\hsize}{!}{\includegraphics[angle=0,width=8cm]{goad_fig13.ps}}
%now fig15
\resizebox{\hsize}{!}{\includegraphics[angle=0,width=8cm]{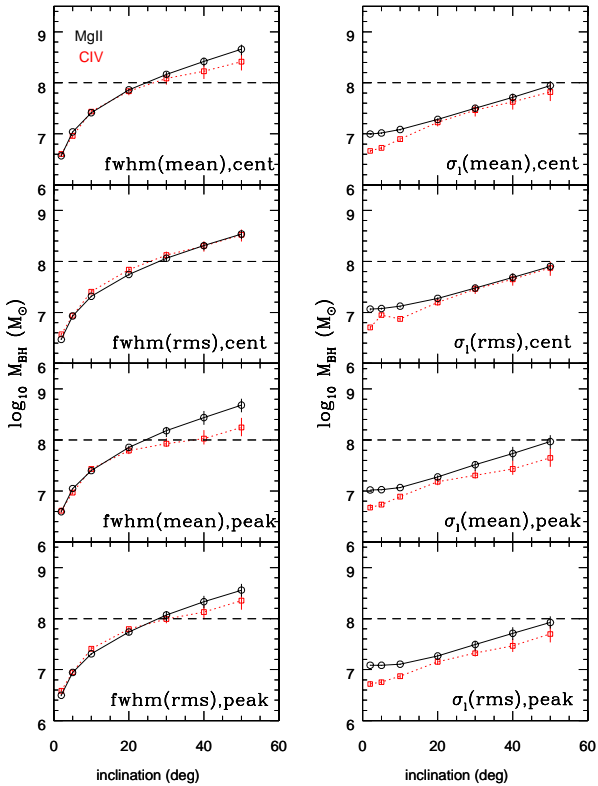}}
\caption{As for figure~\ref{plot_vir_hb+he2}, but now showing the results for
Mg~{\sc ii} $\lambda$2798 (open circles, solid black line), and C~{\sc iv}
$\lambda$1550 (open squares, red dashed line).  }
\label{plot_vir_mg2+c4}
\end{figure}

\subsubsection{Mean or rms profiles}
Figures~\ref{plot_vir_hb+he2}--\ref{plot_vir_mg2+c4} show the derived black
hole mass distribution functions (8 independent estimates) as a function of
line of sight inclination $i$ for all four lines. Because the mass
distribution functions are typically skew the points and their errors here
represent the median and 68\% confidence limits respectively.  The input black
hole mass is indicated by the horizontal dashed line.  As expected, the
derived $M_{\rm BH}$ for all four lines shows a strong dependence on
inclination, with black hole mass estimates systematically lower than the
input mass at low line of sight inclinations.  Mass estimates based on
measurements using the line $fwhm$ (from both the mean and rms profile) show a
greater dependence on inclination at small inclinations, than do mass
estimates based on measurements of the line dispersion $\sigma_{l}$ (ie. as
noted by Collin et al. 2006, $\sigma_{l}$ is a less biased indicator of mass).

For our model, $fwhm/\sigma_{l} \approx 1$ at i=2 degrees, and increases to a
maximum for inclinations $i> 20$ degrees after which time $fwhm/\sigma_{l}$ is
approximately constant (see also figure~\ref{plot_f_new}). Thus the line
shape is largely insensitive to inclination effects at large inclinations but
is a strong function of inclination at small inclinations ($i<$ 20 degrees),
inclinations for which scale-height dependent turbulence may significantly
modify the line profile (see \S~\ref{tei}).
The origin of this effect in our model is due entirely to the implementation
of Gravitational redshift and transverse Doppler shift (see
\S\ref{tei}).  In the absence of these effects, only the line width
(and not line shape) is a strong function of inclination.

Since $fwhm/\sigma_{l}>1$ for all lines (except at the lowest inclinations)
mass estimates based on measurements of the $fwhm$ are generally larger than
those based on $\sigma_{l}$. Black hole mass estimates based on the line
$fwhm$ systematically underestimate the mass at low line of sight
inclinations, while systematically overestimating the mass for line of sight
inclination $i > 20$~degrees (see figures~\ref{plot_vir_hb+he2}--\ref{plot_vir_mg2+c4}).
Mass estimates based on measurements of the line
dispersion, $\sigma_{l}$, systematically under-predict the black hole mass at
all inclinations and for all lines, with the largest discrepancies at the
lowest inclinations. The larger scatter in measurements of the $fwhm$ and
$\sigma_{l}$ of the rms profile is as expected (a reverberation effect), but
does not necessarily lead to a larger scatter in the derived mass, if
variations in the line widths are compensated for by corresponding changes in
the measured delay.

The use of CCF centroid or peak in the virial relation in general makes only a
small difference to the estimated masses (see e.g. left-hand or right-hand
panels of figures~\ref{plot_cent_hb}--\ref{plot_cent_c4}) though the exact
behaviour depends both on the line in question and details of the continuum
variability behaviour during a single model run.  For example, for He~{\sc ii}
the CCF centroid is on average larger than the CCF peak at all inclinations,
while for C~{\sc iv}, the CCF centroid is on average larger than the CCF peak
only for $i > 10$ degrees. Similarly, for H$\beta$, the CCF centroid is on
average larger than the CCF peak for inclinations in the range $10 < i < 50$
degrees, and smaller than the CCF peak otherwise. For Mg~{\sc ii} the CCF
centroid is on average larger than the CCF peak for inclinations in the range
$5< i < 30$ degrees, and smaller than the CCF peak otherwise.  Thus when
considering all lines, we expect to find on average larger black hole masses
at large inclinations when using the CCF centroid, while at low inclinations
the CCF centroid yields lower mass estimates than the CCF peak for all lines
except He~{\sc ii}.

%While steady-state models can be instructive (cf., H$\beta$--Mg~{\sc ii} and
%C~{\sc iv}--He~{\sc ii} in figure~\ref{plot_disc_beta}), they can also be
%misleading since they ignore changes in the virial product due to continuum
%variations.  Therefore, 

\begin{figure}
%\resizebox{\hsize}{!}{\includegraphics[angle=0,width=8cm]{plot_f_cent.ps}}
%\resizebox{\hsize}{!}{\includegraphics[angle=0,width=8cm]{plot_f_all_ll.ps}}
\resizebox{\hsize}{!}{\includegraphics[angle=0,width=8cm]{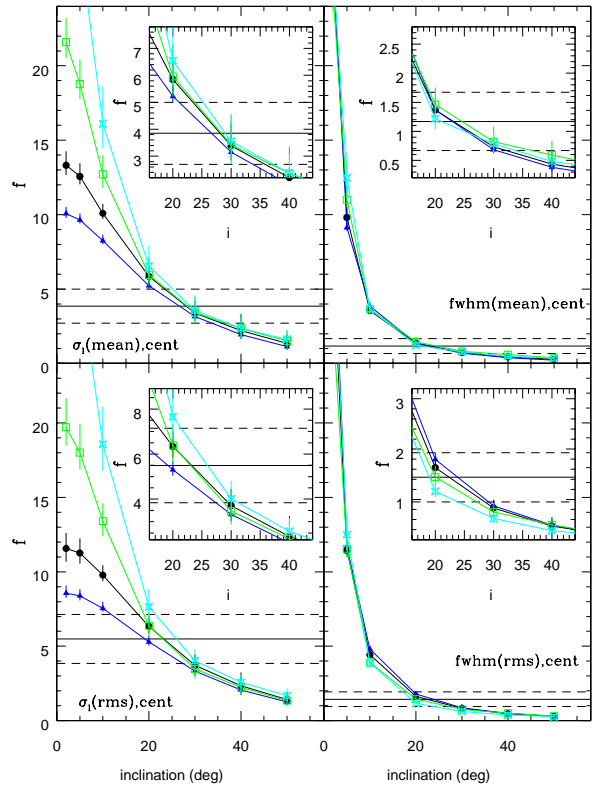}}
\caption{Four estimates of the virial scaling factor f, for four different
lines driven by 1000 continuum light-curves with a characteristic variability
time scale $\tau_{\rm char}$ of 40 days. Black filled circles (H$\beta$), blue
triangles (Mg~{\sc ii}), green squares (C~{\sc iv}), and cyan crosses (He~{\sc
ii}). The horizontal solid and dashed lines represent the average H$\beta$
scale factors, $f$(H$\beta$) and their uncertainties, as derived for the Onken
et al. sample by Collin et~al. 2006.}
% The red open circles are for H$\beta$ as driven by a continuum with
% characteristic timescale of 80 days.}
% figure constructed with plot_f (on angelica)
\label{plot_scale}
\end{figure}

We have recast the derived $M_{\rm BH}$ from our reverberation simulations of
our fiducial BLR model to illustrate the variation in virial scale factor $f$
as a function of inclination for each line.  Here we restrict ourselves to
four estimates only, the four used by Collin et~al.\ 2006, in which velocity
dispersion measurements ($fwhm$, $\sigma_l$) have been taken from the mean and
rms profiles, and the centroid of the CCF is used as a proxy for BLR size.
The results of this exercise are displayed in figure~\ref{plot_scale}, where
the points and their errors once again indicate the median and 68\% confidence
limits respectively. Increasing the characteristic variability timescale for
the continuum variations from 40 to 80 days only marginally alters the virial
scale factor for H$\beta$ (smaller for the line dispersion and larger for the
line $fwhm$).

Continuum variability blurs the differences between the predicted virial scale
factors for each line to the point at which they are virtually
indistinguishable within the errors at inclinations larger than
$\approx$20~degrees. Thus when measurement errors are included {\em we predict
little if any difference between the $f$-factors (or mass estimates) derived
for different lines\/}. Importantly, $f$-factors derived from the $fwhm$ of
the rms and mean profile show significantly less scatter and a closer
correspondence between the lines than those based on $\sigma_{l}$,
particularly at low inclinations.  This consistency between mass estimates for
different emission lines derived using the line $fwhm$ has been confirmed
observationally (Assef et~al. 2011).  Because our light curves are
well-sampled, there is only a small spread in the measured values of $fwhm$
and $\sigma_{l}$ of the mean profile. Thus the dispersion in the computed
virial scale factors from measurements of mean profiles is mostly due to
variations in the location of the centroid of the CCF. For light-curves which
are only sparsely sampled, we would expect larger variations in $fwhm$ and
$\sigma_{l}$ in the mean profiles, though still smaller than those measured
for rms profiles.  Figure~\ref{plot_scale} shows that for mass estimates based
on measurements of the line $fwhm$, the $f$-factors for all lines are
consistent within the errors ($f\pm\delta f$) with those reported by Collin
et~al. (2006) for inclinations in the range $20 <i < 35$ degrees. For
estimates based on measurements of the line dispersion $\sigma_{l}$, the
acceptable range narrows to $20 < i < 30$.

The average virial scale factor for the Decarli et~al. sample is
$f_{d}=1.6\pm1.1$ for H$\beta$ and $f_{d}=2.4\pm 0.16$ for C~{\sc iv}, that is
$f$-factors tend to be larger for C~{\sc iv} than H$\beta$.  For our fiducial
model, in the presence of turbulence $f$(H$\beta$)($\sigma_l$) $<$ $f$(C~{\sc
iv})($\sigma_l$) for $i<20$ degrees for both the mean and rms profile
(figure~\ref{plot_scale}), because of the stronger dependence of the H$\beta$
line width on turbulence due to it forming at preferentially greater distances
and scale height.  By contrast $f$-factors derived from the $fwhm$ of the mean
and rms profile indicate $f$(C~{\sc iv})($fwhm$)$\approx$
$f$(H$\beta$)($fwhm$), though the correspondence is generally worse for the
lowest inclinations.

The larger scatter in the $f$-factor between the low and high ionisation lines
at low $i$ derived using $\sigma_{l}$ (figure~\ref{plot_scale} -- left hand
panels) highlights the reduction in sensitivity of the line-dispersion to
inclination in the presence of strong turbulence (which mainly effects the low
ionisation lines). We suggest that objects for which the lines show large
differences in their virial scale factors (as derived using the line
dispersion) may be indicative of low inclination sources.

%%%%%%%%%%%%%%%%%%%%%%%%%%%%%%%%%%%%%%%%%%%%%%%%%%%%%%%%%%%%%%%%%%%%%%%%%%%%%%%%

\section{Discussion}\label{discussion}

A bowl-shaped BLR geometry provides an elegant solution to the smaller than
predicted dust reverberation sizes by decreasing the measured delays without
altering the dust formation radius. At the same time, since material now lies
away from the observers line of sight, they readily reproduce the observed
lack of response on short timescales evident in the 1-d and 2-d response
functions of the strong optical recombination lines.

Differences in the radial surface emissivity distribution and line anisotropy
among the low and high ionisation lines in the context of a bowl-shaped BLR,
results in large differences in the form of the 1-d response functions at
fixed inclination (see figure A3, appendix).  For geometrically thin discs,
the characteristic variability timescale as determined from the centroid of
their 1-d response functions is independent of inclination for isotropically
emitting clouds, and increases with inclination otherwise. For bowl-shaped
geometries, the centroid of the 1-d response function increases with
inclination even for isotropic emission.  Moreover for the radiation pattern
adopted here, at small inclinations, increased line anisotropy can reduce the
centroid of the response function relative to the isotropic case due to the
reduced contribution of gas at large elevations which lies closer to the
line-of-sight.

While the LILs show an absence of response on short timescales in their 1-d
response function, the 1-d response functions for the HILs, which are formed
at smaller BLR in a more flattened distribution, resemble those of disc-shaped
BLR geometries, with a significant response even on short timescales (though
the response does decline to zero at zero delay). Thus this model {\em
reproduces the observed differences in the location of the peak response among
low and high ionisation lines reported in the literature\/} (e.g., Krolik
et~al.\ 1991, their figure 10 and 11). We note that here we have assumed a
locally linear line response and that the precise form of the 1-d response
function may be modified in the event of significant non-linear effects (this
may include a luminosity-dependent continuum shape as well as the incident
continuum flux-dependent effects already mentioned).

The additional effects of GR and TDS enhance the red-wing response producing
line profiles with extended red-wings, as are sometimes observed in type 1 AGN
(Kollatschny 2003). The strength of the red-blue asymmetry depends primarily
on the line formation radius, being stronger for lines formed at small BLR
radii (i.e., steep radial emissivity distributions).  GR and TDS together
provide significant line width ($\sim$ several hundred km~s$^{-1}$) 
even for face-on geometries with pure-planar Keplerian motion. More
importantly, GR and TDS introduce a strong inclination dependence to the line
profile shape at low inclinations.  In the absence of these effects the shape
of the emission-line profiles are independent of inclination for
flattened/bowl-shaped BLR geometries assuming isotropic emission, though of
course their line widths will show a strong inclination dependence (the
observed line of sight velocity, $v_{obs}$, varies as $v\sin i$).

Mass determinations for flattened/bowl-shaped BLR geometries show a strong
dependence on inclination particularly at small inclinations where the mass
estimates are systematically smaller than the input model. In general, when
reverberation effects are included, mass estimates based on measurements of
the $fwhm$ underestimate the mass at low inclinations, and overestimate the
mass at high inclinations, while those based on measurements of the line
dispersion $\sigma_l$, systematically underestimate the mass at all
inclinations. Mass estimates based on measurements of the emission line $fwhm$
(rms or mean profile) are larger since in general $fwhm/\sigma_{l}>1$.  Our
model also confirms the result of Collin et al.\ 2006, showing that mass
estimates based on measurements of $\sigma_l$ are less biased than those
determined from measurements of the $fwhm$, because of the weaker dependence
of $\sigma_l$ on inclination, particularly at low inclinations.  For our
simulations, we find a better correspondence between mass determinations
derived from different lines particularly at low inclinations, if measurements
of the virial product are performed using the $fwhm$ of the mean or rms
profile.  The form of emission line anisotropy adopted in this model leaves
the profile unchanged, but causes the characteristic response timescale for
the line to increase with increasing inclination. This effect is smaller than
the corresponding increase in velocity with inclination and consequently both
$M_{\rm BH}$ and $f$ are only weakly dependent on line anisotropy
(e.g. figure~\ref{plot_disc_beta}, lower left panel).

Turbulence, as implemented here, modifies the shape of the 2-d response
function and emission-line profile, by moving lower velocity gas that responds
on long timescales to larger line of sight velocities (hence broadening the
lines). The overall effect on line shape is line dependent, and is largely
determined by the line formation radius, so that while turbulence may increase
the ratio $fwhm/\sigma_{l}$ for lines formed at small BLR radii (small scale
heights), the general effect is to reduce $fwhm/\sigma_l$ for lines formed at
large BLR radii (large scale heights), so that at low line of sight
inclinations, the line profiles are characterised by narrow cores and extended
line wings (i.e., Lorentzian), similar to those seen in NLSy1s, see figure 8 and
\S\ref{tei} for a full discussion of these effects.  Because turbulence
randomises the velocity field, turbulence acts to reduce the $v \sin i$
dependence of the line dispersion in flattened BLR geometries.

To summarise, in the absence of turbulence, emission-lines with steeper
emissivity distributions yield : (i) larger estimates for the central black
hole masses, (ii) smaller virial scale factors (f-values) and (iii) smaller
$fwhm/\sigma$ (see e.g. figure~\ref{plot_f_new}).

\begin{figure}
%\resizebox{\hsize}{!}{\includegraphics[angle=0,width=14cm]{plot_suzy_hb.ps}}
%\resizebox{\hsize}{!}{\includegraphics[angle=0,width=14cm]{plot_mike_hb_ll.ps}}%\resizebox{\hsize}{!}{\includegraphics[angle=0,width=14cm]{plot_suzy_hb_ll.ps}}
\resizebox{\hsize}{!}{\includegraphics[angle=0,width=14cm]{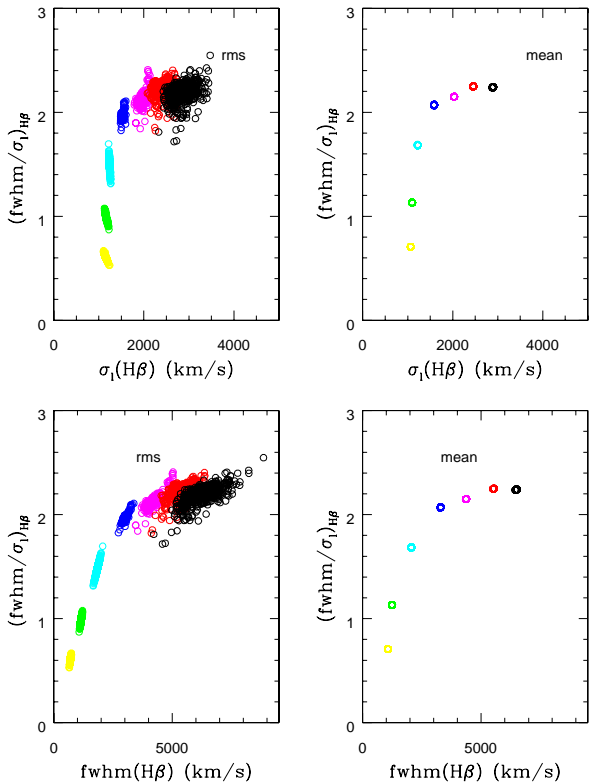}}
\caption{$fwhm/\sigma_{l}$(H$\beta$) versus (i) $\sigma_{l}$(H$\beta$) (top
panels), and (ii) $fwhm$(H$\beta$) (lower panels) for the rms (left hand
panels) and mean profiles (right hand panels) for our fiducial bowl-shaped
model with turbulence parameter $b_{\rm turb}=2$, driven by 1000 simulated continuum
light-curves, and observed at line of sight inclinations 2 (yellow open
circles), 5 (green), 10 (cyan), 20 (blue), 30 (magenta), 40 (red), and 50
(black) degrees. At a given inclination the rms profile shows a larger range
in $fwhm/\sigma_{l}$, a consequence of reverberation within the spatially
extended BLR. A far smaller range is seen for $fwhm/\sigma_{l}$ at fixed
inclination for the mean profile, as measured from the same simulations (same
number of points). For $i > 20$~degrees, the profile shape is largely
independent of inclination.}
\label{plot_suzy}
\end{figure}

\subsection{Line shape as an orientation indicator}

Collin et~al.\ 2006, showed that the shape ($fwhm/\sigma_{l}$) of the broad
H$\beta$ line in reverberation mapped AGN varies by a factor of a few, and
tends to be smaller in Narrow Line objects ($fwhm/\sigma_{l} < 2.35$) than in
Broad Line objects ($fwhm/\sigma_{l} > 2.35$), where $fwhm/\sigma_{l} = 2.35$
is the value appropriate for a Gaussian profile. The boundary separating the
narrow line and broad line objects, hereafter population 1 and population 2
sources, roughly corresponds to a $\sigma_l$ of 2000~km/s, and is broadly
similar to the division of AGN into population A population B sources by
Sulentic et~al.\ (2000)\footnote{Decarli et al.\ (2008) claim that the
correlation found between $fwhm/\sigma_{l}$ and $\sigma_{l}$ is an artifact of
the fitting procedure employed to quantify the velocity dispersion, though as
we have already pointed out differences in $fwhm/\sigma_l$ are to be expected
in BLR geometries bounded by an inner and outer radius.}. In the latter
scheme, population A sources have profiles characterised by narrow cores and
broad wings (ie. are more Lorentzian in shape) while population B sources
appear more flat-topped (i.e., more Gaussian).

%By constructing $H\beta$ mean and rms profiles for all of the reverberation
%mapped AGN, Collin et~al. showed that the line shape as quantified by the
%ratio $fwhm/\sigma_{l}$ tends to be smaller for Narrow Line objects
%($fwhm/\sigma_l < 2.35$) than Broad Line objects ($fwhm/\sigma_{l} > 2.35$),
%where the boundary $fwhm/\sigma_{l} =2.35$ is the value appropriate for a .
%Gaussian profile. This division is broadly similar to the Pop A and Pop B
%samples of Sulentic et~al. Pop 1--Pop A sources have narrow cores and broad
%wings while for Pop 2--Pop B sources the profiles are more flat topped.

By estimating the virial product using $fwhm$ and $\sigma_{l}$ of the mean and
rms profiles of each sample and comparing to the $M_{BH}$ values obtained
using stellar velocity dispersion measurements for the same AGN, Collin et
al.\ (2006) calculated virial scale factors for both Pop 1/A and Pop 2/B
samples, as well as the sample as a whole. A key result of their work is that
the virial scale factor determined from measurements of the line dispersion
(mean or rms profile) is independent of line shape, or AGN type.  By contrast,
virial scale factors based on the line $fwhm$ show significant differences
among the different AGN populations (pop 1/A, pop 2/B) and is thus a more
biased estimator of the velocity dispersion.

Collin et al.\ (2006) speculate that the factor of $\approx3$ difference in
the virial scale factor derived using the $fwhm$ for population 1/A and
population 2/B sources may be due to the increased sensitivity of the $fwhm$
to inclination effects or Eddington ratio, since population 1/A sources tend
to be high Eddington rate sources.  They suggest that the rapid decrease in
$fwhm/\sigma_l$ at small $\sigma_l$ arises in a 2-component BLR, comprising a
disc component producing the emission-line core, and an isotropic (possibly a
wind) component producing the line wings.  Since the core of the emission-line
is more sensitive to the $fwhm$ and arises in a flattened configuration, it
will show a strong dependence on inclination. Conversely, the wings of the
line are more sensitive to $\sigma_l$, and arise in an isotropic component
which is inclination independent. We here offer an alternative
explanation. The rapid decrease in $fwhm/\sigma_l$ at small $\sigma_l$ arises
naturally in planar disc-like (thin or thick) geometries in which the effects
of GR and TDS are taken into consideration. At small inclinations $\sigma_l$
is independent of inclination while the ratio $fwhm/\sigma_{l}$ increases
rapidly (factor of $\sim$ 5). That is, the line $fwhm$ is more sensitive to
changes in inclination for small inclination angles than is the line
dispersion. This effect can be seen more easily in figure~\ref{plot_suzy}
where we show the ratio $fwhm/\sigma_{l}$ for H$\beta$, as measured from the
mean and rms profiles (left and right-hand panels respectively) as a function
of $\sigma_{l}$ (upper panels) and $fwhm$ (lower panels). At low inclinations
$i < 20$~degrees, $\sigma_{l}$ is effectively constant. Variations in line
shape are then entirely due to changes in the line $fwhm$ (compare the upper
and lower left panels of figure~\ref{plot_suzy}). At higher inclinations, the
line shape is nearly independent of inclination. This trend is seen in
measurements from both the mean and rms profiles. The reduced scatter in the
ratio $fwhm/\sigma_{l}$ for the mean profile relative to the rms profile,
suggests that measurements taken from the mean profile are less sensitive to
reverberation effects with the spatially extended BLR.
% new text
Turbulence shifts all velocities to higher values while its effect on
$fwhm/\sigma_{l}$ depends on the scale height at which the line forms. For
large turbulent values $fwhm/\sigma_{l}$ tends toward 2.35 (a Gaussian
profile), as appropriate for a randomised velocity field.  In any event, it is
difficult to see how the bottom right hand corner of figure~\ref{plot_suzy}
can be populated in this model, in agreement with observations (see Collin
et~al. 2006, their figure 3; Kollatschny \& Zetzl 2011, figures 1--3; Peterson
2011).  Remarkably, for our model the break toward smaller $fwhm/\sigma_{l}$
occurs at a $\sigma_{l}$ of between 1500 and 2000~km/s, similar to the
boundary between population 1 and population 2 sources defined by Collin
et~al. 2006 (see also Peterson 2011).  Figure~17 shows the range in profile
shape as a function of $fwhm$ and $\sigma_{l}$ for a single source with fixed
$M_{\rm BH}$, viewed at a range of line of sight inclinations and in the
presence of reverberation effects (ie. approximating the results of many
observing seasons). By comparison, Figure~3 of Collin et~al. (2006) indicates
the mean profile shape observed in AGN which differ in both $M_{\rm BH}$ and
line of sight inclination. Yet our figure 17 and figure 3 of Collin et~al. are
remarkably similar in appearance. Increased mass will tend to shift points in
the upper left panel of Figure 17 toward higher velocity dispersion,
effectively filling in the left hand side, while leaving the line profile
shape at larger velocity dispersions, virtually unchanged, as is observed.
Similarly, the lower left panel of Figure~17 bares a striking resemblance to
Figure~1 of Kollatschny and Zetzl (2011). Yet in their model a Lorentzian
component was input by hand, while here it arises naturally in low inclination
systems for lines formed at large BLR radii and in the presence of scale height
dependent turbulence.

As expected reverberation effects can alter the emission line profile shape,
though for our simulations the range in $fwhm/\sigma_{l}$ at a fixed
inclination is by comparison with NGC~5548, relatively modest
($\sim$10--20\%).
We suggest that the far larger range in $fwhm/\sigma_{l}$ observed over the 13
year monitoring campaign of NGC 5548, which at times crosses the
boundary between population 1 and population 2 sources, is a consequence of
changes in the radial surface line emissivity distribution in response to changes in
the ionising continuum flux.  We also find that at a fixed inclination the
range in $fwhm$ as determined from the rms spectrum is far larger than that
for $\sigma_{l}$ (figure~\ref{plot_suzy} left panels), suggesting that in our
model, the line dispersion is also less sensitive to reverberation effects
within the physically extended BLR.  The far smaller spread in $fwhm$ and
$\sigma_{l}$ for the mean spectra (figure~\ref{plot_suzy} -- right panels) is
mainly due to the absence of contaminating non-variable components and the
fact that the light-curve is uniformly sampled with no gaps between
observations.

%
%Suzy's findings:
%\begin{itemize}
%\item[]{(i) $\sigma_{l}$ is an unbiased indicator of $M_{BH}$.}

%\item[]{(ii) fwhm is sensitive to some undefined physical parameter (Eddington ratio or
%source inclination).}

%\item[]{(iii) $fwhm/sigma_{l}$ is not correlated with $M_{BH} or L.}
%\end{itemize}

\subsection{Anomalous narrow-line quasars}

In a recent study on anomalous narrow-line quasars from the Sloan digital sky
survey, Steinhardt \& Silverman 2011 (arXiv:1109.1554) showed that the virial
masses determined using the familiar scaling relations derived for the broad
H$\beta$ and Mg~{\sc ii} emission-lines differ by up to 0.5 dex in these
objects. ANLs are identified by a broadened narrow H$\beta$ line relative to
Mg~{\sc ii}. This broadened narrow-line component is found to be
well-correlated with broad H$\beta$ in these objects. Since H$\beta$ and
Mg~{\sc ii} are expected to form under similar physical conditions, one would
normally expect a similar relation to be evident in Mg~{\sc ii}. Steinhardt
\& Silverman 2011 argue that the absence of such a correlation in Mg~{\sc ii}
may cast doubt as to the validity of virial mass estimates based on
measurements of the H$\beta$ line width, arguing that some of the broadening
may be due to the presence of a wind. Thermal emission from a wind lacking an
ionisation front would produce an additional H$\beta$ component without
contributing significantly to the Mg~{\sc ii} emission.

Bowl-shaped geometries may provide an alternative solution. If as expected
Mg~{\sc ii} forms at large radial distances, and turbulence is significant at
large scale heights, then one can envisage a situation in which the Mg~{\sc
ii} emission is dominated by a turbulent component. The Mg~{\sc ii} line width
would then be largely independent of inclination effects. By contrast if
H$\beta$ emission arises at smaller radial distances (smaller scale heights),
the lower turbulence and more flattened spatial distribution would ensure a
strong inclination dependence even at modest inclinations (GR and TDS are less
important for these lines due to their larger formation radii). Thus, 
ANLs may represent near face-on objects with a significant turbulent component 
in the Mg~{\sc ii} forming region.

\section{Summary}

We have explored the observational characteristics of a class of broad
emission line region geometries in which the line emitting gas spans the
region between the outer accretion disc and the inner edge of the dusty torus,
by occupying an effective (though not necessarily smooth or continuous)
surface the scale height of which increases with increasing radial distance
(ie. a bowl-shaped BLR geometry), similar to the configuration first proposed
by Gaskell (2009).  A Type~1 AGN spectrum is then observed for line-of-sight
viewing angles which peer over the rim of the bowl. Such a configuration
provides quite naturally the necessary high covering fractions for the
observed emission-line strengths, without obscuring the view of the central
engine (for a reasonable range in line-of-sight viewing angles).  The gas
dynamics are here assumed to be dominated by gravity, and we include in our
model the effects of gravitational redshift, transverse Doppler shift and
scale-height dependent turbulence.  
While we do not exclude possible contributions from infall, outflow, or
line-driven wind contributions to the BLR kinematics within individual AGN, we
consider here what is most likely the underlying and dominant effect of a
gravitationally bound BLR velocity-field consisting of circularised orbital
motion, as might be expected if there are significant dissipative forces
acting within the BLR.  Importantly, we introduce a macroscopic turbulent
component to the cloud motion which imparts substantial scale height to the
BLR at large radii, allowing it to intercept sufficient ionising continuum
radiation to explain the observed emission-line strengths.

By breaking spherical symmetry, a bowl-shaped geometry simultaneously provides
a simple solution to the shorter than predicted dust reverberation timescale
for the inner dusty torus and the absence of a significant response of the
broad optical recombination lines on short timescales.  Gravitational redshift
and transverse Doppler shift offer a mechanism for providing substantial line
width, even in the case of purely transverse motion. They also introduce a
strong red-blue asymmetry into the 2-d response function and emission-line
profile in the form of an enhanced red-wing response at short time-delays
similar to that seen in the recently recovered 2-d emission-line response
function and emission-line profile of the NLSy1 Arp 151 (Bentz et al. 2010,
Brewer et al. 2011).  These effects are most important for lines with steeper
radial emissivity distribution which therefore form at small BLR radii (the
HILs, e.g. He~{\sc ii}, C~{\sc iv}) and for those systems viewed at low line
of sight inclination.  Differences in the line formation radius between the
high- and low-ionisation lines lead to strong differences in the shapes of
their 2-d and 1-d emission-line response functions and emission-line profiles.

While the LILs show a deficit of response on short-timescales, the HILs, which
form toward the base of the bowl in a more flattened configuration, show
significant response on short-timescales, and their 2-d and 1-d response
functions appear similar to those found for geometrically thin discs.
Additionally, GR and TDS for flattened BLR geometries (ie. bowl-shaped, thick
or thin disc geometries), result in line profile shapes which display a strong
inclination dependence at low line of sight inclinations, with $fwhm/\sigma_l$
decreasing with decreasing $i$.  At larger inclinations, the line profile
shape is nearly independent of inclination. Thus inclination, in the presence
of GR and TDS may provide a natural explanation for the observed differences
in line profile shapes among AGN in general and importantly may also explain
the observed differences in the line profile shapes used to discriminate
between pop A and pop B sources (Sulentic et al. 2000, Collin et al. 2006). We
suggest that some of the systems previously identified as high Eddington rate
sources based on their profile shape, may simply be type 1 objects viewed at
low line of sight inclinations. We note that the break between inclination
dependent and inclination independent line profile shapes ($\sigma_l \approx
2000$ km/s, see figure~\ref{plot_suzy}) is remarkably similar to the boundary
separating the high- (pop A) and low- (pop B) Eddington rate sources. Also, we
speculate that high Eddington rate sources may have TORs with larger than
typical covering fractions, necessitating smaller observed viewing angles $i$
for un-obscured lines of sight.

For a bowl-shaped BLR with significant scale-height dependent turbulence, we
expect the low-ionisation lines (e.g. H$\beta$, Mg~{\sc ii}) to display line
shapes which are more Gaussian in shape (larger $fwhm/\sigma_{l}$) than the
high ionisation lines (e.g. C~{\sc iv}, He~{\sc ii}), as is observed (Decarli
et~al. 2008). This arises because the low-ionisation lines form at larger BLR
radii (and hence larger scale heights) where the turbulent (random)
contribution to the velocity field is larger. By contrast the high-ionisation
lines form at small BLR radii (near the base of the bowl) where the turbulent
component is substantially reduced (for example, compare the $fwhm/\sigma_{l}$
as a function of inclination $i$ for the low- and high-ionisation lines with
and without turbulence, figure~\ref{plot_f_new} lower panels).  As can be seen,
as well as significantly broadening the line profile, turbulence softens the
strong dependence of line profile shape on inclination at low inclinations for
the low-ionisation lines.  By moving low velocity gas (at large BLR radii) to
larger line of sight velocities, turbulence removes the strong horns and
shoulders characteristic of pure planar Keplerian motion while reducing any
delays that may be present between the wings and core of the line.  For low
line of sight inclinations, the turbulent contribution to the line of sight
velocity is larger than the Keplerian contribution (for motion confined to the
disc mid-plane) and turbulent broadening acts to produce emission line
profiles which are Lorentzian in form ($fwhm/\sigma_{l} \approx 1$,
e.g. figure~\ref{turb}), with extended broad wings and narrow cores. As far as
we aware, this is the first study to provide physical motivation for the
fitting of emission-line profiles using Lorentzians.  A significant
scale-height dependent turbulent component may help explain the reported
weakness in the correlation between the measured widths of the low- and high-
ionisation lines (e.g., H$\beta$ and C~{\sc iv}). In cases of sufficiently
large contributions from this component, H$\beta$ might even be broader than
C~{\sc iv}, a finding which when taken at face value, appears at odds with
what one would normally expect based on photoionisation calculations of a
gravitationally bound BLR.

A strong motivation for this work was to try to identify which lines and which
of the measured quantities yield the most accurate and most reproducible mass
estimates for $M_{BH}$. The time-stationary model indicates that :

\begin{enumerate} 

\item{Different emission lines predict different virial mass estimates and
hence different virial scale factors (figure~\ref{plot_f_new}).}

\item{Mass estimates made from measurement of the emission line $fwhm$ show a
strong inclination dependence, leading to an underestimate of $M_{BH}$ at low
inclinations, and an overestimate of $M_{BH}$ at large inclination. The
largest discrepancies are for low line of sight inclinations and in the
absence of turbulence for lines formed at large BLR radii (ie. the LILs, see
figure~\ref{plot_f_new}).}

\item{Mass estimates made from measurement of the emission line dispersion
$\sigma_{l}$, underestimate $M_{BH}$ at all inclinations, but show a much
weaker dependence on $i$ than the $fwhm$
(see also figures~\ref{plot_vir_hb+he2}--\ref{plot_vir_mg2+c4}), confirming the results
of Collin et al. (2006), that mass estimates based on $\sigma_{l}$ are less
biased.}

\item{Measurements of the emission line $fwhm$ (mean or rms profile) predict
larger $M_{BH}$ than measurements based on $\sigma_{l}$, since in general
$fwhm/\sigma_{l} > 1$, except at the smallest inclinations.}

\item{In the presence of turbulence, black hole mass estimates based on
measurements of emission lines formed at large BLR radii (ie. H$\beta$,
Mg~{\sc ii}), are less sensitive to inclination effects than those formed at
small BLR radii (ie. C~{\sc iv}, He~{\sc ii}), and for which the turbulent
contribution to the velocity field is weaker.}

\item{In the absence of turbulence, emission lines formed at small BLR radii
(ie. the HILs) yield larger $M_{BH}$ (figure~\ref{plot_f_new}) and therefore
smaller virial scale factors.}
\end{enumerate}

\noindent When the effects of emission-line variability are taken into
account, we find that:

\begin{enumerate}

\item{The cross correlation function (CCF) centroid tracks the continuum
variations more closely than the CCF peak, and shows a better correspondence
with the centroid of the emission line response function in lines whose
variability timescale is more closely matched to the characteristic
variability timescale of the driving continuum (ie. the LILs,
figures~\ref{plot_cent_hb}--\ref{plot_cent_c4}). However, the choice of CCF
centroid over CCF peak has little overall effect on the predicted mass.}

\item{The best correspondence between $M_{BH}$ estimates (and hence $f$-factors)
for different lines, arises from measurements of the $fwhm$ of the mean or rms
profile, consistent with results reported in the literature from observations
(e.g. Assef et~al. 2011). The stability of the mean profile suggests that
measurement of the $fwhm$ of the mean profile is the preferred estimate of
$M_{BH}$ (figure~\ref{plot_scale}).}

\item{The largest discrepancy between the derived virial scale factors
between different emission lines arise from measurements of $\sigma_{l}$ in low
inclination systems ($i<20$ degrees, figure~16). 
We suggest that some of the systems previously identified as high 
Eddington rate sources based on the shapes of their low-ionisation 
emission lines, may simply be more typical type 1 objects viewed at low 
line of sight inclinations.}

%We do not here consider the effects of 
% a substantial wind contribution to the profiles of the high ionisation 
%resonance lines, such as C~{\sc iv}, which may be especially important 
%in these lines in high Eddington rate AGN.}

\end{enumerate}

Our findings at first glance appear at odds with those presented in Peterson
et~al. (2004) who found that the tightest relation between the virial products
derived for different lines in a single source is found for those estimated
using the CCF centroid, and the line dispersion of the rms spectrum. Note that
the difference here is that we calculate the time-delay (centroid or peak) of
the CCF and line width ($\sigma_{l}$ or $fwhm$) of the mean and rms profile
for 1000 realisations of the driving continuum light-curve, and not the
variation in centroid and line width within a single realisation. Due to the
long duration of our light-curves, any biases introduced by windowing effects
are minimised.  In addition, the model presented here utilises a locally
linear response approximation. Gross changes in the radial surface line
emissivity distribution and in the inner and outer radius in response to large
variations in the ionising continuum flux variations will be treated elsewhere
(Goad and Korista, in prep.).

A comparison between the H$\beta$ virial scale factors presented in our
simulations and the empirically estimated virial scale factors based on the
AGN with reverberation data presented in Collin et~al. (2006) suggests a
rather narrow range of typical inclination angles for Type 1 objects,
$i\approx$~20--35 degrees or so, in agreement with the expected range found
via other techniques.

\section{Acknowledgements}
\thanks{We would like to thank the anonymous referee for providing comments
which have led to improved clarity of the work presented here.  Mike Goad and
Kirk Korista would also like to thank the generous hospitality of Keith Horne
and the Department of Physics \& Astronomy at the University of St.\ Andrews
during the initial stages of this work. Kirk Korista would like to thank the
Department of Physics \& Astronomy at the University of Leicester for their
generous hospitality during his stay during the major stage of this work. Mike
Goad would like to thank Dr Graham Wynn for many useful discussions.}

\section{References}
%\begin{thebibliography}{99}
%\bibitem[\protect\citeauthoryear{Baird}{1981}]{b1} Baird S.R., 1981, ApJ, 245, 208

%not used
%\noindent Antonucci, R.R.J., and Miller, J.S. 1985, ApJ 297, 621.

\noindent Antonucci, R. 1993, ARA\&A 31, 473.
%Unified models for active galactic nuclei and quasars

% not used
%\noindent Arribas, S., Mediavilla, E, and Garcia-Lorenzo, B. 1996, ApJ 463, 509.

\noindent Arav, N., Barlow, T.A., Laor, A. and Blandford, R.D. 1997, MNRAS,
288, 105.

\noindent Arav, N., Barlow, T.A., Laor, A., Sargent, W.L.W., and Blandford,
R.D. 1998, MNRAS, 297. 90.

\noindent Assef , R.J., Denney, K.D., Kochanek, C.S., Peterson, B.M.,
Kozlowski, S. et al. 2011, ApJ 742, 93.
%Black Hole Mass Estimates Based on C IV are Consistent with Those Based on the Balmer Lines

\noindent Baldwin, J. Ferland, G. Korista, K.T. and Verner, D. 1995, ApJ 455, L119.
%Locally optimally emitting clouds and the origin of the quasar emission lines.

\noindent Barvainis, R. 1987, ApJ 320, 537.
%Hot dust and the near-infrared bump in the continuum spectra of quasars and active galactic nuclei

\noindent Bentz, M.C., Denney, K.D., Cackett, E.M. et al. 2007, ApJ 662, 205.
%NGC 5548 in a Low-Luminosity State: Implications for the Broad-Line Region

\noindent Bentz, M.C., Horne, K., Barth, A.J. et al. 2010, ApJ 720, l46.
%The Lick AGN Monitoring Project: Velocity-delay Maps from the Maximum-entropy Method for Arp 151

\noindent Bentz, M.C., Walsh, J.L. Baarth, A.J. et al. 2010, ApJ 716, 993.
%The Lick AGN Monitoring Project: Reverberation Mapping of Optical Hydrogen and Helium Recombination Lines

%not used
%\noindent Blandford, R.D. and McKee, C. F. 1982, ApJ 255, 419.
%Reverberation mapping of the emission line regions of Seyfert galaxies and quasars

\noindent Bottorff, M. Korista, K.T. Shlosman, I. and Blandford, R.D. 1997,
ApJ 479, 200. 
%Dynamics of Broad Emission-Line Region in NGC 5548: Hydromagnetic Wind Model
%versus Observations

\noindent Bottorff, M., Ferland, G., Baldwin, J. and Korista, K. 2000, ApJ 542, 644.
% micro-turbulence paper

\noindent Bottorff, M.C., Baldwin, J.A., Ferland, G.J., Ferguson, J.W. and
Korista, K.T. 2002, ApJ 581, 932.
%He II Reverberation in Active Galactic Nucleus Spectra

\noindent Brewer, B.J., Treu, T., Pancoast, A., Bart, A.J., Bennert,
V.N. et~al. 2011, ApJL 733 L33.
%The Mass of the Black Hole in Arp 151 from Bayesian Modelling of Reverberation Mapping Data

%not used
%\noindent Cackett, E. M.; Horne, K. 2006, MNRAS 365, 1180.
%Photoionised H$.1Ž§² emission in NGC 5548: it breathes!

\noindent Chiang, J. and Murray, N. 1996, ApJ 466, 704.
%Reverberation Mapping and the Disk-Wind Model of the Broad-Line Region

%not used
%\noindent Chen, K., and Halpern, J.P. 1989, ApJ 344, 115.
%Structure of line-emitting accretion disks in active galactic nuclei - ARP
%102B

\noindent Clavel, J. Reichert, G.A., Alloin, D. et~al. 1991, ApJ 366, 64.
%Steps toward determination of the size and structure of the broad-line region in active galactic nuclei. I - an 8 month campaign of monitoring NGC 5548 with IUE

\noindent Collier, S., and Peterson, B, M. 2001, ApJ 555, 775.
%Characteristic Ultraviolet/Optical Timescales in Active Galactic Nuclei

%not used
%\noindent Collin, S., and Hur\'{e}, J.-M. 2001, A\&A 372, 50.
%Size-mass-luminosity relations in AGN and the role of the accretion disc

\noindent Collin, S., Kawaguchi, T., Peterson, B.M. and Vestergaard, M. 2006, A\&A 456, 75.
%Systematic effects in measurement of black hole masses by emission-line reverberation of active galactic nuclei: Eddington ratio and inclination

\noindent Corbin, M.R. 1997, ApJ, 485, 517.
%Relativistic Effects in the QSO Broad-Line Region. 

\noindent Crenshaw, D., Kraemer, S.B. Boggess, A. et al. 1999, ApJ 516, 750.
%Intrinsic Absorption Lines in Seyfert 1 Galaxies. I. Ultraviolet Spectra from the Hubble Space Telescope

\noindent Czerny, B., \& Hryniewicz, 2011, A\&A, 525, L8.
%The origin of the broad line region in active galactic nuclei

\noindent Decarli, R., Labita, M., Treves, A., and Falomo, R. 2008, MNRAS 387, 1237.
%On the geometry of broad emission region in quasars

\noindent Denney, K.D., Peterson, B.M. Pogge, R.W. Adair, A. et~al. 2009, ApJ 704, L80.
%Diverse Kinematic Signatures from Reverberation Mapping of the Broad-Line Region in AGNs

\noindent Denney, K.D., Peterson, B.M. Pogge, et~al. 2010, ApJ 721, 715.
%Reverberation Mapping Measurements of Black Hole Masses in Six Local Seyfert Galaxies

\noindent Denney, K.D, Assef, R.J., Bentz, M., et al. 2011,
"Narrow-Line Seyfert 1 Galaxies and their place in the Universe". April 4-6,
2011. Milano, Italy. Editorial Board: Luigi Foschini (chair), Monica Colpi,
Luigi Gallo, Dirk Grupe, Stefanie Komossa, Karen Leighly, Smita
Mathur. Proceedings of Science (PoS, Trieste, Italy), vol. NLS1, Published online at http://pos.sissa.it/cgi-bin/reader/conf.cgi?confid=126, id.34"

%Addressing systematic uncertainties in black hole mass measurements

\noindent Edelson, R. A. and Krolik, J. H. 1988, ApJ 333, 646.

\noindent Elvis, M. 2000, ApJ 545, 63.
% a structure for quasars

\noindent Emmanoulopoulos, D. McHardy, I.M. and Uttley, P. 2010, MNRAS 404,
931.

%not used
%\noindent Emmering, R.T. Blandford, R.D. and Shlosman, I. 1992, ApJ 385, 406.
%Magnetic acceleration of broad emission-line clouds in active galactic nuclei

\noindent Eracleous, M., and Halpern, J.P. 1994, ApJS 90, 1.
%Doubled-peaked emission lines in active galactic nuclei

% not used
%\noindent Eracleous, M., Halpern, J.P., Gilbert, A.M, et~al. 1997, ApJ 490, 216.
%Rejection of the Binary Broad-Line Region Interpretation of Double-peaked Emission Lines in Three Active Galactic Nuclei

\noindent Eracleous, M. Halpern, J.P., 2003 ApJ 599, 886.
%Completion of a Survey and Detailed Study of Double-peaked Emission Lines in Radio-loud Active Galactic Nuclei

%not used
%\noindent Eracleous, M. Halpern, J.P., Storchi-Bergmann, T. et~al. 2004, in : The
%Interplay among Black Holes, Stars and ISM in Galactic Nuclei, (Proc. IAU
%Symposium no.222, 2004)

%not used
%\noindent Eracleous, M., Lewis, K.T., Flohic. H.M.L.G. 2009, NewAR 53, 133. 

%not used
%\noindent Ferguson, J.W., Korista, K.T., Baldwin, J.A. and Ferland, G.J. 1997,
%ApJ 487, 122.
%Locally Optimally Emitting Clouds and the Narrow Emission Lines in Seyfert Galaxies

% not used
%\noindent Ferland, G.J., Hu, C., Wang, J.-M. et al. 2009, ApJ 707, L82.
%Implications of In-falling Fe II-Emitting Clouds in Active Galactic Nuclei:
%Anisotropic Properties

\noindent Ferland, G. J.; Korista, K. T.; Verner, D. A, 1997, ASPC 125, 213.
%Numerical Simulations of Plasmas and Their Spectra

\noindent Ferland, G.J., Korista, K.T., Verner, D.A. et al. 1998, PASP 110, 761.
%CLOUDY 90: Numerical Simulation of Plasmas and Their Spectra

\noindent Ferland, G.J., Peterson, B.M., Horne, K. et~al. 1992, ApJ 387, 95.
%Anisotropic line emission and the geometry of the broad-line region in active
%galactic nuclei

\noindent Ferland, G.J., Shields, G.A. and Netzer, H. 1979, ApJ 232, 382
%Asymmetries of the emission lines of QSOs, Seyfert galaxies, and novae

\noindent Ferrarese, L, Pogge, R.W., Peterson, B.M. et al. 2001, ApJ 555, L79.
%Super-massive Black Holes in Active Galactic Nuclei. I. The Consistency of Black Hole Masses in Quiescent and Active Galaxies

\noindent Fine, S., Croom, S.M. Hopkins, P.F. et al. 2008, MNRAS 390, 1413.
%Constraining the quasar population with the broad-line width distribution

\noindent Fine, S., Croom, S.M., Bland-Hawthorn, J. et~al. MNRAS 2010, 409, 591.
% The CIV linewidth distribution for quasars and its implications for broad-line region dynamics and virial mass estimation

\noindent Fine, S., Jarvis, M. J., Mauch, T., 2011 MNRAS 418 2251.

%Orientation effects in quasar spectra: the broad- and narrow-line regions

\noindent Fromerth, M.J. and Melia, F. 2000, ApJ 533, 172.

\noindent Gaskell and Snedden 1997, in ASP Conf Proc, 159, Emission Lines in Active
Galaxies: New Methods and Techniques ed B.M.Peterson, F.-Z. Cheng \&
A.S. Wilson (San Francisco: ASP), 193.
% binary black hole for double-peaked emitters

\noindent Gaskell, C.M. 2009, New Astronomy Reviews, Volume 53, Issue 7-10, p. 140-148
%What broad emission lines tell us about how active galactic nuclei work

\noindent Gaskell, C.M. and Peterson, B.M. 1987, ApJS 65, 1.

\noindent Goad, M.R. and O'Brien, P.T. and Gondhalekar, P.M. 1993, MNRAS 263, 149.
%Response functions as diagnostics of the broad-line region in active galactic nuclei

\noindent Goad, M.R. 1995, Ph.D. thesis, University College London.
% Mike's thesis

% not used
%\noindent Goad, M. and Wanders, I. 1996, ApJ 469, 113.
%The Effect of a Variable Anisotropic Continuum Source upon the Broad Emission Line Profiles and Responses

\noindent Goad, M. R.; Korista, K. T.; Knigge, C., 2004, MNRAS 352, 277.
%Is the slope of the intrinsic Baldwin effect constant?

\noindent G\"ultelkin, K. et al. 2009, ApJ, 698, 198.

%not used
%\noindent Halpern, J.P. Eracleous, M., Filippenko, A.V., and Chen. K. 1996, ApJ 464, 704.

\noindent Hao, C.N. xIA, x.y. Mao, S. et al. 2005, ApJ 625, 78.
%The Physical Connections among Infrared QSOs, Palomar-Green QSOs, and Narrow-Line Seyfert 1 Galaxies

\noindent H\"{o}nig, S.F., Kishimoto, M. Gandhi, P. et al. 2010, A\&A 515, 23.
%The dusty heart of nearby active galaxies. I. High-spatial resolution mid-IR spectro-photometry of Seyfert galaxies

\noindent H\"{o}nig, S.F., et~al.\ 2006, A\&A, 452, 459
%Radiative Transfer Modelling of Three-Dimensional Clumpy AGN Tori and its Application to NGC 1068

\noindent Horne, K. 1994 ASP Conference Series, 69, 23. Eds P.M. Gondhalekar, K. Horne,
and B.M. Peterson.
% Echo Mapping Problems Maximum Entropy solutions

\noindent Horne, K., Welsh, W.F., and Peterson, B.M. 1991, ApJ 367, L5.
% Echo mapping of broad H-beta emission in NGC 5548

\noindent Horne, K., Korista, K.T. and Goad, M.R. 2003, MNRAS 339, 367.
% Quasar tomography: unification of echo mapping and photoionisation models

\noindent Horne, K., Peterson, B.M. Collier, S.J. and Netzer, H. 2004, PASP, 116, 465.
%Observational Requirements for High-Fidelity Reverberation Mapping

%not used
%\noindent Hu, C.Y., Wang, J.-M, Ho, L.C. et al. 2008, ApJ  687, 78.
%A Systematic Analysis of Fe II Emission in Quasars: Evidence for Inflow to
%the Central Black Hole

\noindent Hu, C.Y., Wang, J.-M, Ho, L.C. et al. 2008, ApJ 683, L115.
%HŽ§² Profiles in Quasars: Evidence for an Intermediate-Line Region

\noindent Hubeny, I., Algol, E., Blaes, O.  and Krolik, J.H. 2000, ApJ 533, 710.
%Non-LTE Models and Theoretical Spectra of Accretion Disks in Active Galactic Nuclei. III. Integrated Spectra for Hydrogen-Helium Disks

\noindent Kaspi, S., and Netzer, H. 1999, ApJ 524, 71.
%Modelling Variable Emission Lines in Active Galactic Nuclei: Method and Application to NGC 5548

\noindent Kaspi, S., Smith, P.S., Netzer, H. et al. 2000, ApJ 533, 631.
%Reverberation Measurements for 17 Quasars and the Size-Mass-Luminosity Relations in Active Galactic Nuclei

\noindent Kawaguchi, T., and Mori, M. 2010, ApJ 724, L183.
%Orientation Effects on the Inner Region of Dusty Torus of Active Galactic Nuclei
\noindent Kawaguchi, T., and Mori, M. 2011, ApJ 737, 105.
%Near-infrared Reverberation by Dusty Clumpy Tori in Active Galactic Nuclei

\noindent Kelly, B.C., Bechtold, J., and Siemiginowska, A. 2009, ApJ 698, 895. 
%Are the Variations in Quasar Optical Flux Driven by Thermal Fluctuations?

\noindent Kishimoto, M., Honig, S.F., Beckert, T. and Weigelt, G. 2007, A\&A
476, 713
%The innermost region of AGN tori: implications from the HST/NICMOS type 1 point sources and near-IR reverberation

\noindent Kishimoto, M., Honig, S.F., Antonucci, R. et al. 2009, A\&A 507, L57.
%Exploring the inner region of type 1 AGNs with the Keck interferometer

\noindent Kishimoto, M., Honig, S.F., Tristram, K.R.W, and Weigelt, G. 2009
a\&a 493, L57.
%Possible evidence for a common radial structure in nearby AGN tori

\noindent Kishimoto, M., Honig, S.F., Antonucci, R. et al. 2011, A\&A 527, 121.
%The innermost dusty structure in active galactic nuclei as probed by the Keck interferometer

\noindent Kollatschny, W., and Bischoff, K. 2002, A\&A 386, L19.
%Geometry and kinematics in the central broad-line region of a Seyfert 1 galaxy

\noindent Kollatschny, W. 2003, A\&A 407, 461.
%Accretion disk wind in the AGN broad-line region: Spectroscopically resolved line profile variations in Mrk 110

\noindent Kollatschny, W. 2003, A\&A 412, L61.
%Spin orientation of super-massive black holes in active galaxies

\noindent Kollatschny, W., and Zetzl, M. 2010, A\&A 522, 36.
%Line profile and continuum variability in the very broad-line Seyfert galaxy Mrk 926

\noindent Kollatschny, W., and Zetzl, M. 2011, Nature 470, 366.
%Broad-line active galactic nuclei rotate faster than narrow-line ones.

\noindent K\"onigl, A. and Kartje, J.F. 1994, ApJ 434, 446.
%Disk-driven hydromagnetic winds as a key ingredient of active galactic nuclei unification schemes

\noindent Koratkar, A., and Gaskell, C.M. 1991, ApJS 75, 719.

\noindent Korista, K.T., Alloin, D., Barr, P. et~al. 1995, ApJS 97, 285.
%Steps toward determination of the size and structure of the broad-line region in active galactic nuclei. 8: an intensive HST, IUE, and ground-based study of NGC 5548

\noindent Korista, K.T., Baldwin, J., Ferland, G.J. and Verner, D. 1997, ApJ
108, 401.

\noindent Korista, K.T., Ferland, G.J. and Baldwin, J. 1997, ApJ 487, 555.
%Do the Broad Emission Line Clouds See the Same Continuum That We See?

\noindent Korista, K.T. and Goad, M.R. 2004, ApJ 606, 749.
%What the Optical Recombination Lines Can Tell Us about the Broad-Line Regions of Active Galactic Nuclei

\noindent Korista, K.T., and Goad, M.R. 2000, ApJ 536, 284.
% Locally Optimally Emitting Clouds and the Variable Broad Emission Line Spectrum of NGC 5548

\noindent Korista, K.T., and Goad, M.R. 2001, ApJ 553, 695.
% The Variable Diffuse Continuum Emission of Broad-Line Clouds

\noindent Koshida, S., Yoshii, Y., Kobayashi, Y. et~al. 2009, ApJL 700, L109.
% Variation of Inner Radius of Dust Torus in NGC4151

\noindent Kozlowski, S., Kochanek, C.S., Udalski, A. et al. 2010, ApJ 708, 927.
%Quantifying Quasar Variability as Part of a General Approach to Classifying Continuously Varying Sources

\noindent Krause, M. Schartmann, ,. and Burkett, A. 2012, MNRAS in press, arXiv:1207.0785v1
% Magnetohydrodynamic stability of broad line region clouds.

\noindent Krolik, J.H., Horne, K., Kallman, T.R. et~al. 1991, ApJ 371, 541.
%Ultraviolet variability of NGC 5548 - Dynamics of the continuum production region and geometry of the broad-line region continuum anisotropy

\noindent Krolik, J.H. and Begelman, M.C. 1988, ApJ 329, 702.
%Molecular tori in Seyfert galaxies - Feeding the monster and hiding it

\noindent Landt, H., Bentz, M.C., Peterson, B.M. et al. 2011, MNRAS 413, L106
%The near-infrared radius-luminosity relationship for active galactic nuclei

\noindent Landt, H., Elvis, M., Ward, M.J. et al. 2011, MNRAS 414, 218.
% The near-infrared broad emission line region of active galactic nuclei - II. The 1-$(B&L(Bm continuum

\noindent Landt, H., Bentz, M.C., Peterson, B.M. et al. In :Proceedings of the conference "Narrow-Line Seyfert 1 Galaxies and their place in the Universe". April 4-6, 2011. Milano, Italy.
% The near-IR broad-emission line region of AGN

\noindent Landt, H., Bentz, M.C., Ward, M.J. et al. 2008, ApJS 174, 282.
%The Near-Infrared Broad Emission Line Region of Active Galactic Nuclei. I. The Observations

%not used
%\noindent Lewis, K.T., Eracleous, M. and Storchi-Bergmann, T. 2010, ApJS 187, 416.
%Long-term Profile Variability in Active Galactic Nucleus with Double-peaked Balmer Emission Lines

\noindent Laor, A. 2006, ApJ 643, 112.
% Evidence for line broadening by electron scattering in the BLR of NGC~4395.

\noindent Liu, Y., \& Zhang, S.N., 2011, ApJL, 728, L44.
%Dusty Torus Formation by Anisotropic Radiative Pressure Feedback of Active Galactic Nuclei

\noindent Livio, M., and Chun, X. 1997, ApJL 486, L835.
%On the Observational Evidence for Accretion Disks in Active Galactic Nuclei

\noindent MacLeod, C. L., Ivezic, Z., Kochanek, C. S., Kozlwski, S., Kelly,
B. et al. 2010, ApJ 721, 1014.
%Modelling the Time Variability of SDSS Stripe 82 Quasars as a Damped Random Walk

\noindent Mannucci, F.; Salvati, M.; Stanga, R. M. 1992, ApJ 394, 98.
%Line profile and variability data to probe the broad-line region geometry - Of disks and nests

% not used
%\noindent Marziani, P., Sulentic, J.W., Negrete, C. A. et al. 2010, MNRAS 409 1033.
%Broad-line region physical conditions along the quasar eigenvector 1 sequence

\noindent Mathews, W.G., and Ferland, G.J. 1987, ApJ 323, 456. 
%What heats the hot phase in active nuclei?

\noindent McHardy, I.M., Papadakis, I.E., Uttley, P. et al. 2004, MNRAS 348, 783.	
%Combined long and short time-scale X-ray variability of NGC 4051 with RXTE and XMM-Newton

\noindent Melnikov, A. V., Shevchenko, I. I., 2008, MNRAS 389, 478.
% On reverberation and cross-correlation estimates of the size of the broad-line region in active galactic nuclei

\noindent Minezaki, T.Yoshii, Y., Kobayashi, Y. et al. 2004, ApJL 600 L35.
%Inner Size of a Dust Torus in the Seyfert 1 Galaxy NGC 4151

\noindent Mor, R., and Netzer, H. 2012, MNRAS 420 526
%Hot graphite dust and the infrared spectral energy distribution of active galactic nuclei

\noindent Mor, R., and Trakhtenbrot, B. 2011, ApJL 737 L36
%Hot Dust Clouds with Pure Graphite Composition around Type-I Active Galactic Nuclei

\noindent Mullaney, J.R. and Ward, M.J. 2008, MNRAS 385, 43.
%Optical emission-line properties of narrow-line Seyfert 1 galaxies and comparison active galactic nuclei

\noindent Murray, N. and Chiang, J. 1997, ApJ, 474, 91.
% Disk Winds and Disk Emission Lines

\noindent Nandra, K., George, I.M., Mushotzky, R.F., Turner, T.J. and Yaqoob,
T. 1999 ApJ 523, L17.
%The Properties of the Relativistic Iron K-Line in NGC 3516

\noindent Nandra, K., George, I.M., Mushotzky, R.F., Turner, T.J. and Yaqoob, T. 1997 ApJ 477, 602
%ASCA Observations of Seyfert 1 Galaxies. II. Relativistic Iron K alpha Emission orientation indicators

\noindent Nemmen, R.S. and Brotherton, M.S. 2010, MNRAS 408, 1598.
% Quasar bolometric corrections: theoretical considerations

\noindent Nenkova, M., Sirocky, M.M., Nikutta, R. et al. 2008, ApJ 685, 160. 
%AGN Dusty Tori. II. Observational Implications of Clumpiness

%\noindent Nenkova, M., Sirocky, M.M., Ivezic, Z., and Elitzur, M. 2008, ApJ 685, 147. 
%AGN Dusty Tori. I. Handling of Clumpy Media

\noindent Netzer, H. 1987, MNRAS 225, 55.
%Quasar discs. II - A composite model for the broad-line region

\noindent Netzer, H., and Laor, A. 1993, ApJL 404, L51.
%Dust in the narrow-line region of active galactic nuclei

\noindent O'Brien, P.T., Goad, M.R. and Gondhalekar, P.M. 1994, MNRAS 268, 845.
%Response Functions as Diagnostics of the Broad Line Region in Active Galactic Nuclei - Part Two - Anisotropic Line Emission

\noindent O'Brien, P.T., Goad, M.R. and Gondhalekar, P.M. 1995, MNRAS 275, 1125.
%The luminosity-dependent broad-line region in active galactic nuclei

\noindent Onken, C. A.,  Ferrarese, L., Merritt D. et~al. 2004, ApJ 615, 645.

\noindent Paltani, S. 1999, In : BL Lac Phenomenon, a conference held 22-26 June, 1998 in Turku, Finland, p. 293 
%Constraining BL Lac Models using Structure Function Analysis

\noindent Pancoast, A., Brewer, J.B., Treu, T. et al. 2012, ApJ in press.
% The Lick AGN monitoring project 2011: Dynamical modelling of the BLR in Mrk 50

\noindent Park, D., et~al.\ 2012, ApJ, 737, 30.
%The Lick AGN Monitoring Projection: Re-calibrating Single-Epoch Virial Black Hole
% Mass Estimates

\noindent P\'{e}rez, E., Robinson, A., and de la Fuente, L. 1992, MNRAS 255, 502.

\noindent Peterson, B.M., Balonek, T.J. Barker, E.S. et~al. 1991, ApJ 368, 119.
%Steps toward determination of the size and structure of the broad-line region
%in active galactic nuclei. II - an intensive study of NGC 5548 at optical
%wavelengths

\noindent Peterson, B. M., Wanders, I., Horne, K., Collier, S., Alexander, T.,
\& Kaspi, S. 1998, PASP, 110, 660.

%not used
%\noindent Peterson, B.M., Barth, A.J., Berlind, P. et al. 1999, ApJ 510, 659.
%Steps toward Determination of the Size and Structure of the Broad-Line Region in Active Galactic Nuclei. XV. Long-Term Optical Monitoring of NGC 5548

\noindent  Peterson, B.M., and Wandel, A. 1999, ApJL 521, L95.
%Keplerian Motion of Broad-Line Region Gas as Evidence for Super-massive Black Holes in Active Galactic Nuclei

\noindent  Peterson, B.M., and Wandel, A. 2000, ApJL 540, L13.
%Evidence for Super-massive Black Holes in Active Galactic Nuclei from Emission-Line Reverberation

%not used
%\noindent Peterson, B.M., Berlind, P., Betram, R. et al. 2002, ApJ 581, 197.
%Steps toward Determination of the Size and Structure of the Broad-Line Region in Active Galactic Nuclei. XVI. A 13 Year Study of Spectral Variability in NGC 5548

\noindent Peterson, B.M., Ferrarese, L., Gilbert, K.M. et al. 2004, ApJ 613, 682.
%Central Masses and Broad-Line Region Sizes of Active Galactic Nuclei. II. A Homogeneous Analysis of a Large Reverberation-Mapping Database

%not used
%\noindent Peterson, B.M., Bentz, M.C., Desroches, L.-B. et al. 2005, ApJ 632, 799.
%Multi-wavelength Monitoring of the Dwarf Seyfert 1 Galaxy NGC 4395. I. A Reverberation-based Measurement of the Black Hole Mass

\noindent Peterson, B.M. 2010 IAUS, 267, 151.
%Toward Precision Measurement of Central Black Hole Masses

\noindent Peterson, B.M. 2011 IAUS, 267, 151.
"Narrow-Line Seyfert 1 Galaxies and their place in the Universe". April 4-6,
2011. Milano, Italy. Editorial Board: Luigi Foschini (chair), Monica Colpi,
Luigi Gallo, Dirk Grupe, Stefanie Komossa, Karen Leighly, Smita
Mathur. Proceedings of Science (PoS, Trieste, Italy), vol. NLS1, Published online at http://pos.sissa.it/cgi-bin/reader/conf.cgi?confid=126, id.34"
% Masses of Black Holes in Active Galactic Nuclei : Implications for NLS1s

\noindent Pier,E.A. \& Voit, G.M. 1995, ApJ 450, 628.
%Photoevaporation of Dusty Clouds near Active Galactic Nuclei

\noindent Raban, D.J., Jaffe, W., Rottgering, Huub, et al. 2009, MNRAS 394, 1325.
%Resolving the obscuring torus in NGC 1068 with the power of infrared interferometry: revealing the inner funnel of dust

\noindent Ramos Almeida, C., et~al.\ 2009, ApJ, 702, 1127
%The Infrared Nuclear Emission of Seyfert Galaxies on Parsec Scales: Testing the Clumpy Torus Model

\noindent Richards, G.T., Kruczek, N.E., Gallagher, S.C. et al. 2011, AJ 141, 167. 
%Unification of Luminous Type 1 Quasars through C IV Emission

\noindent Robinson, A. 1995, MNRAS 272, 647.
% On the diversity of the broad emission-line profiles in active galactic
% nuclei

\noindent Robinson, A. 1995, MNRAS 276, 933.
% The profiles and response functions of broad emission-lines in AGN.

%not used
%\noindent Robinson, A., Axon, D.J., P\'{e}rez, E., and Vila-Vilar\'{o},
%B. 1996, in Vistas in Astronomy, Vol 40, pp77-81.

%not used
%\noindent Rokaki, E. Boisson, C. and Collin-Souffrin, S. 1992, A\&A 253, 57. 
%Fitting the broad line spectrum and UV continuum by accretion discs in active galactic nuclei

\noindent Schmitt, H.R. Antonucci, R.R.J. Ulvestad, J.S. et al. 2001, ApJ 555, 663.
%Testing the Unified Model with an Infrared-selected Sample of Seyfert Galaxies

\noindent Shakura, N. I., Sunyaev, R. A. 1973 A\&A 24, 337.
%Black holes in binary systems. Observational appearance.

%not used
%\noindent Shields, J.C., and Ferland, G.J. 1993, ApJ 402, 425.
%Implications of Lyman-alpha equivalent widths in active galactic nuclei

\noindent Shields, G.A., Ludwig, R.R., and Salviander, S. 2010, ApJ 721 1835.
%Fe II Emission in Active Galactic Nuclei: The Role of Total and Gas-Phase Iron Abundance

\noindent Sparke, L.S. 1993, ApJ 404, 570.
%Does the inner broad-line region DIM down when the power turns up?

\noindent Steinhardt, L.S. and Silverman, J.D. 2011, arXiv:1109.1554v1.
%Do Anomalous Narrow Line Quasars Cast Doubt on Virial Mass Estimation?

%not used
%\noindent Strateva, I.S., Strauss, M.A., Hao, L. t al. 2003, AJ 126, 1720.
%Double-peaked Low-Ionisation Emission Lines in Active Galactic Nuclei

%not used
%\noindent Stockton, A., and Farnham, T. 1991, ApJ 371, 525.
% binary black hole

\noindent Suganuma, M. Yoshii, Y., Kobayashi, Y. et al. 2004, ApJL 612, L113.
%The Reverberation Radius of the Central Dust Hole in NGC 5548

\noindent Suganuma, M. Yoshii, Y., Kobayashi, Y. et al. 2006, ApJ 639, 46.
%Reverberation Measurements of the Inner Radius of the Dust Torus in Nearby Seyfert 1 Galaxies

\noindent Sulentic, J.W., Zwitter, T. Marziani, P. and Dultzin-Hacyan D. 2000,
ApJ 536 L5.
%Eigenvector 1: An Optimal Correlation Space for Active Galactic Nuclei

\noindent Tanaka, Y., Nandra, K., Fabian, A.C. et~al. 1995, Nature 375, 659.
%Gravitationally redshifted emission implying an accretion disk and massive black hole in the active galaxy MCG-6-30-15

\noindent  Tristram, K.R.W., Meisenheimer, K., Jaffe, W. et al. 2007, A\&A
474, 837.
%Resolving the complex structure of the dust torus in the active nucleus of the Circinus galaxy

\noindent Tristram, K.R.W., Raban, D., Meisenheimer, K., Jaffe, W. et
al. 2009, A\&A 502, 67.
%Parsec-scale dust distributions in Seyfert galaxies. Results of the MIDI AGN snapshot survey

%not used
%\noindent Tsvetanov, Z. I.; Kriss, G. A.; Ford, H. C. 1996, in Vistas in Astronomy, Vol 40, pp71-75.  
% ionisation cones

\noindent Ulrich, M.-H. Boksenberg, A., Penston, M. ET AL. 1991, ApJ 382, 483.
%The ultraviolet spectrum of NGC 4151 from 1978 to 1990 - General characteristics and evolution

\noindent Ulrich, M-H. and Horne, K. 1996, MNRAS 283, 748.
%A month in the life of NGC 4151: velocity-delay maps of the broad-line region

\noindent Urry, C.M. and Padovani, P. 1995, PASP 107, 803.
%Unified Schemes for Radio-Loud Active Galactic Nuclei

%not used
%\noindent Walter, R., Orr, A., Courvoisier, T.J.-L. et al. 1994, A\&A 285, 119.
%Simultaneous observations of Seyfert 1 galaxies with IUE, ROSAT and GINGA

%not used
%\noindent Wandel, A., Peterson, B.M., and Malkan, M.A. 1999, ApJ 526, 579.
%Central Masses and Broad-Line Region Sizes of Active Galactic Nuclei. I. Comparing the Photoionisation and Reverberation Techniques

%not used
%\noindent Wanders, I. and Peterson, B.M. 1996, ApJ 466, 174.
%A Long-Term Study of Broad Emission Line Profile Variability in NGC 5548

\noindent Welsh, W.F., and Horne, K. 1991, ApJ 379, 586.
%Echo images of broad-line regions in active Galactic nuclei

\noindent Welsh, W.F. 1999, PASP 111, 1347.
%On the Reliability of Cross-Correlation Function Lag Determinations in Active Galactic Nuclei

%\noindent Wilson, A. in Vistas in Astronomy, Vol 40, pp63-70.
% ionisation cones

\noindent Wilson, A. S., Tsvetanov, Z. I., 1994 AJ 107, 1227.
%Ionisation cones and radio ejecta in active galaxies

\noindent Woo, J-H., Treu, T. Barth, A.J., et al. 2010, ApJ 716, 219.
% The lick AGN monitoring project: The MBH-sigma relation for
% reverberation mapped AGN

%not used
%\noindent Wu, S-M., Wang, T-G., and Dong, X-B., 2008, MNRAS 389, 213.
%Broad reprocessed Balmer emission from warped accretion discs

\noindent Yoshii, Y. Kobayashi, Y. and Minezaki, T. 2004, AN 325, 540.
%The dust distribution in the central region of AGNs: New results from the MAGNUM telescope

\noindent Zamfir, S., Sulentic, J.W., Marziani, P., and Dultzin, D. 2010,
MNRAS 403, 1759.
%Detailed characterisation of H$(B&B(B emission line profile in low-z SDSS quasars

\noindent Zhu, J, Zhou, X., Xue, S.J. et al. 2009, ApJ 700, 1173.
%Evidence for an Intermediate Line Region in Active Galactic Nuclei's Inner Torus Region and its Evolution from Narrow to Broad Line Seyfert I Galaxies

\appendix

\section{Response functions and emission-line profiles from our simulations}
\begin{figure}
%\resizebox{\hsize}{!}{\includegraphics[angle=0,width=14cm]{2d_fiducial.ps}}
%\resizebox{\hsize}{!}{\includegraphics[angle=0,width=14cm]{2d_fiducial_dark.ps}}
%\resizebox{\hsize}{!}{\includegraphics[angle=0,width=14cm]{2d_local_all.ps}}
\resizebox{\hsize}{!}{\includegraphics[angle=0,width=14cm]{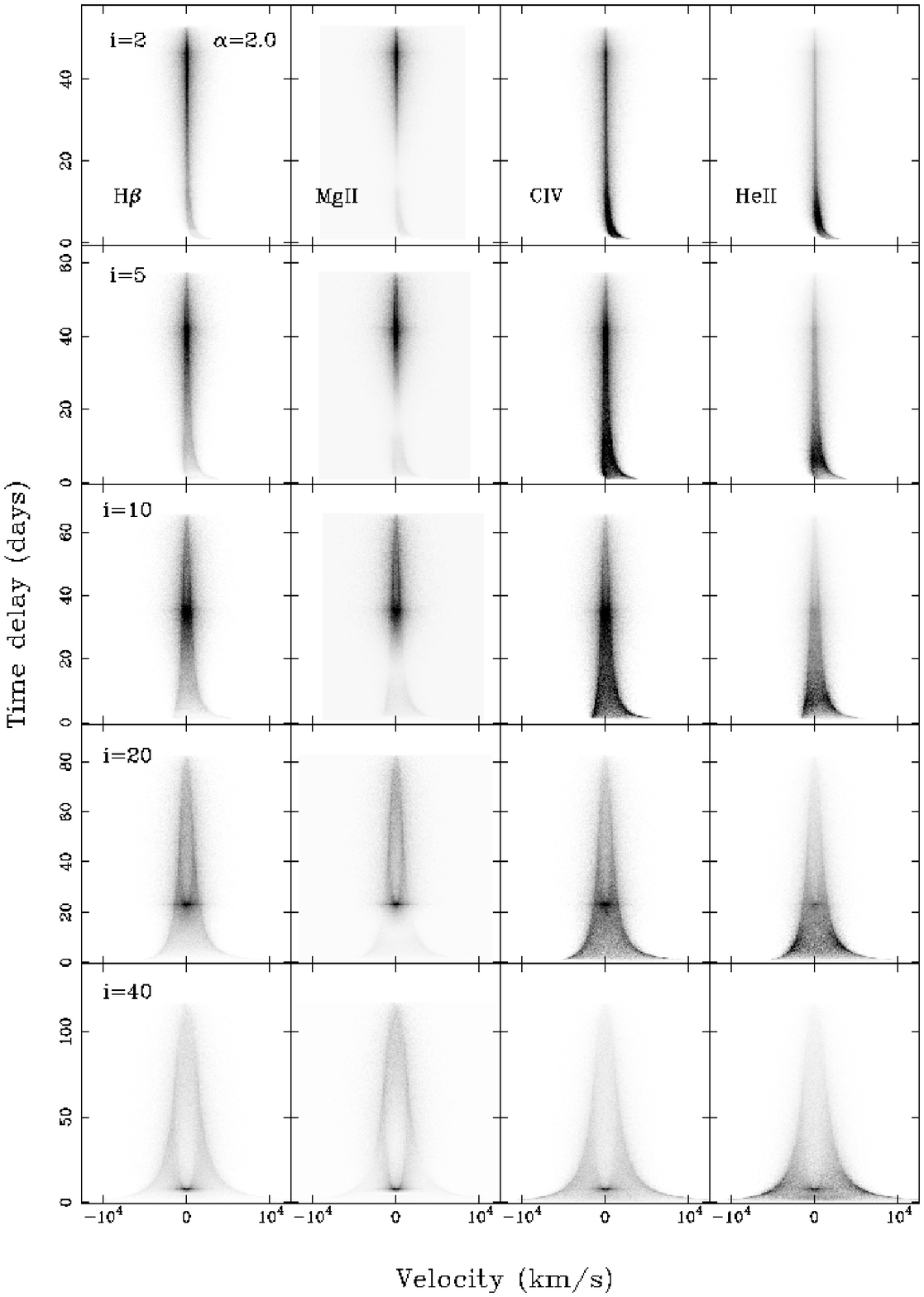}}
\caption{The responsivity-weighted 2-d response functions for our fiducial
  model for viewing angles in the range 2--40 degrees. TDS and GR are both
  included, and the turbulence parameter is set to $b_{\rm turb}=2$.}
\label{plot_2d_fiducial}
\end{figure}

\begin{figure}
%\resizebox{\hsize}{!}{\includegraphics[angle=0,width=14cm]{prof_merge.ps}}
\resizebox{\hsize}{!}{\includegraphics[angle=0,width=14cm]{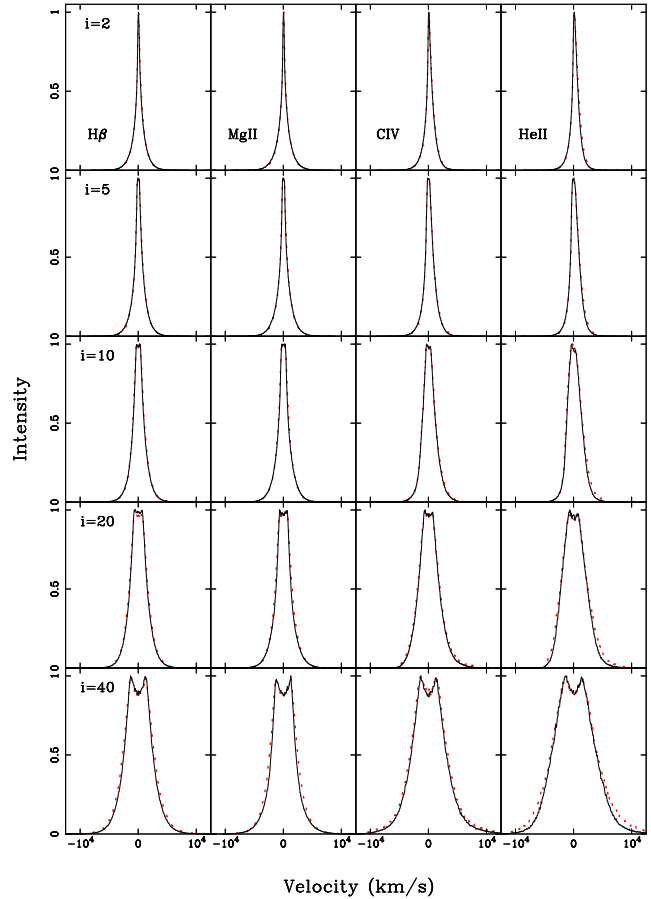}}
%\resizebox{\hsize}{!}{\includegraphics[angle=0,width=14cm]{prof_fiducial.ps}}
\caption{The corresponding variable emission-line profiles (solid black line)
  for figure~\ref{plot_2d_fiducial}. Shown in red is the equivalent variable
  line profile for emissivity-weighting only ($\eta(r)=constant$). All profiles have been normalised to their peak intensity to aid comparison.}
\label{plot_prof_fiducial}
\end{figure}

\begin{figure}
%\resizebox{\hsize}{!}{\includegraphics[angle=0,width=14cm]{1d_fiducial.ps}}
%\resizebox{\hsize}{!}{\includegraphics[angle=0,width=14cm]{resp_merge.ps}}
\resizebox{\hsize}{!}{\includegraphics[angle=0,width=14cm]{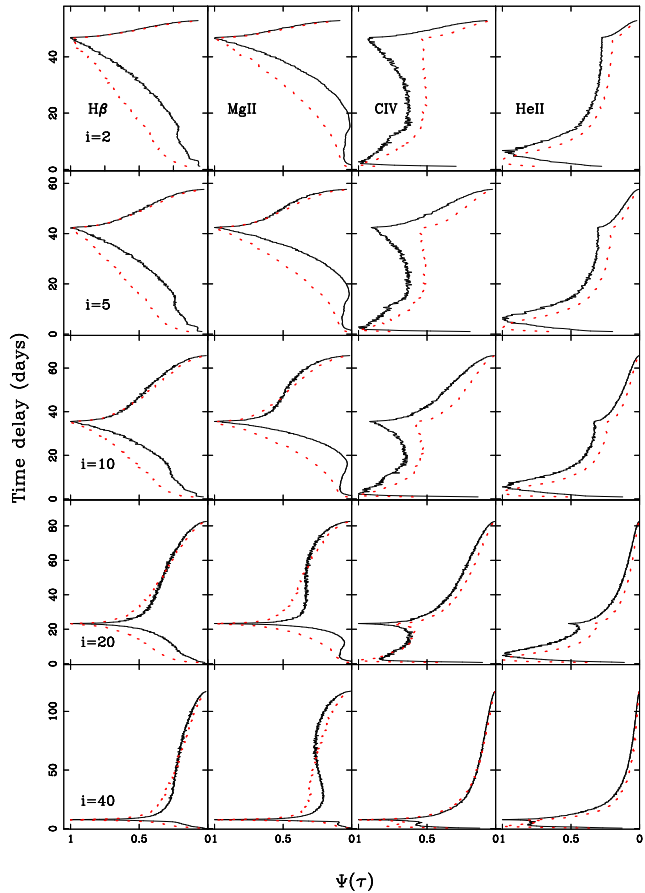}}
\caption{The corresponding 1-d responsivity-weighted response functions
  (solid black line) for figure~\ref{plot_2d_fiducial}. Shown in red is the
  equivalent emissivity-weighted (ie. $\eta(r)=constant$) 1-d response
  function. Response functions have been normalised to their peak
  intensities to aid comparison. Responsivity-weighting significantly
  modifies both the form and amplitude of the 1-d response.}
\label{plot_1d_fiducial}
\end{figure}

In figures~\ref{plot_2d_fiducial}--\ref{plot_1d_fiducial} we show the model
2-d and 1-d responsivity-weighted response functions and variable
emission-line profiles for our fiducial BLR geometry for each of the four
lines described in the text and line of sight viewing angles in the range
2--40~degrees. The models include the effects of TDS and GR and have been
calculated assuming a turbulence parameter $b_{\rm turb}=2$.

There are notable differences between the 2-d and 1-d responsivity-weighted
response functions and variable emission-line profiles for all of the
lines. Broadly speaking the lines fall into two categories with the dividing
line set by the scale height at which the line forms. H$\beta$ and Mg~{\sc ii}
are formed at large scale heights. Consequently at low inclinations they
display broader 2-d response functions at large time-delays and Lorentzian
profiles. The effects of TDS and GR are less prominent for these lines and the
line profiles appear more symmetric. The 2-d response functions at low
inclination indicate a large deficit in response at short time-delays. The
peak in the response function is located at the outer edge on the side of the
bowl nearest the observer, and moves to smaller delays as the inclination
increases. At small inclinations, the turbulence is large enough to fill in
the gap between the horns normally present in the emission-line profiles for
bowl-shaped geometries. At larger inclinations, the horns re-appear because of
their large separation in velocity space. Cranking up the turbulence would
once again remove the peaks from the emission-line profile.

C~{\sc iv} and He~{\sc ii} have steeper emissivity distributions and
consequently form at smaller BLR radii, which in our model implies small scale
heights. Since these lines originate in a more flattened distribution, the
effects of turbulence on the 2-d and 1-d response functions are less
significant.  The 2-d response functions therefore display a strong red-blue
asymmetry on short time-delays with an enhanced red-wing response.  Their
emission-line profiles are broader in comparison to H$\beta$ and Mg~{\sc ii}
at large inclinations, and the enhanced red-wing response is still evident
even at large inclinations. Large differences are also notable in the 1-d
response functions, where at large inclinations the response functions of
C~{\sc iv} and He~{\sc ii} show a strong resemblance to those found for
geometrically thin discs. Their small line formation radii give rise to
enhanced response at small time-delays when compared to those of H$\beta$ and
Mg~{\sc ii}.

\section{Anisotropic illumination}\label{continuum_shape}

A strong assumption of our model is that the gas on the bowl-surface is
illuminated by a continuum whose shape is independent of scale height. Yet, in
the TOR model of Kawaguchi and Mori (2010,2011) the bowl-shaped geometry is
formed because the disc emission is strongly anisotropic, thereby allowing
dust grains to form at much smaller radii at low elevations (near to mid-plane
of the disc). If an anisotropic continuum source is indeed responsible for
shaping the bowl geometry, then it will almost certainly have a strong effect
on the shape of the 1-d and 2-d responsivity-weighted response function and
variable emission-line profile. Here we look at 3 alternative models for the
disc continuum emission : strong, intermediate and weak anisotropy. For each
we model the broad band continuum as the sum of a varying UV component and a
constant X-ray component, the relative fractions of which are determined by
the dependence of the intensity of the UV component on the polar angle.

To represent the strong anisotropy dependence we adopt the disc-illumination
function first proposed by Netzer (1987), ie.

\begin{equation}
I(\phi) = \frac{I_{\rm uv}}{3}\cos \phi\left(1 + 2\cos \phi\right) + I_{\rm x} \, ,
\end{equation}

\noindent where $I_{uv}$ and $I_{x}$ are the relative intensities (or ionising
photon fluxes) of the UV and X-ray components, and $\phi$ is the polar angle.
The weak anisotropy function is taken from Nemmen and Brotherton (2010), for which we
assume
\begin{equation}
I(\phi) = I_{\rm uv}\sin\left[
  \left(\frac{\pi}{2}\times\frac{\phi_{0}}{\phi}\right)^{10} \right ]+ I_{\rm x} \, , 
\end{equation}

\noindent for polar angles $\phi \ge \phi_{0} = 72$ degrees, and

\begin{equation}
I(\phi) = I_{\rm uv} + I_{\rm x} \, , 
\end{equation}

\noindent otherwise.  Finally, we model an intermediate anisotropic
illumination function with a simple cosine dependence, ie.

\begin{equation}
I(\phi) = I_{\rm uv}\cos\phi+ I_{\rm x} \, .
\end{equation}

\begin{figure}
%\resizebox{\hsize}{!}{\includegraphics[angle=0,width=14cm]{illum_func.ps}}
\resizebox{\hsize}{!}{\includegraphics[angle=0,width=14cm]{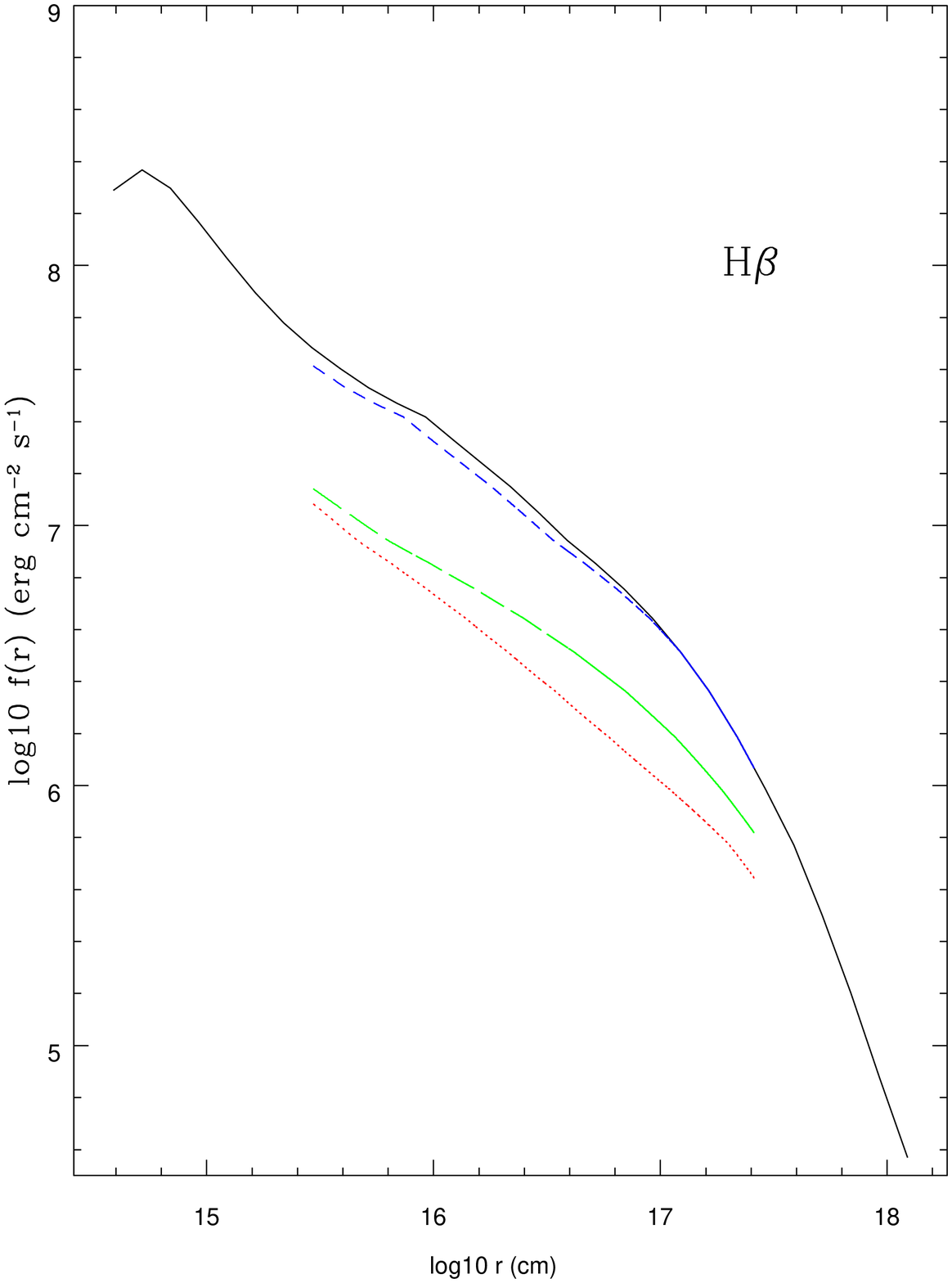}}
\caption{The radial surface line emissivity distributions for our disc
illumination functions: Isotropic continuum -- black solid line, Netzer's disc
illumination function -- red dotted line, a disc illumination pattern
approximating that of Nemmen and Brotherton (2010) -- blue dashed line, and a
disc illumination function with simple cosine dependence -- green long-dashed
line.}
\label{plot_illum_func}
\end{figure}

\begin{figure}
%\resizebox{\hsize}{!}{\includegraphics[angle=0,width=8cm]{plot_illum_ll.ps}}
\resizebox{\hsize}{!}{\includegraphics[angle=0,width=8cm]{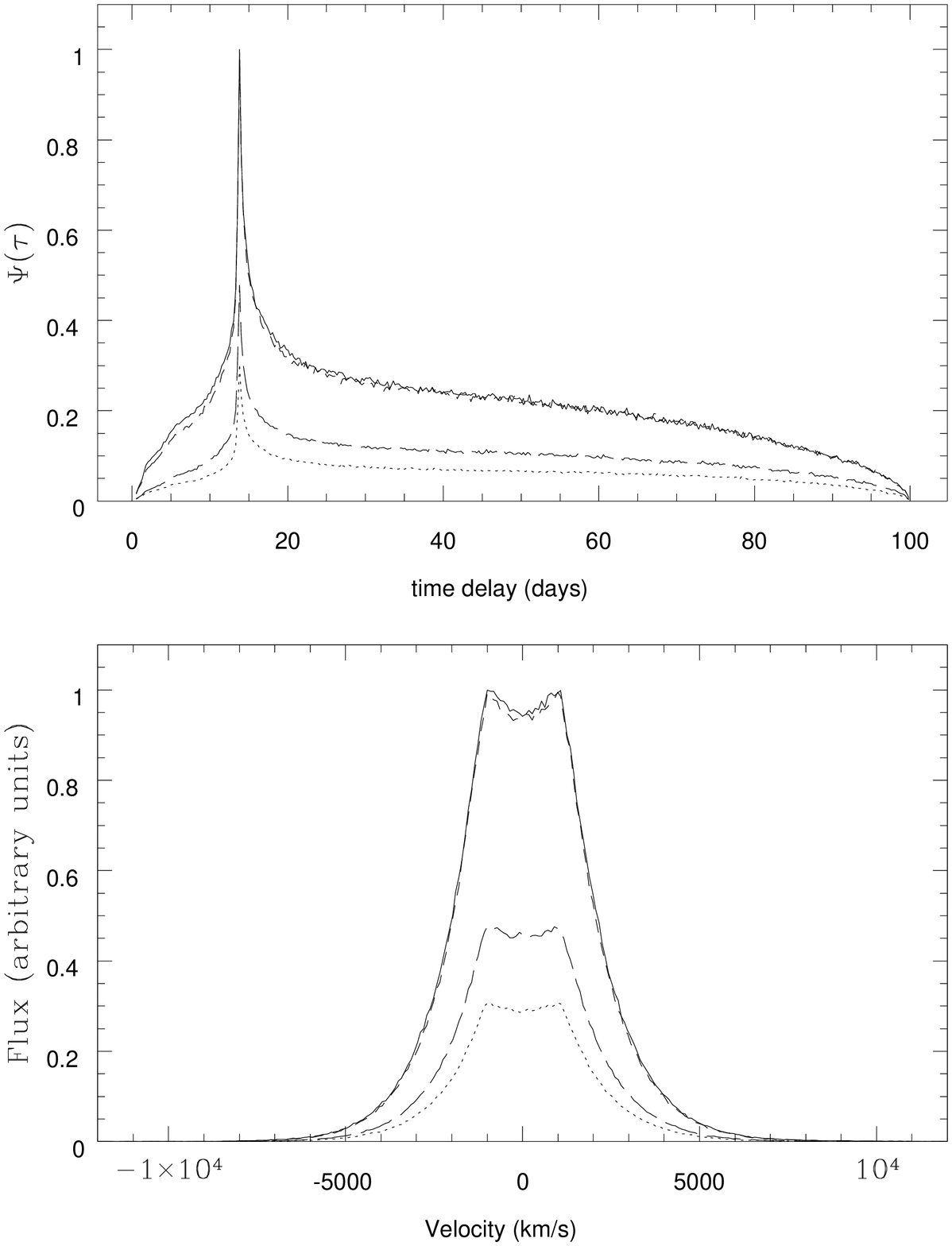}}
\caption{1-d responsivity-weighted response functions and variable
emission-line profiles for the disc illumination functions described in the
text for our fiducial bowl-shape BLR geometry, turbulence parameter
$b_{turb}=2$, and observed at inclination $i=30$~degrees.  Isotropic continuum
-- solid line, Netzer's disc illumination function -- dotted line, a disc
illumination pattern approximating that of Nemmen and Brotherton (2010) --
dashed line (nearly coincident with the solid line), and a disc illumination
function with a simple cosine dependence -- long-dashed line.}
\label{illum_resp}
\end{figure}

\noindent Note we do not calculate new photoionisation model grids for a broad
range of continuum shapes. Instead as a first approximation, we assume that we
can use our original emissivity grids (line flux as a function of radial
distance $R$ and hydrogen ionising photon flux $\Phi_{\rm H}$) as a look-up
table. That is we use the line flux at a given radial distance to find the
corresponding hydrogen ionising photon flux $\Phi_{H}$.  $\Phi_{\rm H}$ is
then modified according to equations B1--B4, and a new flux determined from
the same model grid.  In figure~\ref{plot_illum_func} we illustrate the
modified H$\beta$ radial emissivity distribution for each of the disc
illumination functions.  Figure~\ref{illum_resp} illustrates the 1-d
responsivity-weighted response functions (upper panel) and variable
emission-line profiles (lower panel) resulting from the modified radial
emissivity distributions as described above for our fiducial model and
$i=30$~degrees. For each model we assume that the UV continuum comprises 99\%
of the total ionising continuum intensity, ie. $I_{\rm uv} = 0.99 I_{\rm
totl}$.

\section{The driving continuum light-curve}\label{continuum_shape}

The X-ray light curves of AGN show correlated variability over a broad range
in timescales. This variability is normally quantified in terms of the
variability power, $P$, as a function of temporal frequency $\nu$, the power
spectral density (PSD) distribution. For AGN, the X-ray PSD is approximated by
a power-law in frequency ($P\propto \nu^{-\alpha}$) with slope $\alpha=1$ at
low frequencies, breaking to a slope of $2$ at the highest frequencies
(McHardy et~al. 2004) appropriate for a red-noise process.

At optical wavelengths, poor temporal sampling has often negated the use of
the PSD in determining the optical continuum variability. Instead, variability
is generally characterised in terms of the first order structure function
$S(\tau)$ (Collier and Peterson 2001), where
\begin{equation}
S(\tau)=\frac{1}{N(\tau)}\sum_{i<j}[f(t_{i})-f(t_{j})]^{2} \, ,
\end{equation}

\noindent where $f$($t_{i}$) is the flux measured at time $t_{i}$, $N(\tau)$ is
the number of pairs of points, and the sum is over all pairs for which
$\tau=t_{j}-t_{i}$. Paltani (1999) showed that the first order structure
function is related to the one-sided power density spectrum $P(f)$ via:

\begin{equation}
S(\tau)=2\left[\int_{0}^{\infty} P(f)df-\int_{0}^{\infty}\cos(2\pi f \tau)df
  \right] \, .
\end{equation}

\noindent The form of the structure function can also be approximated
by a power-law on intermediate timescales, breaking to a flatter slope
on both short timescales ($\tau_{min}$), where its value approaches
twice the noise variance ($\sigma_{n}^{2}$), and on long-timescales
($\tau_{max}$), where its value approaches twice the signal variance
($\sigma_{var}^{2}$).

Collier and Peterson (2001) performed a structure function analysis of the UV
and optical light-curves of a small sample (13) of AGN, including NGC~5548,
observed as part of the 'AGN Watch' and OSU AGN monitoring programmes. They
showed that the UV and optical SFs are similar on timescales of 5--60~days. In
particular, for NGC~5548, the UV and optical power-law slope $b$ of the SF on
intermediate timescales is approximately 1.5, and flattens on a characteristic
timescale $\tau_{char}=40$~days. For stationary time series, the power-law
slope of the structure function $b$ is related to the power-law slope of the
PSD $\alpha$, by $\alpha=b+1$. Thus, for NGC~5548, the slope of the UV and
optical SF suggests $\alpha \approx 2.5$.  However, we caution that
non-stationary effects can steepen the derived power-law slope while
deviations from a strictly power-law slope may be introduced by bias resulting
from binning and irregular sampling of the data (Collier and Peterson
2001). For a stationary random process $S(\tau)$ is simply related to the
autocorrelation function $ACF(\tau$), such that
\begin{figure}
%\resizebox{\hsize}{!}{\includegraphics[angle=0,width=8cm]{plot_sf_feb2012.ps}}
\resizebox{\hsize}{!}{\includegraphics[angle=0,width=8cm]{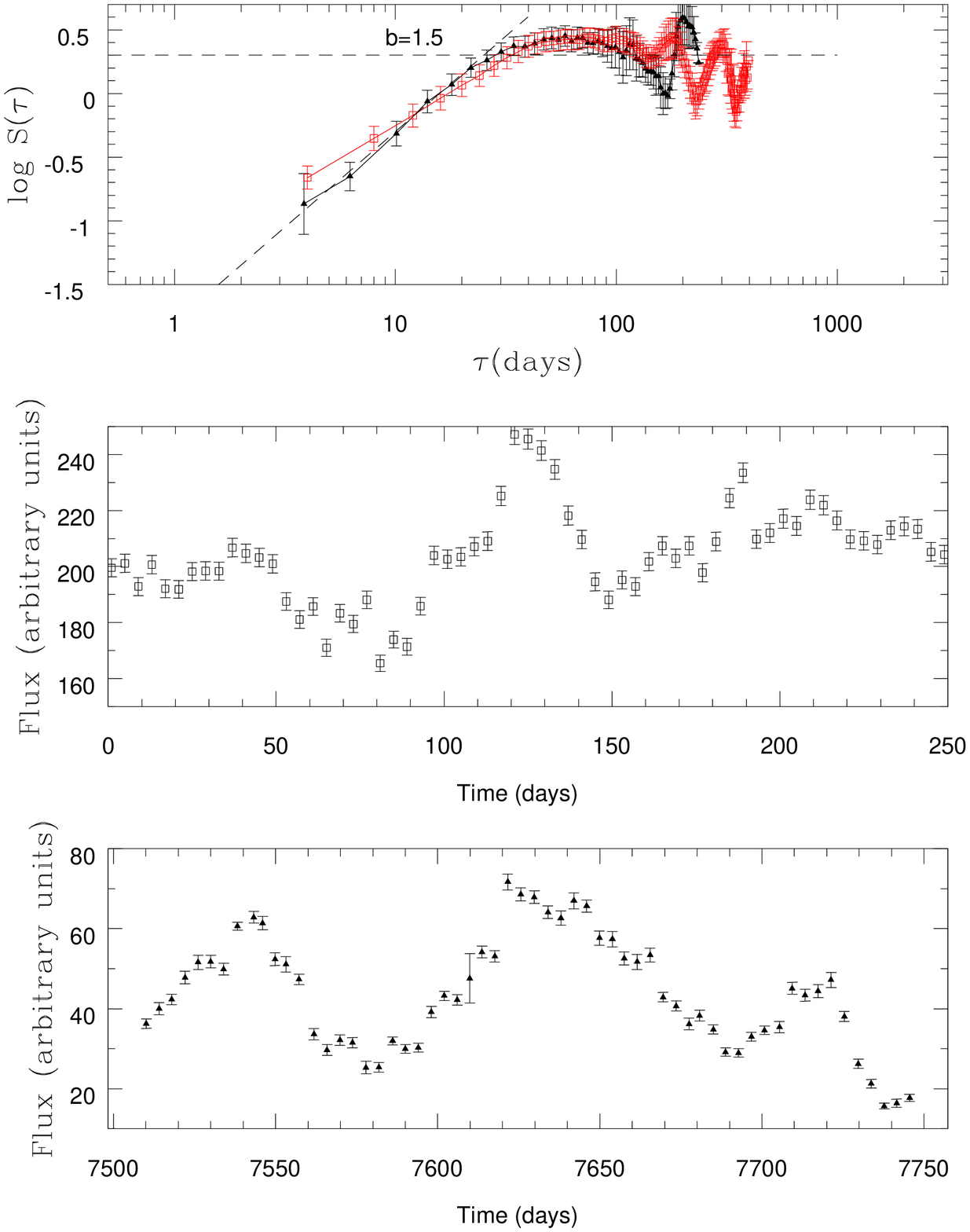}}
\caption{Upper panel -- a typical structure function for one of our simulated
  light-curves (open squares). Also shown is the structure function derived
  for the GEX extracted 1337\AA continuum from the 1989 IUE monitoring
  campaign of NGC~5548. Middle-panel -- the corresponding simulated light-curve
  sampled at 4 day intervals. Lower-panel -- the 1337\AA continuum light-curve
from the 1989 monitoring campaign of NGC~5548.}
\label{plot_sf}
\end{figure}

\begin{equation}
S(\tau) =  2 \left[\sigma^{2}-ACF(\tau)\right] \,\ .
\end{equation}

\noindent Thus the form of the SF can vary from one observing season to the
next even when the process responsible for the variability remains the
same. Similar critiques have been made concerning the use of structure
functions in quantifying Blazar variability (Emmanoulopoulos et al. 2010), and
in particular when assigning characteristic timescales to features observed in
the SF on long timescales where the SF is poorly defined.

In figure~\ref{plot_sf} (top panel) we indicate a typical structure function
calculated from one of our model light-curves (open squares). Also shown is
the structure function derived from the GEX extracted 1337\AA\ continuum
light-curve (filled triangles) of NGC~5548 taken from the 1989 IUE monitoring
campaign (Clavel et~al. 1991). The dashed diagonal line indicates the slope of
the structure function over the region of interest. In the middle panel we show the simulated
light-curve sampled at 4 day intervals (to match the sampling of the
light-curve in the IUE campaign).  The lower panel, shows the continuum
light-curve from the 1989 IUE monitoring campaign.

\label{lastpage}

\end{document}